\documentclass[12pt, preprint]{aastex}
\usepackage{epsfig}

\shorttitle{Survey of Be Star Disks}
\shortauthors{Touhami et al.}

\newcommand{\noprint}[1]{}

\begin{document}

\accepted{}
\title{A CHARA Array Survey of Circumstellar Disks around\\
          Nearby Be-type Stars}

\author{
Y. Touhami\altaffilmark{1}, D. R. Gies\altaffilmark{1}, G. H. Schaefer\altaffilmark{2},
H. A. McAlister\altaffilmark{1}, S. T. Ridgway\altaffilmark{3}, N. D. Richardson\altaffilmark{4},
R. Matson\altaffilmark{1}, E. D. Grundstrom\altaffilmark{5}, T. A. ten Brummelaar\altaffilmark{2},
P. J. Goldfinger\altaffilmark{2}, L. Sturmann\altaffilmark{2}, J. Sturmann\altaffilmark{2},
N. H. Turner\altaffilmark{2}, C. Farrington\altaffilmark{2}
}

\email{yamina@chara.gsu.edu,
gies@chara.gsu.edu, schaefer@chara-array.org, hal@chara.gsu.edu, ridgway@noao.edu,
richardson@astro.umontreal.ca, rmatson@chara.gsu.edu, erika.grundstrom@vanderbilt.edu,
theo@chara-array.org, pj@chara-array.org, sturmann@chara-array.org, judit@chara-array.org,
nils@chara-array.org, farrington@chara-array.org
}

\altaffiltext{1}{Center for High Angular Resolution Astronomy, Georgia State University, P.O. Box 3969,
Atlanta, GA 30302-3969, USA}
\altaffiltext{2}{The CHARA Array, Mount Wilson Observatory, Mount Wilson, CA 91023, USA}
\altaffiltext{3}{National Optical Astronomy Observatory, P.O. Box 26732, Tucson, AZ 85726-6732, USA}
\altaffiltext{4}{D\'epartement de physique and Centre de
Recherche en Astrophysique du Qu\'ebec (CRAQ), Universit\'e de Montreal,
CP 6128 Succ. A., Centre-Ville, Montr\'eal, Qu\'ebec H3C 3J7, Canada}
\altaffiltext{5}{Physics and Astronomy Department,
Vanderbilt University, 6301 Stevenson Center, Nashville, TN 37235, USA}

\slugcomment{Accepted to ApJ}

\begin{abstract}
We report on a high angular resolution survey of
circumstellar disks around 24 northern sky Be stars.
The $K$-band continuum survey was made using the CHARA Array
long baseline interferometer (baselines of 30 -- 331 m).
The interferometric visibilities were corrected for the flux
contribution of stellar companions in those cases where the Be star
is a member of a known binary or multiple system.
For those targets with good ($u, v$) coverage, we used a four-parameter
Gaussian elliptical disk model to fit the visibilities
and to determine the axial ratio, position angle,
$K$-band photospheric flux contribution,
and angular diameter of the disk's major axis. For the other targets
with relatively limited ($u, v$) coverage, we constrained the axial ratio,
inclination angle, and/or disk position angle where necessary in order to
resolve the degeneracy between possible model solutions.
We also made fits of the ultraviolet and infrared spectral energy
distributions to estimate the stellar angular diameter and infrared
flux excess of each target.  The mean ratio
of the disk diameter (measured in $K$-band emission) to stellar diameter (from SED
modeling) is 4.4 among the 14 cases where we reliably
resolved the disk emission, a value which is generally lower
than the disk size ratio measured in the higher opacity H$\alpha$ emission line.
We estimated the equatorial rotational velocity from the projected
rotational velocity and disk inclination for 12 stars, and most of
these stars rotate close to or at the critical rotational velocity.
\end{abstract}


\setcounter{footnote}{5}


\section{Introduction}                              

Classical Be stars are non-supergiant, B-type stars that
are surrounded by hot gaseous disks. This circumstellar gas is responsible for
many observational characteristics such as hydrogen Balmer emission
lines, IR flux excess, and short- and long-term flux variability \citep{por03}.
Optical and infrared interferometry has become an important tool in
characterizing Be stars and their disks \citep{ste11}.
The first interferometric survey of Be stars was made by \citet{qui97} to
resolve the H$\alpha$ emission in seven Be stars.
Their survey showed that the emitting regions are flattened,
which is strong observational evidence of a disk-like geometry.
\citet{qui97} combined optical interferometry and spectropolarimetry
to derive the disk position angle on the sky, and they
found good agreement between these techniques.
\citet{tyc04,tyc05,tyc06,tyc08} used the
Navy Precision Optical Interferometer (NPOI) to observe the
H$\alpha$ emission from the disks of seven Be stars.
Their observations showed that a direct correlation exists between
the disk sizes and the net H$\alpha$ luminosities.

Infrared observations have begun to reveal the spatial properties
of the continuum and line emission of Be star disks.
\citet{gie07} made the first CHARA Array long-baseline
interferometric observations in the $K$-band of four bright Be stars,
$\gamma$~Cas, $\phi$~Per, $\zeta$~Tau, and $\kappa$~Dra,
and they were able to resolve the disks and to
constrain their geometrical and physical properties.
\citet{mei07b} studied the geometry and kinematics of the Be star $\kappa$~CMa
in the Br$\gamma$ emission line and in the nearby continuum using the VLTI/AMBER instrument.
\citet{mei11} observed the Be binary system $\delta$~Sco
using spectrally-resolved interferometry with the VLTI/AMBER and CHARA/VEGA instruments.
Their observations show that the disk varies in size from 4.8 mas in H$\alpha$,
to 2.9 mas in Br$\gamma$, and to 2.4 mas in the $K$-band continuum.
\citet{mei12} completed a survey of eight Be stars with VLTI/AMBER and
measured the disk extensions in the Br$\gamma$ line and the nearby continuum.
Their study suggests that the disk kinematics are dominated
by Keplerian rotation and that the central stars have a mean ratio
of angular rotational to critical velocity of $\Omega_{\rm rot}/\Omega_{\rm crit} = 0.95$.
In addition, \citet{mei09} used the VLTI/MIDI instrument to determine the
$N$-band (10 $\mu$m) disk angular size for seven Be stars.

Interferometry offers us the means to explore Be star disks in large
numbers and to begin to understand their properties as a whole.
Here we present results from such a survey that we conducted
in the $K$-band continuum using the CHARA Array
long-baseline interferometer. In Section 2, we list our sample stars,
present our observational data sets, and describe the data reduction process.
In Section 3, we describe a method that we implemented to correct
the interferometric measurements for the flux of stellar companions.
We discuss in Section 4 the spectral energy distributions and their use
in estimating the stellar angular diameter and infrared excesses of Be stars.
In Section 5, we present fits of the interferometric visibilities using
simple geometrical models, and in Section 6, we discuss the results with
a particular comparison of the $K$-band and H$\alpha$ disk sizes.
Finally, we summarize our results and draw our conclusions in Section 7.


\section{Observations and Reduction}                           


We selected 24 Be stars as targets for this
interferometric survey. The main selection criteria were that the stars are
nearby and bright, well within the limiting magnitude of the CHARA Classic
tip-tilt servo system ($V <$ 11) and the near-IR fringe
detector ($K <$ 8.5). The selected Be stars had to have
declinations north of about $-$15$^{\circ}$ to be accessible
with the interferometer at low air-mass values.
Furthermore, most of the targets have recently shown
hydrogen emission and a near-IR flux excess.  We relied
particularly on spectrophotometric and H$\alpha$ observations conducted
by \citet{tyc06}, \citet{gru07}, \citet{gie07}, and \citet{tou10}.
The targets and their adopted stellar parameters are
presented in Table~\ref{param}. Columns 1 and 2 list the star names,
columns 3 to 5 list the spectral classification from the compilation by \citet{yud01}
and the stellar effective temperature $T_{\rm eff}$ and gravity $\log g$ from
\citet{fre05} (see their Table 9 ``Apparent parameters'').
The stars HD~166014 and HD~202904 are not listed by \citet{fre05},
so we used the parameters for these two from \citet{gru07}.
Columns 6 and 7 list predictions for the position angle $PA$ of
the projected major axis of the disk that should be $90^\circ$ different
from the intrinsic polarization angle \citep{mcd99,yud01} and for
$r$, the ratio of the minor to major axis sizes according to the estimated
stellar inclination from \citet{fre05}.

\placetable{param}      


Measuring the instrumental transfer function of the CHARA Array
interferometer is performed by observing calibrator
stars with known angular sizes before and after each target
observation. The calibrator stars are selected to be close to
the targets in the sky, unresolved with the interferometer's largest baseline,
and without known spectroscopic or visual binary companions.
We collected photometric data on each calibrator star in order to construct
their spectral energy distribution (SED) and to determine
their angular diameter. The collected $UBVRIJHK$ photometry
(available from \citealt{tou12}) is transformed into
calibrated flux measurements using procedures described by \citet{col96} and
\citet{coh03}.  The stellar effective temperature $T_{\rm eff}$ and
the gravity $\log g$ (generally from the compilation of \citealt{sou10})
are used to produce a model
flux distribution that is based on Kurucz stellar atmosphere models.
Note that we generally used Johnson $U$ magnitudes compiled by \citet{kar06}
and $B,V$ magnitudes from \citet{amm06}, who list Tycho $B$ and $V$ magnitudes
that are slightly different from Johnson $B$, $V$ magnitudes.
The photographic $R$ and $I$ magnitudes were collected from \citet{mon03},
which are only slightly different from Johnson $R$, $I$ magnitudes.
The near-IR $J,H,K$ photometry was collected from the 2MASS survey \citep{skr06}.
The use of non-standard photometry introduces errors in
the best-fit, limb-darkened angular diameter of the calibrators
that are comparable to or smaller than the estimated uncertainties
given in Table~2.  We also collected measurements of $E(B-V)$
and applied a reddening correction to the model SED before
fitting the SED of the calibrators.
We then computed a limb-darkened angular diameter $\theta_{LD}$ by direct comparison of
the observed and model flux distributions.  Based upon the limb-darkening coefficients
given by \citet{cla00}, we transformed the limb-darkened angular
diameter to an equivalent uniform-disk angular diameter $\theta_{UD}$ assuming
a projected baseline of 300~m.

Columns 1 and 2 of Table~\ref{calib}
list the calibrator star and its corresponding target, respectively,
columns 3 and 4 list the calibrator effective temperature $T_{\rm eff}$
and reference source, columns 5 and 6 give the surface
gravity $\log g$ and reference source, column 7 gives the spectral classification,
and columns 8 and 9 list the adopted interstellar reddening $E(B-V)$
and the reference source, respectively. Column 10 of Table~\ref{calib} lists the best-fit
limb-darkened angular diameter $\theta_{LD}$ derived from fitting
the calibrator SED,  and column 11 lists the
corresponding uniform-disk angular diameter $\theta_{UD}$.


The observations were conducted
between 2007 October and 2010 November using the
CHARA Classic beam combiner operating in the $K$-band
(effective wavelength = 2.1329~$\mu$m; \citealt{ten05}).
We need a minimum of two interferometer baselines at
substantially different projected angles on the sky
to map the circumstellar disks around our targets, and we generally used
the South-West baseline of length $\sim$ 278~m oriented at 39$^{\circ}$ west
of north (S1/W1) and the South-East baseline of length $\sim$ 330~m
oriented at 22$^{\circ}$ east of north (S1/E1). Each target and its
calibrator were observed throughout a given night in series of 200
scans recorded with a near-IR detector on a single
pixel at a frequency that ranged between 500 - 750 Hz, depending on the seeing conditions
of each particular night of observation.
The interferometric raw visibilities were then estimated by performing an
integration of the fringe power spectrum.  We used the CHARA
Data Reduction Software (ReduceIR; \citealt{ten05})
to extract and calibrate the target and the calibrator interferometric
visibilities.  Then the raw visibilities were calibrated by
comparing them to the time-interpolated calibrator visibilities
and rescaling them according to the predicted calibrator visibility
for the given projected baseline and stellar angular diameter.
The resulting calibrated visibilities are listed in Table~\ref{calibvis}
(given in full in the electronic version).
Column 1 of Table~\ref{calibvis} lists target HD number,
column 2 lists the heliocentric Julian date of mid-observation,
column 3 lists the telescope pair used in each observation,
columns 4 and 5 list the $u$ and $v$ frequencies, respectively,
column 6 lists the projected baseline,
column 7 lists the effective baseline (see \S~5.2),
columns 8 and 9 list the calibrated visibility and
its corresponding uncertainty, respectively, and lastly,
columns 10 and 11 list the visibility measurements and uncertainties
corrected for the flux of stellar companions for those cases with known
binary parameters.  We will discuss this correction in detail in the next section.

\placetable{calibvis}      

The internal uncertainties from fitting individual fringes are generally
smaller than $<5\%$. The scatter in the
data depends mostly on the target magnitude and seeing
conditions at the time of the observations, which usually varies
with a Fried parameter in the range $r_0 \simeq 2.5 - 14$~cm.
The visibility uncertainties for the brightest
targets range between 2$\%$ and 5$\%$, while those for
the faintest ones reach up to 8$\%$.

Overall we obtained a relatively good set of observations at different
hour angles for each star in our sample, with the
exceptions of HD~58715 and HD~148184 where the position angle coverage
was limited to only one projected baseline.  Figure Set 1 shows the
distribution of the observations in the $(u,v)$ plane for our sample.

\placefigure{uv1}     
\placefigure{uv2}     
\placefigure{uv3}     


\section{Correction for the Flux of Nearby Companions}             

\subsection{Influence of Binary Companions on Interferometric Measurements} 

Our measurements of the sizes and orientations of Be star disks are based upon
the observed decline in visibility caused by the extended angular distribution
of disk flux.  However, a drop in visibility can also occur if a stellar
companion is within the effective field of view of the interferometer,
and if the binary component is ignored, then the disk dimensions will be
overestimated (in the extreme case apparently implying the presence of
a large disk where none is present).  Binary systems are relatively common
among B-stars in general and Be stars in particular \citep{abt87,mas97,gie01},
so it is important to investigate what role they play in the interpretation
of our interferometric measurements.

Fortunately, the signatures of binary companions in interferometric observations
are well understood \citep{her71,arm92,dyc95,bod00}, and given the projected separation
and magnitude difference we can determine how the companion will affect our measurement.
According to the van Cittert-Zernicke theorem, the complex visibility
is related to the Fourier transform of the angular spatial distribution in the sky,
and the measured fringe amplitude is proportional to the real part of the
visibility \citep{dyc95,bod00}.  In the ideal case of a binary star consisting
of two point sources, the visibility varies according to a cosinusoidal term with a
frequency that depends on the projected separation and the baseline and wavelength
of observation.  The interferometric fringe observed will display an amplitude
that depends on the real part of the visibility according to the projected separation
and binary flux ratio.  For observations like ours that record the flux over a wide
filter band, the fringe pattern is only seen over a range in optical path delay
that is related to the coherence length.   Binaries with separations smaller than
the coherence length will display the full amplitude variation expected from
the visibility dependence on binary separation and flux ratio \citep{bod00,rag09},
while those with separations larger than the coherence length will appear as
separated fringe packets \citep{dyc95,obr11,rag12}.

Thus, there are several separation ranges that are key to this discussion.  First, we may
ignore those binary companions that are outside of the field of view of the interferometer
($0.8 \times 0.8$ arcsec; \S 3.6) and separated by more than atmospheric seeing disk.
Second, there are binary companions that are effectively within the field of view, but
whose projected separations are large enough that the fringe packets of the components
do not overlap ($\gtrsim 9$ mas for an observation with a 300~m baseline).
This is by far the most common case for our observations, and indeed the fringe packet
of the companion is usually located far beyond the recorded scan.
In this situation the flux of the companion will act to dilute the measured visibility
of the fringe but will not change its morphology (\S 3.3).
Third, there are binaries that have such small projected separations
($\lesssim 9$ mas for an observation with a 300~m baseline)
that their fringe packets overlap and create an oscillatory pattern in the
observed visibility due to the interference between fringe packets (\S 3.3).
This probably occurred in only a few cases among our observations (\S 3.7).

Ideally, we should model the fringe visibility in terms of the binary
projected separation and flux ratio together with a parametrization
of the Be disk properties (\S 5.1).  However, because this is a
survey program, our observational results are generally too sparse in
coverage of baseline and position angle range to attempt such a general solution.
For example, most of the known companions have angular separations that are
relatively large, and the fringe packet of a companion would have only been
recorded in short baseline observations.  However, such short baseline data
would not resolve the Be disks, which is our primary scientific goal.
There are a few cases where the projected separations are small and the
binary creates oscillations in visibility with projected baseline,
but again our baseline coverage is generally too sparse to measure these
together with the disk properties.  Consequently, we made the decision to rely
solely on published data on the binary companions of our targets (\S 3.2) in order
to perform the corrections to the measured visibility where it was necessary
to do so (i.e., the known companion was sufficiently bright and close enough
to influence our visibility measurements).

The visibility correction procedure we adopted is outlined in the following
subsections.  The method is built upon the scheme described by \citet{dyc95}
and considers how a binary companion influences the appearance of the
combined fringe packets.  We use estimates of the flux ratio (\S 3.5, 3.6)
and the orbital projected separation (\S 3.2) for each observation to build a
numerical relation between the Be star visibility and net observed visibility.
Then we interpolate within these relations to determine the Be star visibility
alone (\S 3.3).  We also extend this approach to correct the visibilities for
two multiple star systems (\S 3.4).

\subsection{Binary Stars in the Sample} 

Many Be stars in our sample are known binaries or multiple systems.
We checked for evidence of the presence of companions
through a literature search with frequent
consultation of the Washington Double Star Catalog (\citealt{mas01}),
the Fourth Catalog of Interferometric Measurements of Binary Stars
(\citealt{har01}), and the Third Photometric Magnitude Difference
Catalog\footnote{http://www.usno.navy.mil/USNO/astrometry/optical-IR-prod/wds/dm3}.
We only considered those companions close enough to influence the
interferometry results (i.e., those with separations less than a few arcsec).
We show the binary search results in Table~\ref{binary}, which includes visual binaries
(typical periods $> 1$~y) and spectroscopic binaries (typical periods $< 1$~y).
Columns of Table~\ref{binary} list the star name, number of components, reference code
for speckle interferometric observations, and then in each row,
the component designation, orbital period, angular semimajor axis,
the estimated $K$-band magnitude difference between the components
$\triangle K$, a Y or N for whether or not a correction for the flux
of companions was applied to the data, and a reference code
for investigations on each system. Entries appended with a semi-colon
in Table~\ref{binary} indicate parameter values with large
uncertainties. These include the single-lined spectroscopic
binaries, where we simply assumed a primary mass from the spectral
classification and $1 M_\odot$ for the companion to
derive the semimajor axis $a$, which was transformed
to an angular semimajor axis using the distance from
{\it Hipparcos} \citep{van07}.

\placetable{binary}      

We find that ten of the 24 Be stars in our sample have
no known companion, and five others have companions that
are too faint (all single-lined spectroscopic binaries)
to influence the interferometric measurements.
Thus, no corrected visibilities are listed for these 15 targets in Table~3.
However, the companions were bright enough ($\triangle K < 3.2$) and
close enough (separation $< 1\arcsec$) for the remaining nine targets that
we had to implement the corrections method outlined below.

We need to determine the companion's separation and position
angle at the time of each observation in order to find
the angular separation projected along the baseline we used.
This was done by calculating the binary relative separation
for the time of observation from the astrometric orbital
parameters using the method outlined by \citet{rag09}.
We show in Table~\ref{orbital} the adopted orbital parameters
for those binaries where we made visibility corrections.
Note that no entry is present for HD~166014, where only
one published measurement exists (see Appendix), and
we simply assumed that the projected separation was larger
than the recorded scan length.
There are four cases where we present new orbital elements
that are based upon published astrometric measurements, and
we caution that these are preliminary and used only to
estimate the separations and position angles at the times
of the CHARA Array observations.  Details about these preliminary
fits are given in the Appendix.

\placetable{orbital}

\subsection{Fringe Visibility for Be Stars in Binaries} 

The changes in visibility caused by a binary companion can be
described equivalently in terms of the real part of the complex
visibility or the amplitude of interfering fringe packets
\citep{dyc95,bod00}.  Here we develop a fringe packet approach
to the problem that simulates the binary changes and that
is directly applicable to the fringe amplitudes that we measure.
We begin by considering how the fringe patterns of binaries
overlap in order to assess the changes in visibility caused
by a binary consisting of two unresolved stars,
and then we extend the analysis to the situation where one star
(Be plus disk) is partially resolved.  The fringe packet for star $i$
observed in an interferometric scan of changing optical path length
has the form
\begin{eqnarray}
F_{i} = {{\sin x} \over x} \cos(2\pi \frac{a}{\lambda} + \phi)
\end{eqnarray}
where $x=\pi a / \wedge_{coh}$,
$a$ is the scan position relative to the center of the fringe,
$\wedge_{coh}$ is the coherence length given by $\lambda^2 /\delta \lambda$
(equal to  13 $\mu$m for the CHARA Classic $K^\prime$ filter),
and $\phi$ is a phase shift introduced by atmospheric fluctuations
\citep{obr11}.  If two stars are present in the field of view (\S 3.6)
with a projected separation along the scan vector of $x_2$, then
their fringe patterns may overlap and change the composite appearance
of the fringe according to \citep{bod00}
\begin{equation}
F_{tot} = \frac{1}{1 + f_2/f_1} F_1 + \frac{f_2/f_1}{1 + f_2/f_1} F_2
\end{equation}
where $f_1$ and $f_2$ are the monochromatic fluxes of the stars.
Then the fringe visibility is calculated by
\begin{equation}
V=\frac{\max (F_{tot})-\min (F_{tot})}{2 + \max (F_{tot}) + \min (F_{tot})} .
\end{equation}
In this discussion we will ignore the very small decline in fringe amplitude
related to the angular diameters of the stars themselves, because their
angular diameters are all very small (see Table 6 below).

We show a series of such combined
fringe patterns in the panels of Figure~\ref{app1}
for an assumed flux ratio of $f_2/f_1 = 0.5$.
In the top panel, the projected separation is zero, and the two
patterns add to make the fringe pattern of a single unresolved star.
The associated visibility that we would measure equals one in this case.
However, in the second panel from the top,
we show how a projected separation of 1 $\mu$m results in a much
lower visibility because the peaks associated with star 1 are
largely eliminated by the troughs associated with star 2.
In the third panel from the top, the separation is just large enough
(comparable to the coherence length) that the fringe pattern of the
companion emerges from the blend, and the lower panel shows
a separated fringe packet in which both fringe patterns are
clearly visible.  Note that the relationship between the projected angular separation
(mas = milli-arcsecond) and the separation of fringe centers ($\mu$m) is given by
\begin{equation}
\rho_{\rm mas} = \frac{206.265 \rho_{\rm \mu m}} {B}
\end{equation}
where $B$ is the projected baseline in meters. For instance,
using a 330~m baseline, the longest scan length is 150 $\mu$m,
which corresponds to a separation of $\sim$ 94 mas.

\placefigure{app1}

We show in Figure~\ref{app2}
the net visibility that would be measured as a function of
projected separation $x_2$.  This shows that in general
the observed visibility will be less than that of
a single star.  As expected, for very close separations
the visibility varies cosinusoidally with $x_2$ with a
frequency of $2\pi / \lambda$.  On the other hand, for projected
separations larger than the coherence length $\wedge_{coh}$,
the visibility approaches the value $1/(1 + f_2/f_1)$ equal
to the semiamplitude of the flux diluted fringe pattern of the target.

\placefigure{app2}

Now suppose that star 1 is a Be star with a disk that is
partially resolved, so that if it were observed alone,
it would show a visibility $V=V_c < 1$. Consequently,
its fringe pattern would have an semiamplitude given by
$V_c/(1 + f_2/f_1)$.  We show a selection of
model binary fringe patterns in Figure~\ref{app3}
again for $f_2/f_1 = 0.5$ and a specific separation of
$x_2 = 10$ $\mu$m.  The panels show from top to bottom
the progressive appearance of the combined fringe patterns
as $V_c$ drops from 1 to 0.25. Now the visibility drops
in tandem until $V_c = 0.50$ where the maximum and minimum
are set by the fringe pattern of the companion.
We show the relationship between the Be star visibility $V_c$
and the net observed visibility $V_o$ in Figure~\ref{app4}
(solid line for $x_2 = 10$ $\mu$m).  At this separation,
there is some slight destructive interference between the
fringe patterns that decreases the maximum amplitude for
star 2, and as $V_c$ declines to zero, the net visibility
attains the amplitude of star 2 alone $(f_2/f_1)/(1 + f_2/f_1)$.
Figure~\ref{app4} also shows the $(V_c, V_o)$ relationship for two other
separations.  The dotted line shows the case of zero separation for maximum
constructive interference, and here the visibility declines
linearly to $(f_2/f_1)/(1 + f_2/f_1)$ as $V_c$ tends to zero.
Finally, the dashed line shows the case for a very large separation
in which the fringe pattern of star 2 falls beyond the recorded
portion of the scan.  Here the visibility starts at its diluted
value of $1/(1 + f_2/f_1)$ at $V_c=1$ and declines to near
zero at $V_c=0$.

\placefigure{app3}
\placefigure{app4}

Thus, to correct the observed visibilities for the presence of
a companion, we need a diagram like Figure~\ref{app4} 
for each observation of a target.  We calculated the projected
separation of the stars at the time of the observation from the
dot product of the relative position vector (from the
angular orbital elements given in Table~5) and the $(u,v)$ spatial
frequencies for the baseline used.  Then we created an associated
$(V_c, V_o)$ diagram for each observation based upon the projected
separation and effective flux ratio.  The corrected visibility $V_c$
was then found by interpolating in the relationship at the
observed $V_o$ value.  In some rare circumstances, we encountered
a double-valued $(V_o, V_c)$ relation, so no correction was
attempted because of this ambiguity.

\subsection{Fringe Visibility for Be Stars in Multiple Systems} 

There are six systems in our sample with two or more close companions.
Both companions of HD~5394 ($\gamma$~Cas) are faint, so no correction was made.
In the cases of HD~23862 (Pleione) and HD~200120 (59~Cyg), the inner companion
is very faint, so corrections were made for only the outer, brighter companion.
The spectroscopic pair that comprises the B component of HD~4180 ($o$~Cas) has
such a small angular semimajor axis that it was treated as a single object,
and thus this system was also corrected as a binary star (\S 3.3).  This left two
systems, HD~198183 (the triple $\lambda$~Cyg) and HD~217675 (the quadruple $o$~And),
that required corrections for the flux of additional components.
The very close B pair of $o$~And was regarded as a single object (see Appendix),
so both $\lambda$~Cyg and $o$~And were treated as triple star systems.

We made visibility corrections for these two systems in
much the same way as for the binaries, except in this case
the fringe normalizations were assigned by
\begin{equation}
F_{tot}=\frac{F_1}{1 + f_2/f_1 + f_3/f_1}+
\frac{(f_2/f_1)F_2}{1 + f_2/f_1 + f_3/f_1}+\frac{(f_3/f_1)F_3}{1 + f_2/f_1 + f_3/f_1}.
\end{equation}
Again, we formed model visibilities from the coaddition
of the fringe patterns, determined the $(V_c, V_o)$
relationships for the time and baseline configuration of
each observation, and then used the inverted relation
$(V_o, V_c)$ to determine the corrected visibility.

Unfortunately, there are significant uncertainties surrounding
both the magnitude differences and orbital elements for the
companions of HD~198183 and HD~217675, and these introduce
corresponding uncertainties in the amounts of visibility
correction.  Our results on these two systems must therefore be
regarded as representative visibility solutions rather than
definitive ones.  However, the corrected interferometry visibilities
are all close to one for these two Be stars, which suggests
that their disks are only marginally resolved if at all.
On the other hand, the much lower uncorrected visibilities
of these two show that the influence of the companions is clearly present.
Both targets will be important subjects for future, multiple baseline
observations with the CHARA Array to determine their orbital properties.

\subsection{$K$-band Magnitude Difference} 

Our visibility correction procedure requires a knowledge of both the
projected separation of the stars (\S 3.2) and their monochromatic flux ratio.
Unfortunately, the magnitude differences between the Be star
primary and the companion are generally available only in the $V$-band,
and we need to estimate the magnitude differences in the $K$-band.
We must consider the color difference between the components and how
much brighter the Be star plus disk appears in the $K$-band
compared to the $V$-band.  The predicted magnitude difference is given by
\begin{equation}
\triangle K_{\rm obs} = -2.5 \log_{10} \frac{F_2}{{F_1+F_d}}\\
=-2.5 \log_{10} \frac{F_2}{F_1}+ 2.5 \log_{10} \left(1 + \frac{F_d}{F_1}\right)
\end{equation}
where $F_2,~F_1, ~F_d$ are the monochromatic
$K$-band fluxes for the Be companion, the Be star, and the Be disk, respectively.
We can estimate the first term from the color differences of the Be star
and companion using
\begin{equation}
\triangle K_{\rm bin} = -2.5 \log_{10} {F_2\over F_1}\\
=\triangle V_{\rm bin} + (V-K)_{1} - (V-K)_{2}
\end{equation}
where we will assume that the disk contribution is negligible in the
$V$-band so that $\triangle V_{\rm bin} = \triangle V_{\rm obs}$.
In the absence of other information, we estimated the color differences
$(V-K)$ by assuming that both the Be star and companion are
main sequence objects, and we used the relationship between
$(V-K)$ and magnitude difference from a primary star of effective
temperature $T_{\rm eff}$(Be) \citep{fre05} for main sequence stars
from \citet{lej01} to find $(V-K)$ for both stars.

We determined the infrared flux excess term
$(1 + {F_d / F_1})$ from our observed estimate
of $E^\star (V^\star - K)$ \citep{tou11}, which are related by
\begin{equation}
E^\star (V^\star - K) = 2.5 \log{{F_{\rm tot}^K} \over {F_{\rm tot}^V (F_1^K/F_1^V)}}\\
=2.5 \log_{10}{{[1 + F_d/F_1 + F_2/F_1]^K} \over {[1 + F_d/F_1 + F_2/F_1]^V}}
\end{equation}
where the superscripts indicate the filter band.  If we
assume that the disk contributes no flux in the $V$-band, then we can
rearrange this equation to find the $K$-band flux excess relative
to that of the Be star alone.  Then we can combine the results
from the two equations above to predict the observed
the $K$-band magnitude difference
\begin{equation}
\triangle K_{\rm obs}=\triangle K_{\rm{bin}}+E^\star (V^\star-K)
+ 2.5 \log_{10}
\left(1 + 10^{-0.4 \triangle V_{\rm{bin}}}
- 10^{-0.4 (\triangle K_{\rm{bin}}+ E^\star (V^\star - K))}\right).
\end{equation}

Our estimates for $\triangle K_{\rm obs}$ are listed in Tables 4 and 5.
Note that there are several instances where we give a range for the
magnitude difference; these are Be stars that are single-lined
spectroscopic binaries with a companion of an unknown type.
We consider two hypothetical cases.  First, we assume that the companion is a main
sequence star of one solar mass, and we use the \citet{lej01} main
sequence relation to obtain the magnitude and color differences of the
companion. The second case is to assume that the companion is a hot subdwarf
(similar to the case of $\phi$~Per; see \citealt{gie98}) with a
typical effective temperature of 30~kK and a stellar radius of $1
R_\odot$. We then estimate $\triangle K_{\rm obs}$ by adopting the main sequence
radius for the Be star according to its effective temperature and by using the Planck
function for both stars in order to estimate the monochromatic $K$-band flux ratio.
Table~\ref{binary} lists those cases with a hyphen in the $\triangle K$
column giving the magnitude difference range between that for
a hot subdwarf (smaller) and for a solar-type companion (larger).
However, we made no visibility corrections in most of these cases
because the nature of the companion is so uncertain (with the
exception of $\phi$~Per where the companion's spectrum was detected
and characterized by Gies et al.\ 1998).

\subsection{Seeing and Effective Flux Ratio} 

The CHARA Classic observations were recorded on a single
pixel of the Near Infrared Observer camera (NIRO),
and the physical size of the pixel corresponds to a
square of dimensions $0.8 \times 0.8$ arcsec on the sky.
The flux of companions with separations small compared to 0.8 arcsec
will be more or less completely included in the observations,
but the flux of companions with larger separations from the central
Be star may be only partially
recorded according to the separation and seeing conditions
at the time of observation.  Therefore, we need to calculate
the effective flux ratio of companion to target based upon
the relative amounts of flux as recorded by this one pixel.

Seeing is usually computed in real time
according to the CHARA tip-tilt measurements of the $V$-band
flux of the targets. By assuming that the $K$-band seeing
varies with wavelength by $\lambda^{-1/5}$ \citep{you74},
the $K$-band seeing disk is about $\sim$0.76 times that in $V$.
The effective flux ratio is given by
\begin{equation}
f_2 / f_1 = I_2 / I_1
\end{equation}
where $I_1$ and $I_2$ are the net intensity contributions of the
primary and secondary component, respectively, recorded by the pixel.
We assume a Gaussian profile for the point spread function as
projected on the detector,
\begin{equation}
I(x, x_0, y, y_0) = \frac{1}{2\pi \sigma^2} \exp \left[ -{{1}\over{2}} \frac{(x-x_0)^2 + (y-y_0)^2}{\sigma^2} \right]
\end{equation}
where $(x_0, y_0)$ are the coordinates of the central position
of the star on the detector chip, and $\sigma$ is related to the
seeing ($\sigma = 2.355^{-1} ~\theta_{seeing}$).
The intensity distributions of the primary and the secondary
components integrated over one pixel on the detector are given by

\begin{eqnarray}
I_1 = Q  \int \int I(x, 0, y, 0) ~dx ~dy
\end{eqnarray}
\begin{eqnarray}
I_2 =    \int \int I(x, 0, y, \rho ) ~dx ~dy
\end{eqnarray}
where $\rho$ is the separation of the binary, and $Q$ is
the actual ratio of primary to secondary flux,
which is derived from the magnitude difference of
the two components in the $K$-band,
\begin{eqnarray}
Q = 10^{ 0.4~\triangle K_{\rm obs}}.
\end{eqnarray}

\subsection{Visibility Corrections and Their Uncertainties} 

We used the visibility correction scheme described in \S 3.3 and 3.4
with the predicted projected angular separation (from the orbital elements
in Table 5 and the observed $(u,v)$ spatial frequencies in Table 3)
and the effective flux ratio (\S 3.5 and 3.6) to derive estimates
of the Be star visibility alone.  The corrected visibilities and
their corresponding uncertainties are listed in the last two columns of
Table~\ref{calibvis} for the nine cases with significant companions.
These uncertainties do not include the contributions to the error
budget from uncertainties in projected separation and flux ratio.
In most cases the projected separations are much larger than the
coherence length (typically 9 to 27 mas for 300 m to 100 m baselines),
which corresponds to the separated fringe packet case (see lower panel of
Fig.~2).  Thus, the correction depends mainly on the flux dilution term,
$1/(1+f_2/f_1)$, and typical uncertainties of 0.2 mag for $\triangle K_{\rm obs}$
only amount to correction uncertainties of $\approx 0.02$ in visibility,
i.e., generally smaller than the measurement errors.  On the other hand,
binaries with projected separations less than the coherence length
(mainly observations of $o$~Cas and $\phi$~Per; see Table 5) have corrections that reflect
the fringe amplitude oscillation caused by beating between the fringe patterns
of the components (Fig.~3).  These corrections may amount to $(f_2/f_1)/(1+f_2/f_1)$
in visibility, $\approx 0.10$ for the relatively faint companions considered
here.  Figure~3 shows that the full range of the correction varies
over a projected separation difference of $\triangle x_2 = \lambda / 2$, which corresponds to
a projected angular separation difference of approximately 0.7 to 2.2~mas
for 300~m and 100~m baselines, respectively.  The predictive accuracy
of the orbital separation probably has comparable uncertainties for the
cases of $o$~Cas and $\phi$~Per, so it is possible that the uncertainties
in the corrections may be similar to the corrections themselves in these close
separation instances.  Nevertheless, we found that implementing the corrections
even in these close cases tended to reduce the scatter in the fit of the
Be disk visibilities (\S 5.1), so we will adopt the corrected visibilities
in the subsequent analysis of all nine targets with significant companions.


\section{Spectral Energy Distributions}                    

The spectral energy distributions (SEDs) of our sample stars can help us
to estimate the photospheric angular diameter of the Be star
and the IR flux excess (from a comparison of the observed
and extrapolated stellar IR fluxes). The observed IR flux excess can thus be
directly compared with the disk flux fraction derived from fits
of the visibility measurements.  One difficulty with this approach
is that the IR flux excess is usually derived assuming that the disk
contributes no flux in the optical range, so the stellar SED can
be normalized to the optical flux.  However, this may lead to an
underestimate of the IR excess if the disk flux emission is significant in
the optical, which may be the case for stars with dense and large
circumstellar disks.  The disk flux fraction declines at lower
wavelength because of the drop in free-free opacity, and the
disk contribution is negligible in the ultraviolet (UV) part of the spectrum.
Therefore, we decided to make fits of the UV flux to set
the photospheric flux normalization, which will provide
a more reliable estimate of the IR flux excess.

We collected UV spectra for all our targets from the archive of
the {\it International Ultraviolet Explorer (IUE) Satellite},
maintained at the NASA Mikulski Archive for Space Telescopes
at STScI\footnote{http://archive.stsci.edu/iue/}.
In most cases, we used the available low dispersion, SWP (1150 - 1900 \AA)
and LWP/LWR (1800 - 3300 \AA) spectra, and in cases where there were fewer than
two each of these, we used high-dispersion spectra for these
two spectral ranges. In the case of HD~217891, all but one of the spectra
were made with the small aperture, so we formed an average spectrum
and rescaled the flux to that measured by the TD-1 satellite (\citealt{tho78}).
In the case of HD~203467, there were no long-wavelength spectra available
in the {\it IUE} archive, so we used a combination of {\it IUE} SWP spectra (1150 --
1900 \AA), SKYLAB objective-prism spectrophotometry (1900 -- 2300 \AA; \citealt{hen79}),
and OAO-2 spectral scans (2300 -- 3200 \AA; \citealt{mea80}). Fluxes
from each spectrum were averaged into 10 \AA~bins from 1155 to 3195 \AA,
and then the fluxes from all the available spectra for a given target
were averaged at each point in this wavelength grid.

We created model spectra to compare with the {\it IUE} observations by
interpolating in the grid of model LTE spectra obtained from
R.\ Kurucz.  These models were calculated for solar abundances and a microturbulent
velocity of 2 km~s$^{-1}$.  The interpolation was made in effective
temperature and gravity using estimates for these parameters (see Table 1)
from the Be star compilation of \citet{fre05}. For those
sample stars in binaries, we formed a composite model spectrum by
adding a model spectrum for each companion that was scaled according
to the $K$-band magnitude difference listed in Table~\ref{orbital}.

We then made a non-linear, least-squares fit of the observed UV spectrum
with a model spectrum transformed according to the extinction curve
of \citet{fit99} and normalized by the stellar, limb-darkened
angular diameter $\theta_{LD}$, and we assumed a standard ratio of total to
selective extinction of $R=3.1$ for the interstellar extinction curve.
Finally, we considered an extension of the fitted photospheric SED
into the $K$-band, and we determined an IR flux excess from
\begin{equation}
E^\star (UV - K) = 2.5 \log (1 + {F_{tot}\over F_1} {{F_{obs}-F_{tot}}\over {F_{tot}}})
\end{equation}
where the monochromatic $K$-band fluxes are $F_1$ for the
Be star component, $F_{tot}$ for the sum of the photospheric fluxes of the
Be and all companions (if any), and $F_{obs}$ is the observed
flux from 2MASS \citep{coh03,cut03}.

\placetable{SEDinfo}

Our results are listed in Table~\ref{SEDinfo} that gives the HD number
of the star, the derived reddening $E(B-V)$ and its uncertainty,
the limb darkened stellar angular diameter
$\theta_{s}$ and its uncertainty, and the infrared excess from the disk $E^\star (UV - K)$
and its uncertainty.  The final columns list the photospheric fraction of flux
$c_p$ and its uncertainty that are related to the IR flux excess by
\begin{equation}
E^\star (UV - K) = -2.5 \log c_p .
\end{equation}
Our derived interstellar reddening values are generally smaller
than those derived from the optical colors because the disk flux
contribution increases with increasing wavelength through the optical band
which mimics interstellar reddening \citep{dou94}.
Furthermore, our derived angular diameters
may be smaller in some cases from previous estimates because of
the neglect of the flux of the companions in earlier work.
Note that we have neglected the effects of rotation (oblateness and
gravity darkening) on the SED, and these may influence the
flux normalization \citep{fre05}.


\section{Gaussian Elliptical Fits}                          

\subsection{Method}

Here we show how we can use the interferometric visibility
measurements to estimate the stellar and disk flux contributions
and to determine the spatial properties of the disk emission component.
We used a two-component geometrical disk model
to fit the CHARA Classic observations and to measure the
characteristic sizes of the circumstellar disks.
The model consists of a small uniform disk representing the
central Be star and an elliptical Gaussian component
representing the circumstellar disk \citep{qui97,tyc04}.
Because the Fourier transform function is additive, the total visibility of
the system is the sum of the visibility function of the
central star and the disk,
\begin{equation}
V_{tot} = c_p V_s + (1-c_p) V_d
\end{equation}
where $V_{tot}$, $V_s$, and $V_d$ are the total, stellar,
and disk visibilities, respectively, and $c_p$ is the
ratio of the photospheric flux contribution to the total
flux of the system.  The visibility for a uniform disk star
of angular diameter $\theta_s$ is
\begin{equation}
V_s = 2 J_1(\pi x \theta_s) / (\pi x \theta_s)
\end{equation}
where $J_1$ is the first-order Bessel function of the first kind
and $x$ is the spatial frequency of the interferometric observation,
$x = \sqrt{u^2 + v^2}$.  The central Be star is usually mostly unresolved
even at the longest baseline of the interferometer, which brings
its visibility close to unity, $V_s \simeq 1$.
The disk visibility is given by a Gaussian elliptical distribution
\begin{equation}
V_d = \exp \left[ -\frac{(\pi \theta_{\rm maj} s)^2}{4 \ln 2} \right]
\end{equation}
where $\theta_{\rm maj}$ is the full-width at half-maximum (FHWM)
of the spatial Gaussian distribution along the major axis and $s$ is given by
\begin{equation}
s = \sqrt{r^2(u \cos PA - v \sin PA)^2 + (u \sin PA + v \cos PA)^2}
\end{equation}
where $r$ is the axial ratio and $PA$ is the position angle of
the disk major axis. Thus, the
Gaussian elliptical model has four free parameters: the
photospheric contribution $c_p$, the axial ratio $r$, the
position angle $PA$, and the disk angular size $\theta_{\rm maj}$.

\subsection{The Effective Baseline}

The brightness distribution of the disk projected onto the sky
is a function of the inclination and position angle of the major axis.
Interferometric observations with a given baseline $B_p$
will sample the disk elliptical distribution according to
the angle between the projection of the baseline on the sky
and the position angle of the disk major axis.
It is helpful to consider rescaling the baseline to account
for the changes in the disk size with direction.
For a flat disk inclined by an angle $i$ and oriented with the major axis
at a position angle $PA$ (measured from north to east), we can rescale using the
effective baseline $B_{\rm eff}$ \citep{tann08}
\begin{equation}
B_{\rm eff} =
B_p \sqrt{\cos^2 (\phi_{\rm obs}-PA) + \cos^2 i \sin^2 (\phi_{\rm obs}-PA)}
\end{equation}
where $\phi_{\rm obs}$ is the baseline position angle at the time of the observations.
This new quantity, the effective baseline, takes into consideration
the decrease in the interferometric resolution due to the
inclination of the disk in the sky, and thus for the purposes of analysis, it
transforms the projected brightness distribution of the disk into a nearly circularly
symmetric brightness distribution.  Thus, the disk part of the visibility
can be considered as a function of $B_{\rm eff}$ alone, and below we
will use this parameter to present the interferometric results.
However, if there is also a stellar flux
contribution, than its projection and visibility will be a function of the
projected baseline $B_p$ only (if the star appears spherical in the sky).
Consequently, models with both stellar and disk contributions should be
presented for both the major and minor axis directions in order to show the
range in visibility with baseline direction in a single plot.
Along the minor axis, for example, $\phi_{\rm obs}-PA = 90^\circ$, so
that the relation becomes $B_{\rm eff} = B_p \cos i$, and consequently
for edge-on systems with $i \sim 90^\circ$ a given $B_{\rm eff}$ may
correspond to a $B_p$ far larger than available with the Array.
Such observations (if possible) would start to resolve the stellar disk
and lead to low net visibility (see the dotted line in Fig.\ 8.3).

\subsection{Model Degeneracy} 

The simple geometrical representation of the Be star system may
in some cases present a solution family or
degeneracy that exists between two fundamental parameters of the Gaussian
elliptical model: the Gaussian full width at half maximum $\theta_{\rm maj}$
and the stellar photospheric contribution $c_p$. In order to explain
the ambiguity in the model, we consider the case of a locus
of ($c_p$, $\theta_{\rm maj}$)
that produces the same visibility measurement at a particular
baseline. We illustrate this relation in Figure~\ref{test1}, which shows
an example of a series of ($c_p$, $\theta_{\rm maj}$) Gaussian elliptical
visibility curves that all produce a visibility point $V$ = 0.8
at a projected baseline of 200 m (for $\lambda$ = 2.1329~$\mu$m
and a Be star angular diameter $\theta_{\rm UD}$ = 0.3~mas).
The plot shows models for
($c_p$, $\theta_{\rm maj}$) = (0.156, 0.6 mas) (solid line),
($c_p$, $\theta_{\rm maj}$) = (0.719, 1.2 mas) (dotted line), and
($c_p$, $\theta_{\rm maj}$) = (0.816, 2.4 mas) (dashed line).
Figure~\ref{test2} shows the relationship between the
Gaussian elliptical full width at half maximum and the stellar
photospheric contribution for the family of
curves that go through the observed point $V$ = 0.8~ at a 200~m baseline.
Larger circumstellar disks are associated with larger
stellar flux contributions, and vice-versa, which demonstrates that
a single interferometric measurement does not discriminate between
a bright small disk and a large faint one.  Additional measurements at
different baselines are necessary to resolve this ambiguity.

\placefigure{test1}     
\placefigure{test2}     

Note that if the interferometric observations are all located on
one projected baseline in the ($u,~v$) plane, then the disk
properties are defined in only
one dimension. Thus, in such circumstances it is not possible to
estimate the axial ratio $r$ or the position angle of the disk major
axis $PA$.  Only a lower limit for $\theta_{\rm maj}$ can be set in
such cases.

\subsection{Fitting Results} 

The circumstellar disk is modeled with a Gaussian elliptical
flux distribution centered on the Be star.  This
disk model has four independent parameters ($r$, $PA$, $c_p$, $\theta_{\rm maj}$),
and one assumed parameter, the stellar diameter $\theta_s$ (Table 6).
The fitting procedure consists of solving for the model parameters
using the IDL non-linear least squares curve fitting routine MPFIT (\citealt{mar09}),
which provides a robust way to perform multi-parameter surface fitting.
Model parameters can be fixed or free depending on the ($u, v$) distribution of
the observations.

For the cases with limited coverage in the ($u, v$) plane,
setting some model parameters to fixed values was
necessary in order to fit successfully the data.
We adopted additional constraints on some model
parameters that are well defined from studies
such as inclination estimates from \citet{fre05}
and intrinsic polarization angles from \citet{mcd99} and \citet{yud01}.
Also, we occasionally used values of the IR flux excess derived from the SEDs of Be stars
to estimate the stellar photospheric contribution $c_p$ when needed.

Our fitting results are summarized
in Table~\ref{visfit}, where the cases with a fixed parameter are identified by
a zero value assigned to its corresponding uncertainty.
Column 1 of Table~\ref{visfit} lists the HD number of the star, columns 2 and 3
list the axial ratio $r$ and its uncertainty, columns 4 and 5 list
the disk position angle along the major axis $PA$ and its uncertainty,
columns 6 and 7 list the photospheric contribution
$c_p$ and its uncertainty, columns 8 and 9 list the
angular FWHM of the disk major axis $\theta_{\rm maj}$ and its uncertainty,
column 10 lists the reduced $\chi_{\nu}^2$ of the fit, columns 11 and 12 list
the corrected photospheric contribution $c_p$(corr) and its uncertainty (see \S~5.5),
respectively, columns 13 and 14 list the disk-to-star
radius ratio $R_d/R_s$ and its uncertainty, and finally,
column 15 indicates the cases of fully resolved disks (Y), marginally
resolved disks (M), and unresolved disks (N) in our Be star sample.

Plots of the best-fit solutions showing the visibility curves
of the system disk-plus-star as a function of the effective baseline
in meters along with our interferometric data are presented
in Figure Set 8. 
The solid lines in Figure Set 8 represent
the best-fit visibility model of the disk along the major axis,
the dotted lines represent the best-fit visibility model of the
disk along the minor axis, and the star signs represent the
interferometric data.

\placetable{visfit}                    

\placefigure{vis1}                     
\placefigure{vis2}                     
\placefigure{vis3}                     

We find that the circumstellar disks of the four Be
stars, HD 23630, HD 138749, HD 198183, and HD 217675 were
unresolved at the time of our CHARA Array observations,
while the circumstellar disks of
HD 23862, HD 142926, HD 164284, HD 166014, HD 200120,
and HD 212076 were only marginally resolved.
Those are the cases where we had to fix $r$, $PA$, and/or $c_p$
in order to make Gaussian elliptical model
fits to the data.
On the other hand, we successfully resolved the circumstellar disks around
the other 14 Be stars and were able to perform
four-parameter Gaussian elliptical fits for most of them as listed in
Table~\ref{visfit}.

\subsection{Corrections to the Gaussian Model}  

Modeling a circumstellar disk with an elliptical Gaussian
intensity distribution is convenient but not
completely realistic.  The flux distribution in the model
assumes that light components from the
circumstellar disk and the central star are summed and
that no mutual obscuration occurs.
It is important to note that in the case of small disks
most of the model disk flux is spatially coincident with
the photosphere of the star, so the assignment of the flux
components becomes biased.

To illustrate this effect, we show in
Figure~\ref{gaucorr} the model components for a case with a faint disk.
The dotted line in Figure~\ref{gaucorr} shows the
assumed form of the intensity of the uniform disk of the star
(angular diameter of 0.68~mas), the dashed line shows the
Gaussian distribution of the circumstellar disk along the
projected major axis (FWHM = 0.55 mas),
and the solid line shows the sum of the two
intensity components. In the case where the FWHM of the circumstellar emission
is similar to or smaller than the stellar diameter, most of the disk
flux occurs over the stellar photosphere where the sum produces a
distribution similar to that of a limb-darkened star.

\placefigure{gaucorr}                               

The interpretation of the results obtained from the Gaussian elliptical fits
must be regarded with caution in situations
where the derived disk radius is smaller
than the star's radius and a significant fraction of the disk flux is spatially
coincident with that of the star. Consequently, we decided to correct
the Gaussian elliptical fitting results in two ways.
First, the disk radius was
set based upon the relative intensity decline from
the stellar radius, and we adopted the disk radius to be that
distance where the Gaussian light distribution along the
major axis has declined to half its value at the stellar equator.
The resulting ratio of disk radius to star radius is then
given by
\begin{equation}
{{R_d}\over{R_s}} = (1 + ({\theta_{\rm maj}} / \theta_s)^2)^{1/2}
\end{equation}
where $\theta_{\rm maj}$ is the Gaussian full-width at half maximum along
the major axis derived from the fits, and
$\theta_s$ is the angular diameter of the central star.
Secondly, the model intensity over the photosphere of
the star from both the stellar and disk components was
assigned to the flux from the star in an optically thin approximation.
The fraction of the disk flux that falls
on top of the star is $f (1-c_p)$, where $f$ is found by integrating
over the stellar disk the Gaussian spatial distribution given by
\citep{tyc04}
\begin{equation}
I_{\rm env}(x,y) = \frac{4 \ln 2}{\pi r \theta_{\rm maj}^2}
\exp \left[ -\frac{(x^2/r^2 + y^2)}{\theta_{\rm maj}^2/ 4 \ln 2} \right]
\end{equation}
where $r$ is the axial ratio and $(x,y)$ are the sky coordinates
in the direction of the minor and major axes.
Consequently, this fraction of the model disk flux
should be reassigned to the star and removed from the disk
contribution. Then the revised ratio of total disk to
stellar flux is
\begin{equation}
{{F_d}\over{F_\star}} = {{(1-c_p)(1-f)}\over{c_p + (1- c_p) f}}
 = {{1-c_p({\rm corr})}\over{c_p({\rm corr})}}
\end{equation}
which is lower than the simple estimate of $(1-c_p)/c_p$.
We list in columns 11 to 14 of Table~\ref{visfit} the revised
values of the stellar flux contribution $c_p$ and the Be disk radius
${{R_d}/{R_s}}$ (along with the corresponding uncertainties) obtained by applying
this correction using the stellar angular diameters from Table 6.


\section{Discussion}                                          

\subsection{Comparison with Other Results} 

It is important to validate our results against other published data
wherever possible.  However, since this is the first large scale study
of the $K$-band emission of northern Be stars, it is difficult to make
direct comparisons.  We list in Table~8 the available measurements
of Gaussian elliptical model parameters of Be star disks for the
stars in our sample (excluding the work of \citealt{gie07} that is
included in our analysis).  There is only one other $K$-band measurement
available by \citet{pot10}, but even in this case a direct comparison
of $\theta_{\rm maj}$ is difficult because of the different assumptions
made about the remaining parameters.  Measurements of the disk emission
in other bands will likely yield different diameter estimates (\S6.2).
Nevertheless, we can compare our results on the geometry of the disks
with previous work, and we can consider the parameter estimates resulting
from other kinds of observations.

We begin by comparing the stellar and disk flux contributions
that we derived from the Gaussian elliptical fits (Table~7) with
those estimated from the IR-excess in the spectral energy distributions
(Table~6).  Among the 14 stars in our sample with reliable detections
of disk emission, we fit for the stellar flux contribution in six cases.
The corrected parameter $c_p ({\rm corr}) = F_s / (F_s + F_d)$ is plotted
together with the SED estimate of this ratio in Figure~10.  The uncertainties
are significant, but these two estimates of stellar flux in the $K$-band
appear to be consistent (perhaps surprisingly so, because of the
known temporal flux variations and large time span between the 2MASS
photometry and our interferometric observations).

\placefigure{cpcomp}                                

Next we consider the disk axial ratio $r$ that is related approximately to
the disk normal inclination $i$ by \citep{gru06}
\begin{equation}
r \approx \cos i + 0.022 \sqrt{\frac{R_d}{R_s}} \sin i
\end{equation}
where the second term accounts for the increase in the minor axis
size caused by the increase in disk thickness with radius.
We expect that the disk inclination will be the same as the stellar
spin inclination because the disk is probably fed by equatorial mass loss.
\citet{fre05} estimated the stellar inclinations for most of our targets
by comparing the projected rotational velocity $V\sin i$ with the
critical rotational velocity $V_{\rm crit}$, for which the equatorial
centripetal and gravitational accelerations balance, and by assuming
that Be stars as a class share a common ratio of angular rotational to
critical velocity.  The axial ratios derived from their estimates of
spin inclination are given in the final column of Table~1.
There are seven targets with fitted estimates of $r$ among our 14 reliable
detections, and we plot these values together with $r(i)$ from the
inclinations from \citet{fre05} in Figure~11 (indicated by square symbols).
Figure~11 includes other $r$ estimates from previous interferometry
(Table~8).   With two exceptions ($\upsilon$~Cyg = HD~202904 above and
$\gamma$~Cas = HD~5394 below), the estimates from interferometry and
rotational line broadening are in broad agreement.  Three of our
targets have prior interferometric estimates of $r$, and our results
agree within the uncertainties.  Note that the increase in disk thickness
with radius implies that $r$ will appear larger in bands where the
disk emission extends to larger radius (H$\alpha$), and in two cases
($\psi$~Per = HD~22192 and $\zeta$~Tau = HD~37202) $r$(H$\alpha$) is
larger than $r$($K$-band), while they are the same in the third case
($\gamma$~Cas = HD~5394).

\placefigure{rcomp}                                 

The disk normal is expected to have the same position angle in the
sky as that for the intrinsic polarization (from scattering in
the inner disk; \citealt{qui97}).  We show in Figure~12 a comparison
of the position angles derived from interferometry with those
from polarimetric studies \citep{mcd99,yud01}.  These generally
agree within the uncertainties, but there are some exceptions
($\upsilon$~Cyg = HD~202904 and several cases with large axial
ratio $r$ where it is difficult to determine position angle).
Our results are fully consistent for four targets with
previous interferometric estimates of $PA$.   All these comparisons
lend support to our strategy of fixing the $c_p$, $r$, and $PA$ parameters
in those cases where there were ambiguities with the full four-parameter,
Gaussian elliptical fit.

\placefigure{pacomp}                                

\subsection{Disk Diameters} 

We clearly detected the visibility decline due to the disk in 14 stars
in our sample.  The ratio of disk to stellar radius (Table~7, column 13)
varies from 1.5 to 10 with a mean value of 4.4 among this subsample,
but our detections are probably biased towards those cases with larger
and brighter disks.  We compare these $K$-band diameters with those
measured for H$\alpha$ (Table~8) in Figure~13, and this demonstrates
that in most cases the H$\alpha$ disk diameters are much larger.
\citet{gie07} attributed this difference to the larger opacity of H$\alpha$
compared to the free-free and bound-free opacities that dominate the disk emission
in the near-infrared.  There are three targets that fall below the trend:
$\upsilon$~Cyg = HD~202904 (but with an uncertainty that may be consistent
with the trend), $\phi$~Per = HD~10516, and $\kappa$~Dra = HD~109387.
It is curious that the latter two are both short-period binaries,
and it may be that the usually larger H$\alpha$ emitting regions are
truncated by tidal forces, so that their $K$-band and H$\alpha$ disk sizes
are comparable.  After setting aside these three discrepant cases,
the unweighted slope of the relation starting from an origin at
$(R_d(K{\rm -band})/R_s, R_d({\rm H}\alpha)/R_s) = (1,1)$ is
$\triangle R_d({\rm H}\alpha) / \triangle R_d(K{\rm -band}) = 4.0 \pm 0.5$
(s.d.\ of the mean), and this slope is indicated by a dotted line in Figure~13.
Such a correlation is helpful
in predicting the size of the circumstellar disk in the $K$-band
from H$\alpha$ observations, and vice-versa, especially since Be star
disks are generally highly variable and simultaneous, multiwavelength
observations are usually difficult to obtain.

\placefigure{Kha}                              

Next, we consider the connection between disk size and brightness.
If the disk gas near the central Be star has a temperature
that is some fraction of the stellar temperature,
then we might expect to find a correlation between
the surface intensity $I$ of the inner disk and star
in the limit of high disk optical depth,
\begin{equation}
I_{\rm env}(x=0,y=\theta_s/2) = b I_s
\end{equation}
for some constant $b$.
We can explore this relationship by considering
the Gaussian elliptical model prediction for the
disk brightness at the stellar equator \citep{tyc04},
\begin{equation}
I_{\rm env}(x=0,y=\theta_s/2) = (1 - c_p)
 {{4 \ln 2}\over{\pi r \theta_{\rm maj}^2}}
 2^{-(\theta_s / \theta_{\rm maj})^2}.
\end{equation}
The mean stellar intensity in the model is
\begin{equation}
I_s = {{c_p(\rm corr)} \over {\pi (\theta_s / 2)^2}}.
\end{equation}
Then we can equate these with proportionality constant $b$
to obtain
\begin{equation}
{{1 - c_p}\over{c_p({\rm corr})}} =
 b r (\theta_{\rm maj}/\theta_s)^2 ~2^{(\theta_{\rm maj}/\theta_s)^{-2}}.
\end{equation}
This relation predicts that the disk to star flux ratio (left hand side)
is approximately proportional to the ratio of disk to star projected area
(right hand side).

\placefigure{fxratio} 

The Gaussian elliptical parameters from Table~7
and the stellar angular diameters from Table~6 were used to calculate
the terms on both sides of the equation, and they are plotted together
in Figure~\ref{fxratio} (where we assumed a minimum uncertainty of $5\%$
for those cases where the formal uncertainties may be underestimated).
We see that the Be stars with the largest disk to stellar flux ratio
are often those with a large ratio of circumstellar to stellar angular
diameter.  However, a counter example is the case of $\kappa$~Dra
= HD~109387 that is plotted with a large but faint disk (lower right
location in Fig.~\ref{fxratio}).  The fit of the Gaussian elliptical
model parameters in this case is constrained by a set of
100~m baseline measurements from \citet{gie07} (see Fig.~8.10),
and if a fit was made from the new measurements alone, then the disk size
would be much smaller (as also suggested by the smaller H$\alpha$
disk size shown in Fig.~13).  Given these uncertainties, our
measurement for $\kappa$~Dra may be set aside from consideration here.
An unweighted, nonparametric, correlation test for the remaining 13 Be stars
yields a Kendall's statistic of $\tau = 0.205$, which has null rejection
probability of $37\%$, i.e., a value which is not small enough to reject
with confidence the null hypothesis of no correlation.
We suspect that the poor correlation results from a break down
among the faint disk stars of our simplifying assumption that the
disks are optically thick, which may only be applicable to the
brightest disk cases.  We show for completeness in Figure~\ref{fxratio}
a linear fit of the sample of 13 stars that has a constant of
proportionality of $b=0.18 \pm 0.04$ (s.d.\ of the mean).
We caution again that such a relation may only
hold for dense, optically thick disks and that our results may be
biased against detection of fainter disks in general.
We are planning to investigate further the question of disk surface intensity
in another paper that will apply physical models and radiative transfer
calculations to fit the observed visibilities.

\subsection{Detection Limits} 

We found that the criterion for a confident detection of the circumstellar disk
is usually a decline in visibility below $V=0.8$ at the longest
baselines available.  We can use this criterion to estimate the limitations
on disk sizes that we can detect for Be stars at different distances.
The visibility measured along baselines aligned with the projected
major axis of the disk is a function of the photospheric flux
fraction $c_p$, the ratio of disk to stellar radius $R_d/R_s$,
and the stellar angular diameter $\theta_s$ (eq.\ 17, 18, 19).
Thus, given $R_d/R_s$ and $\theta_s$ we can find $\theta_{\rm maj}$,
and the remaining parameter to estimate is $c_p$.
In practice this could be estimated from
a simultaneous analysis of the SED of the Be star.
However, for our purpose here, we adopted the relationship
between the disk to star flux ratio and projected surface areas
(eq.\ 29 and Fig.\ 14) to estimate $c_p$ from $r$ and $R_d/R_s$.
This relationship is poorly defined for fainter disks
($(1-c_p)/c_p < 1$), where the disks become optically
thin and the ratio of areas argument no longer applies.

The stellar angular diameter is found from the assumed
stellar radius and distance, and we made estimates for
two cases, B0~V and B8~V types for the Be star, and for
three distances corresponding to visual magnitudes 3, 5, and 7.
We used stellar radii and absolute magnitudes for these
classifications from the compilation of \citet{gra05}, and
we neglected any extinction in the calculation of distance
from the magnitude difference.  Figure~15
shows the resulting predicted visibilities for a $K$-band measurement
with a projected baseline of 300~m as a function of $R_d/R_s$
for these different cases.  Each plot shows how the visibility
at this baseline declines as the disk size increases, and
we can use these to estimate the smallest disk detectable.
For example, we see in Figure~15a for a B0~V star of
apparent magnitude 5 that the curve dips below $V=0.8$ at
$R_d/R_s = 3.4$ from which we would infer that only
disks larger than this would be detected with the CHARA Array.
As expected, we can detect smaller disks in nearer (brighter)
Be stars. Figure~15b shows the case for a later B8~V type
that is somewhat more favorable for detection of smaller disks.
Note that at small disk radii
we simply assume that all the flux is stellar, so the
limiting visibility near $R_d/R_s = 1$ corresponds to the
stellar visibility.

\placefigure{limit1} 
\placefigure{limit2} 

\subsection{Rotational and Critical Velocities}

Be stars are generally fast rotators, but it is difficult to determine
the actual equatorial rotational velocity $V_{\rm rot}$ from the
observed projected rotational velocity $V\sin i$ (measured from the
rotational Doppler broadening of the lines; \citealt{tow04}) because
of the unknown stellar inclination $i$.  If we assume that the
stellar spin and disk rotational axes are coaligned, then we can use
the disk axial ratio to set the stellar inclination (eq.\ 25).
We collected the seven $r$ values derived from our fits (Table~7)
together with estimates from previous work (Table~8) to find
mean $\sin i$ factors for 12 Be stars in our program.
We adopted the projected photospheric rotational velocities $V \sin i$
and the critical rotational velocities $V_{\rm crit}$ from \citet{fre05},
who derived these values after making corrections for the effects of
gravity darkening.  Then we divided $V \sin i$ by $\sin i$
to find the equatorial rotational velocity $V_{\rm rot}$.
Our results appear in Table~9 that lists the HD number of the star,
$V \sin i$ and $V_{\rm crit}$ \citep{fre05}, $V_{\rm rot}$, and two ratios,
${V_{\rm rot}}/{V_{\rm crit}}$ and ${\Omega_{\rm rot}}/{\Omega_{\rm crit}}$.
The ratio of angular rotational velocities is given for the Roche approximation
using expressions from \citet{eks08},
\begin{equation}
\frac{\Omega_{\rm rot}}{\Omega_{\rm crit}} =
 {3 \over 2} \frac{V_{\rm rot}}{V_{\rm crit}}
 \left[1 - {1\over 3}\left(\frac{V_{\rm rot}}{V_{\rm crit}}\right)^2\right].
\end{equation}

\placetable{tabq}                       

We find that most of the stars in this subsample are
rotating very quickly, and two targets may have
attained the critical rate ($\gamma$~Cas = HD~5394 and
48~Lib = HD~142983).  However, there are two targets
with much more moderate rotational velocities,
$\beta$~Psc = HD~217891 and $\upsilon$~Cyg = HD~202904.
The former has a rather large axial ratio $r=0.70\pm0.15$,
and it is possible that the uncertainties would allow
a small inclination and hence large rotational velocity.
However, the case of $\upsilon$~Cyg is more difficult
to understand.  \citet{nei05} argue on the basis of the
narrow lines in the spectrum that the axial ratio
should be close to $r\approx 0.9$ if it is actually
a rapid rotator.  This is much larger than the value
we derive from the Gaussian elliptical fits, $r=0.26 \pm 0.13$.
Additional interferometric observations are needed to
confirm or remove this discrepancy.

We find a mean angular velocity ratio of
${\Omega_{\rm rot}}/{\Omega_{\rm crit}} = 0.88 \pm 0.17$ (s.d.)
for the sample of 12 Be stars or $0.95 \pm 0.04$ (s.d.) if
we remove $\beta$~Psc and $\upsilon$~Cyg from the sample.
These results support earlier Be star angular velocity ratio
estimates of 0.88 \citep{fre05}, 0.7 -- 1.0 \citep{cra05},
0.8 -- 1.0 \citep{hua10}, and $0.95 \pm 0.02$ \citep{mei12}.
The consistency of these results suggests that rapid rotation
in Be stars plays a key role in the Be phenomenon.


\section{Conclusions}

Our CHARA Array interferometric survey of northern sky Be stars
has led to the resolution of extended $K$-band continuum emission
from the disks of 14 stars among of our sample of 24 stars.
We interpreted the visibility measurements with a simple geometrical
model that assumes a Gaussian elliptical brightness distribution
for the circumstellar disks.  The model fits yield estimates
of the disk angular diameter, axial ratio, position angle, and
the photospheric and disk flux contributions in the $K$-band.
We demonstrated that the results are consistent with earlier
interferometric studies of the disk H$\alpha$ emission and
with other estimates for the disk parameters from spectroscopic
and polarimetric studies.

We determined estimates for the stellar angular diameters and
infrared flux excesses from fits of the ultraviolet and near-infrared
spectral energy distributions.  We find that the mean ratio of
the $K$-band disk to stellar diameter is 4.4 with a range from
1.5 to 10 for the detected cases.  The ratio is similar among
both early- and late-type Be stars, which suggests that Be star
disks tend to scale in size and flux with the properties of the
central star (although we caution that fainter, smaller disks
may have escaped detection).  The diameters of the $K$-band
emitting region are much smaller than those for the H$\alpha$
emission in most cases, and the difference is probably due to
the higher opacity of H$\alpha$ that extends the emission to
the lower density gas found at larger radii.  The apparent
axial ratio of the disk emission is related to the disk inclination,
and we can reliably assign the derived inclination to the stellar
spin axis also.  We used the interferometrically observed axial ratios
for a sub-sample of 12 Be stars to convert the projected rotational
velocity from spectroscopy to the equatorial rotational velocity,
and we found that these stars generally rotate close to their critical
velocities.

We were surprised to find that so many of our targets (14 of 24) are
members of binary and multiple star systems, and we developed methods
to account for the influence of the flux of the companions on the
interferometric visibilities.  Some Be stars may have been spun up through
mass transfer in a close binary system \citep{pol91}, and we have identified
the spectral features of the hot, stripped-down, mass donor star in
several cases \citep{gie98,mai05,pet08}.  \citet{tok06} has found that a
large fraction of close, spectroscopic binaries among solar-type stars have
distant third companions, and it is possible that the tertiary received
the excess angular momentum from the natal cloud that needed
to be removed from the central region in order to form a close binary
during star formation.  Thus, the presence of a wide companion
around a Be star may indicate that the central object was originally a
close binary that subsequently experienced interaction, mass transfer,
and spin-up of the gainer.  The derivation of orbits for these wide
companions will be especially useful to measure masses of Be
stars and to see, for example, if they are overluminous for their
mass because of large scale internal mixing caused by mass transfer
and fast rotation \citep{gie98}.

Our survey provides a first epoch of diameter measurements that will
serve in future studies of the growth and dissipation of disks that
are frequently observed in Be stars \citep{por03}.  The Gaussian
elliptical models provide a useful but first-order description of the
disks, and it will be important to explore the development of
disk asymmetries related to disk instabilities (such as one-armed
spiral features; \citealt{car09}).  Such studies will require closure
phase measurements in addition to interferometric visibility, and
instruments at the CHARA Array like the Michigan Infrared Combiner
\citep{mon10} and VEGA \citep{mou11} that use multiple telescopes and
baselines will eventually allow us to reconstruct images of
Be star disks.  Such images will be vitally important to test
physical models of the disks and to determine the nature of the
companion stars.

\acknowledgments
We are grateful to Vincent Coud\'{e} du Foresto, William Hartkopf, Carol Jones, Robert Kurucz,
Brian Mason, Rafael Millan-Gabet, John Monnier, David O'Brien, Deepak Raghavan, Lewis Roberts,
Philippe Stee, and Christopher Tycner for important contributions to this work.
This material is based upon work supported by the National Science Foundation
under Grants AST-0606861 and AST-1009080 (Gies) and AST-0606958 and AST-0908253 (McAlister).
STR acknowledges partial support from NASA grant NNH09AK731.
Yamina Touhami also gratefully acknowledges the support
of a Georgia Space Grant Consortium Fellowship and
a NASA/NExScI visiting fellowship to the California Institute of Technology.
Institutional support has been provided from the GSU College
of Arts and Sciences and from the Research Program Enhancement
fund of the Board of Regents of the University System of Georgia,
administered through the GSU Office of the Vice President
for Research and Economic Development.  The {\it IUE} spectra presented
in this paper were obtained from the Mikulski Archive for Space Telescopes
(MAST) at STScI.  STScI is operated by the Association of
Universities for Research in Astronomy, Inc., under NASA
contract NAS5-26555.  Support for MAST for non-HST data is
provided by the NASA Office of Space Science via grant
NNX09AF08G and by other grants and contracts.


Facilities: \facility{CHARA Array}


\appendix
\section{Notes on Individual Stars}

\noindent{\sl HD 4180.}
\citet{kou10} present a single-lined spectroscopic
orbit for HD 4180 plus interferometric observations from NPOI of the
resolved system A,B.
The components of this binary system are too close for speckle resolution,
and the object appeared single in observations by \citet{mas97}.
\citet{gru07} observed spectral features corresponding to two similar
late-B or early-A stars that showed Doppler shifts on a timescale of
approximately 4~days, and these probably form in the close (Ba, Bb) system
that was suspected by \citet{kou10}.  We have estimated the
$K$-band magnitude difference using the magnitude difference from
NPOI $\triangle R = 2.9$ mag \citep{kou10},
the estimated spectral types from \citet{kou10},
the near-IR color calibration from \citet{weg94}, and the flux excess
$E^\star (V^\star -K) = 0.13$ mag from \citet{tou11}.

\noindent{\sl HD 5394.}
$\gamma$ Cas is a single-lined spectroscopic binary with
a faint (undetected) companion \citep{har00,mir02,nem12,smi12}.
Although this binary could be resolved in our CHARA visibility observations,
the expected large magnitude difference makes detection very difficult.
The distant and faint B companion \citep{rob07} will have no influence
on our measurements.  \citet{smi12} and \citet{ste12} discuss CHARA Array
$H$-band and $R$-band measurements of the disk size and orientation.

\noindent{\sl HD 10516.}
$\phi$ Per is a double-lined spectroscopic system with a
hot subdwarf companion \citep{gie98}.  We used the FUV flux ratio,
temperatures, and gravities from \citet{gie98} and derived
magnitude differences by scaling model spectral energy distributions
from \citet{lan03}.

\noindent{\sl HD 22192.}
No companion is evident in speckle data \citep{mas97}
nor in CHARA VEGA interferometric observations \citep{del11}.

\noindent{\sl HD 23630.}
Alcyone is a bright Pleiades member with seven visual components
listed in the WDS, but all these components have separations greater
than 79 arcsec. The star appears single in speckle \citep{mas97}
and AO observations \citep{rob07}.

\noindent{\sl HD 23862.}
Pleione is a single-lined spectroscopic binary with a low mass
companion, possibly a hot subdwarf or a M-dwarf \citep{nem10}.
The next companion is CHARA 125 that has a separation of $\rho \approx 0\farcs23$
\citep{mas93,rob07}, but it is not always detected
in speckle measurements, indicating a large magnitude difference of
$\triangle V \approx 3.5$ mag.  \citet{lut94} present
a radial velocity study that suggests that the orbital period is $\approx
35$~y, and the eccentricity is found to be large.  We assumed that
the orbital period is the same as that associated with the spectroscopic
shell episodes $P=34.5$~y \citep{sha88,tan07}
and that speckle companion CHARA 125 is the companion in this orbit \citep{gie90}.
We then made a preliminary orbital fit of the speckle data \citep{mas93}
to arrive at the elements presented in Table~5, although we caution that
this is probably only one of a family of possible solutions.
There are five other, fainter but wider ($\rho > 4\farcs6$), visual
companions in the WDS, which should not affect our measurements.

\noindent{\sl HD 25940.}
No companion is evident in the CHARA VEGA observations of \citet{del11}.
No speckle observations of this star are published.

\noindent{\sl HD 37202.}
\citet{ruz09} present an analysis of the
single-lined spectroscopic orbit for $\zeta$~Tau.
No other components are found in speckle \citep{mas93}
or interferometric observations \citep{ste09,sch10}.
Interferometric observations of disk asymmetries are described by
\citet{ste09}, \citet{car09}, and \citet{sch10}.

\noindent{\sl HD 58715.}
\citet{jar89} suggest that $\beta$~CMi is a
single-lined, spectroscopic binary with a period of 218~d, but
this result has not yet been confirmed by other investigators.
Interferometric studies by \citet{mei09} and \citet{kra12}
show no evidence of a close companion.  Furthermore,
no companion is found in speckle data \citep{mas93} and
AO imaging (\citealt{jan11}).
Eight faint and distant companions are listed in the WDS.

\noindent{\sl HD 109387.}
\citet{saa05} show that $\kappa$~Dra is a single-lined,
spectroscopic binary with a faint companion.  \citet{gie07} found
that the addition of a hot companion improved the fit of
the $K$-band interferometry, but \citet{jon08} point out that
density exponent derived by \citet{gie07} is significantly lower
than that determined from H$\alpha$ interferometry.  This discrepancy
casts some doubt about the detection of the companion.
There are no published speckle measurements of this star.

\noindent{\sl HD 138749.}
$\theta$ CrB has a companion whose separation has increased
from $0\farcs642$ in 1976 to $0\farcs813$ in 2010 according to the WDS,
but the position angle varied by only $4^\circ$ over the same interval,
and this suggests a large orbital eccentricity and/or an inclination
$\approx 90^\circ$. The $B,V$-band magnitudes were measured
by \citet{fab00}, and these suggest that the system
consists of a B6~Vnne primary and a A2~V secondary. Assigning
masses for these classifications and assuming that the semimajor
axis is close to the smallest observed ($a = 0\farcs5$), we have
calculated a preliminary period that is given in Table~\ref{orbital}.
There are 28 measurements (from 1976 to 2010) of separation and position
angle in the Fourth Catalog of Interferometric Measurements of Binary Stars
\citep{har01} that we used to derive the preliminary orbital
elements that are given in Table~5.  There are no obvious radial velocity
variations indicative of a spectroscopic binary \citep{riv06}.

\noindent{\sl HD 142926.}
\citet{kou97} present a single-lined
spectroscopic orbit for 4~Her. They argue that the companion star
must be a small and faint object since they see no evidence of its spectral
features.  No other companions are observed with speckle interferometry
\citep{mas97} nor are any companions listed in the WDS.

\noindent{\sl HD 142983.}
The spectrum of 48~Lib is dominated by shell features that
vary on a timescale of a decade, and it is very difficult to study
the photospheric spectrum of the star (B3:~IV:e~shell) to search for
radial velocity variations \citep{riv06,stf12}.
Unfortunately, there are no published speckle observations, and there are no
companions indicated in the WDS.  We assume it is a single object.
Recent $H$-band interferometric observations are discussed by \citet{stf12}.

\noindent{\sl HD 148184.}
\citet{har87} presents a preliminary single-lined
orbit for $\chi$~Oph with a period of 34.121~d.
There are no available speckle observations, and no
companion is indicated in the WDS.  \citet{tyc08}
obtained H$\alpha$ interferometric observations,
and they make no mention of evidence of a companion.

\noindent{\sl HD 164284.}
The visual companion of 66 Oph was first discovered by
\citet{mas09} and confirmed by \citet{tok10}.
We assumed that the current separation corresponds to the
angular semimajor axis, and we estimated the orbital period by
assigning masses assuming main sequence stars, the temperature from
\citet{fre05}, and the measured $\triangle V=2.7$ mag
\citep{tok10}.  There are only four measurements in the
Fourth Catalog of Interferometric Measurements of Binary Stars
\citep{har01}, but we used these to arrive at the
preliminary orbital elements presented in Table~5.
\citet{flo02} discuss spectroscopy of the star and
pulsational behavior, but no mention is made of a spectroscopic
binary companion.

\noindent{\sl HD 166014.}
\citet{tok85} reported a marginal detection of a close
companion to $o$~Her at a separation of 60~mas, but this
was not confirmed in later speckle observations by \citet{mas09}.
However, the interferometric measurements for this star
showed lower visibilities at short baselines, which could not be
fitted by the standard Gaussian elliptical model.
Consequently, we assumed that this decreased visibility is due
to the presence of the companion found by \citet{tok85}.
We applied the binary flux dilution correction to fit the data
by assuming a companion that is 2.5~mag fainter than the Be star
in the $K$-band.
There is no known spectroscopic companion, and \citet{gru07} found no
evidence of radial velocity variability.

\noindent{\sl HD 198183.}
The $\lambda$ Cyg system consists of at least four stars.
Component C is distant and faint, so we ignored its flux.
The AB system has a long period ($\approx 462$ y; \citealt{bai83}), and
we used the orbital elements from \citet{bai83} to estimate the position of A,B.
\citet{bal88} and \citet{bai93} determined an astrometric orbit
for the close pair MCA~63 Aa,Ab that apparently consists
of similar magnitude stars.  However, this close pair was not
detected in recent speckle observations by \citet{mas09},
presumably because their separation was too small at that time.
\citet{gru07} notes the presence of some-short term line profile variability
that might be explained as a composite spectrum consisting of
the Be star (Aa) plus a single-lined spectroscopic binary,
so it is possible that B or Ab has a companion.
The magnitude difference of A,B was determined in the Tycho system
by \citet{fab00}, and we converted this to a Johnson
$\triangle V= 1.46 \pm 0.02$ mag using the formulae from \citet{mam02}.
The magnitude and color differences of A,B suggest that
component B is a late B-type star, and we used $\triangle V$,
the effective temperature of Aa from \citet{hua08},
and the main sequence relations from \citet{lej01}
to estimate the $K$-band magnitude difference.
We encountered two problems related to the visibility correction
for the inner companion Ab.  First, the orbit from \citet{bai93}
predicted that the Aa and Ab fringe patterns would cross each
other for the baseline orientations of observations from JD 2,455,365,
and consequently, large visibility variations were predicted when
none were observed.  We suspect that the orbit needs revision,
and for the purposes of the visibility correction, we altered
the epoch of periastron from BY~1982.668 to BY~1981.526 so
that the predicted, projected separations at the time of our observations
were always larger than half the fringe scan length.
The second problem concerned the flux contribution of Ab.
W.\ Hartkopf (USNO) kindly retrieved his
speckle data on the close Aa,Ab pair and derived an approximate
magnitude difference of $\triangle V= 0.4 \pm 0.3$ mag.
Because we have no color information for the close pair,
we simply assumed that $K$-band magnitude difference was the same.
We found that with Ab this bright, the dilution correction
was too large and led to corrected visibilities larger than one.
By increasing this estimate to $\triangle V= 0.4 + 1\sigma = 0.7$ mag,
the mean of the corrected visibilities was approximately one.
This indicates that the Be star disk was probably unresolved,
unless the flux contributions of the companions are actually
significantly lower than these estimates.

\noindent{\sl HD 200120.}
59~Cyg has a nearby B companion \citep{mas09} plus
three other very distant and faint components.  B.\ Mason (USNO) kindly provided us with a
preliminary orbit for A,B (Table 5) that we used to estimate the position and separation at the
times of our CHARA Array observations.  The Be star is also a spectroscopic binary
with a hot subdwarf companion \citep{mai05}, and consequently we
assume that the smaller $\triangle K$ (brighter) estimate is more reliable in Table~4.

\noindent{\sl HD 202904.}
$\upsilon$~Cyg has four faint and distant companions listed in
the WDS, but there is no close companion detected in speckle interferometric
observations \citep{mas97}.  \citet{nei05} discuss spectroscopic
radial velocities that may be consistent with binary motion for a period
of 11.4~y, but further measurements are required to verify their suggestion.

\noindent{\sl HD 203467.}
There are no companions of 6~Cep listed in the WDS, and,
unfortunately, there are no published speckle observations of this star.  Spectroscopic
observations are discussed by \citet{kou03} who show that the profiles
vary with a 1.621~d cycle, a period that is probably related to pulsation or rotation.

\noindent{\sl HD 209409.}
There are no companions of $o$~Aqr listed in the WDS, and no
companions were found by \citet{oud10} using adaptive optics observations with
VLT/NACO.  \citet{riv06} discuss spectroscopy of this Be-shell star and
note no evidence of a binary companion.  \citet{mei12} obtained preliminary
$K$-band interferometry with VLTI/AMBER, but they did not resolve the disk
in the continuum.

\noindent{\sl HD 212076.}
No companions of 31~Peg are listed in the WDS.
\citet{riv03} describe the short term spectroscopic
variations related to pulsations, but there is no evidence
of a spectroscopic companion.

\noindent{\sl HD 217675.}
\citet{zhu10} present a re-analysis of all the
existing plus new astrometric and radial velocity measurements for $o$~And.
They show that the system has a $2+2$ hierarchy and the pairs
share a wide orbit with a period of 117~y.  The A component is
probably a spectroscopic binary consisting of the Be star and
late-B star companion in a 5.7~y orbit, while the B component
consists of a pair of similar late-B stars in a 33~d spectroscopic orbit.
We treated the B component as a single object because the binary
separation is so small, and then we corrected the visibilities
using the orbits from \citet{zhu10} and the magnitude
differences of $\triangle V({\rm A,B})=2.21$ \citep{zhu10}
and $\triangle V({\rm Aa,Ab})=1.90$ \citep{hor04}.
The flux ratios in the $K$-band were estimated using the $V-K$ color
calibration from \citet{weg94} and spectral types from \citet{zhu10}.

\noindent{\sl HD 217891.}
No companions of $\beta$~Psc are listed in the WDS and none
were found in adaptive optics observations by \citet{rob07}.
\citet{dac86} discuss radial velocity measurements that appear
to be relatively constant.





\clearpage
\setcounter{figure}{0}
\begin{figure*}
\begin{center}
{\includegraphics[angle=0,height=5.5cm]{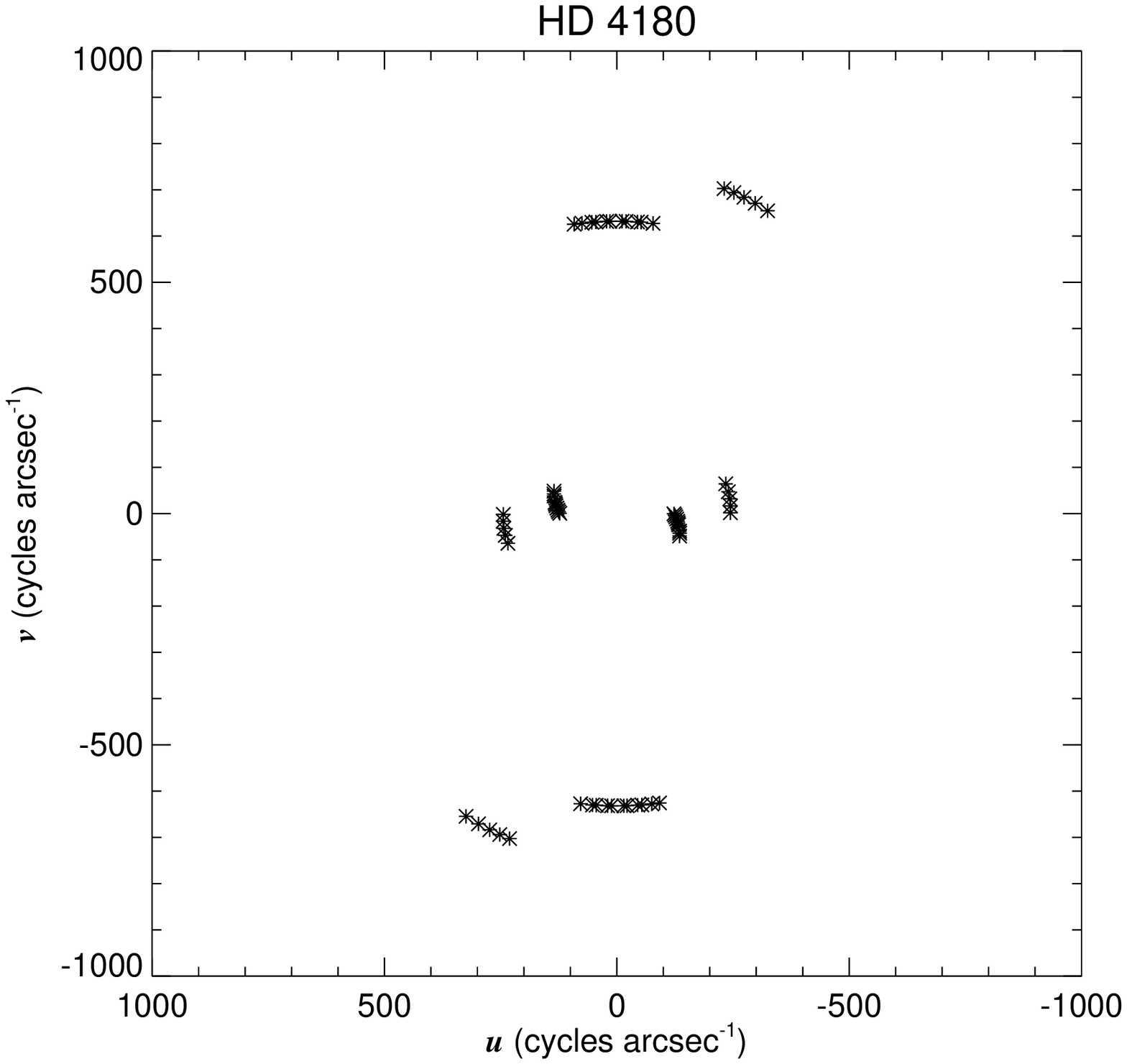}}
{\includegraphics[angle=0,height=5.5cm]{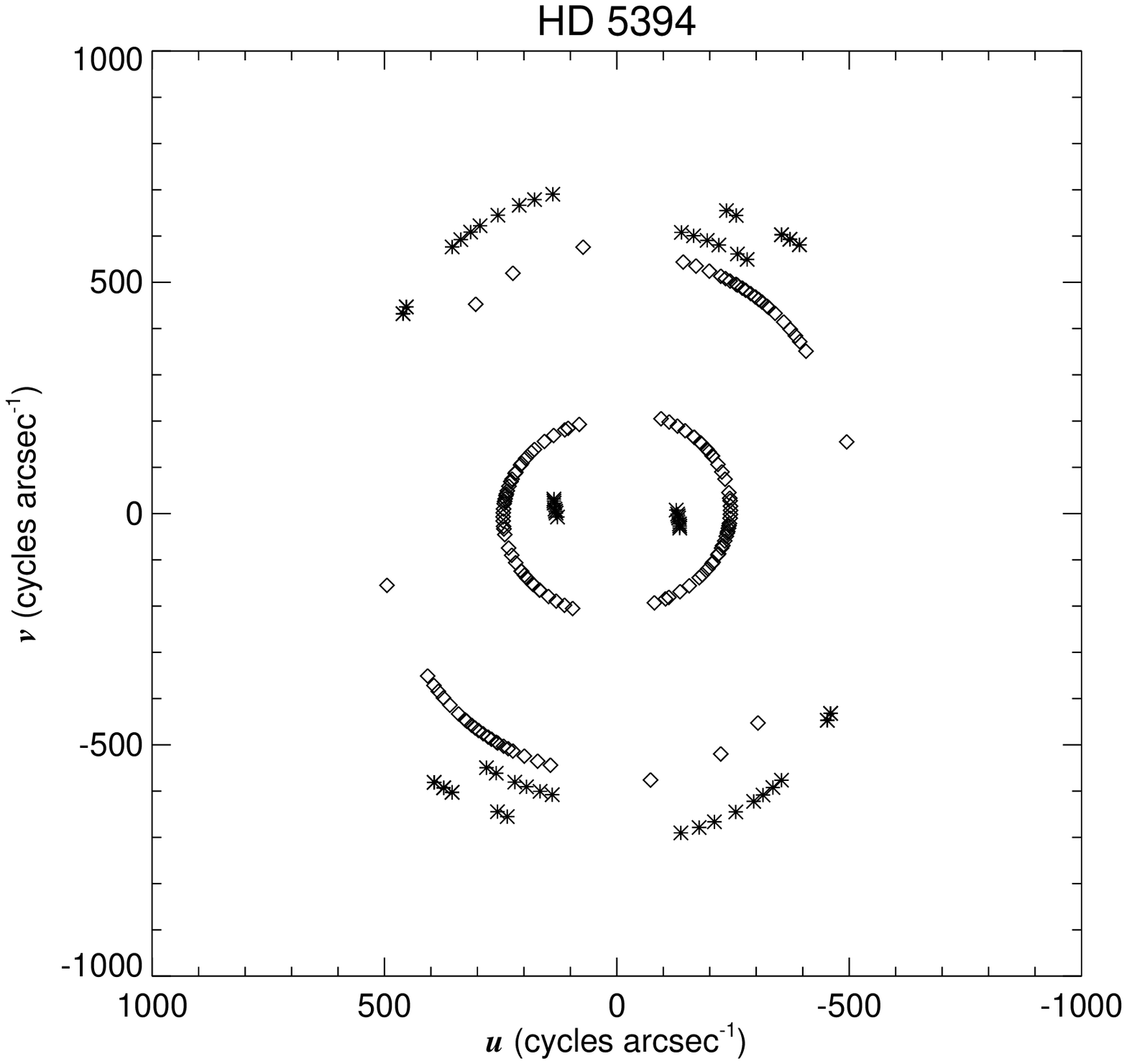}}
{\includegraphics[angle=0,height=5.5cm]{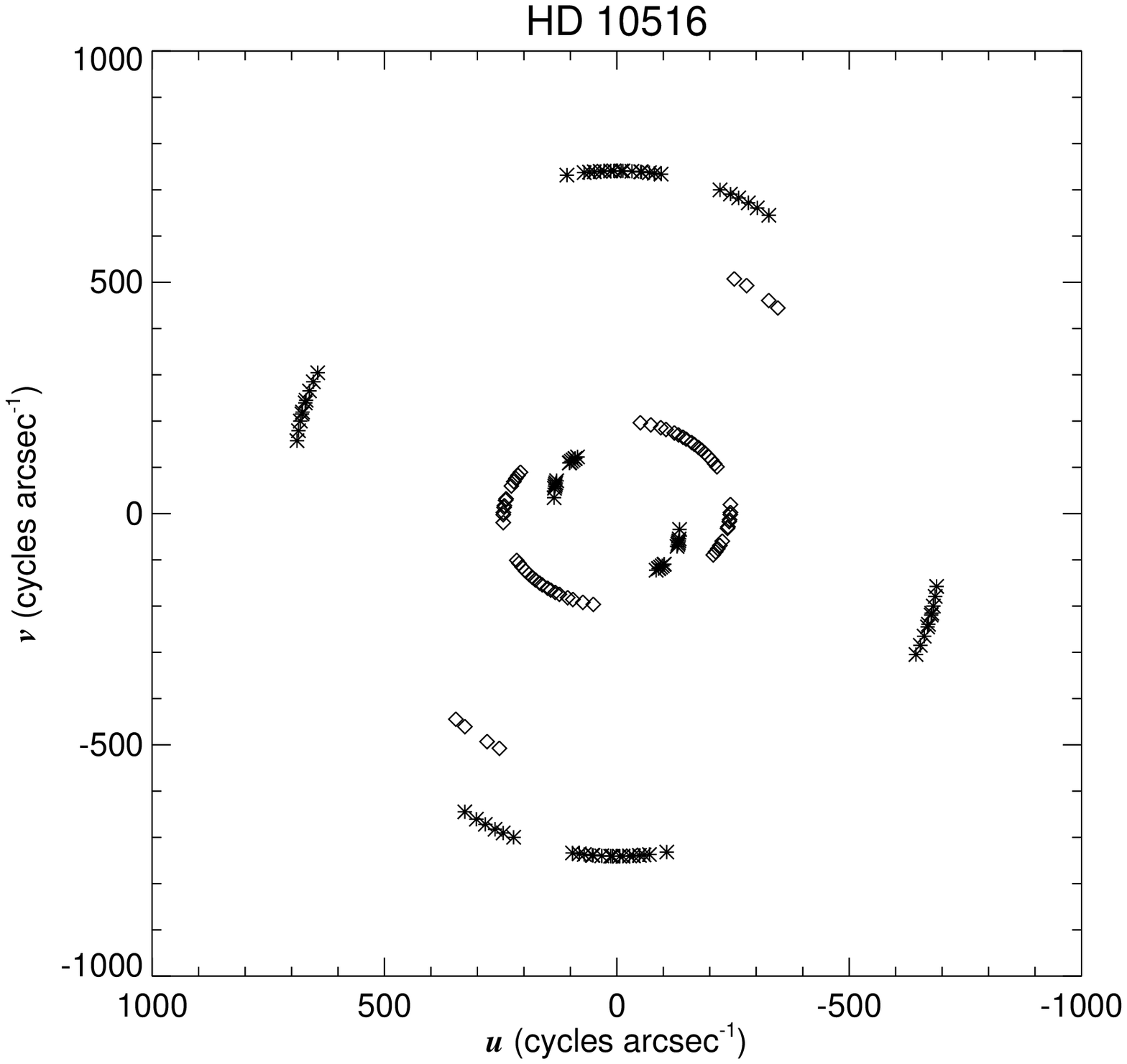}}
{\includegraphics[angle=0,height=5.5cm]{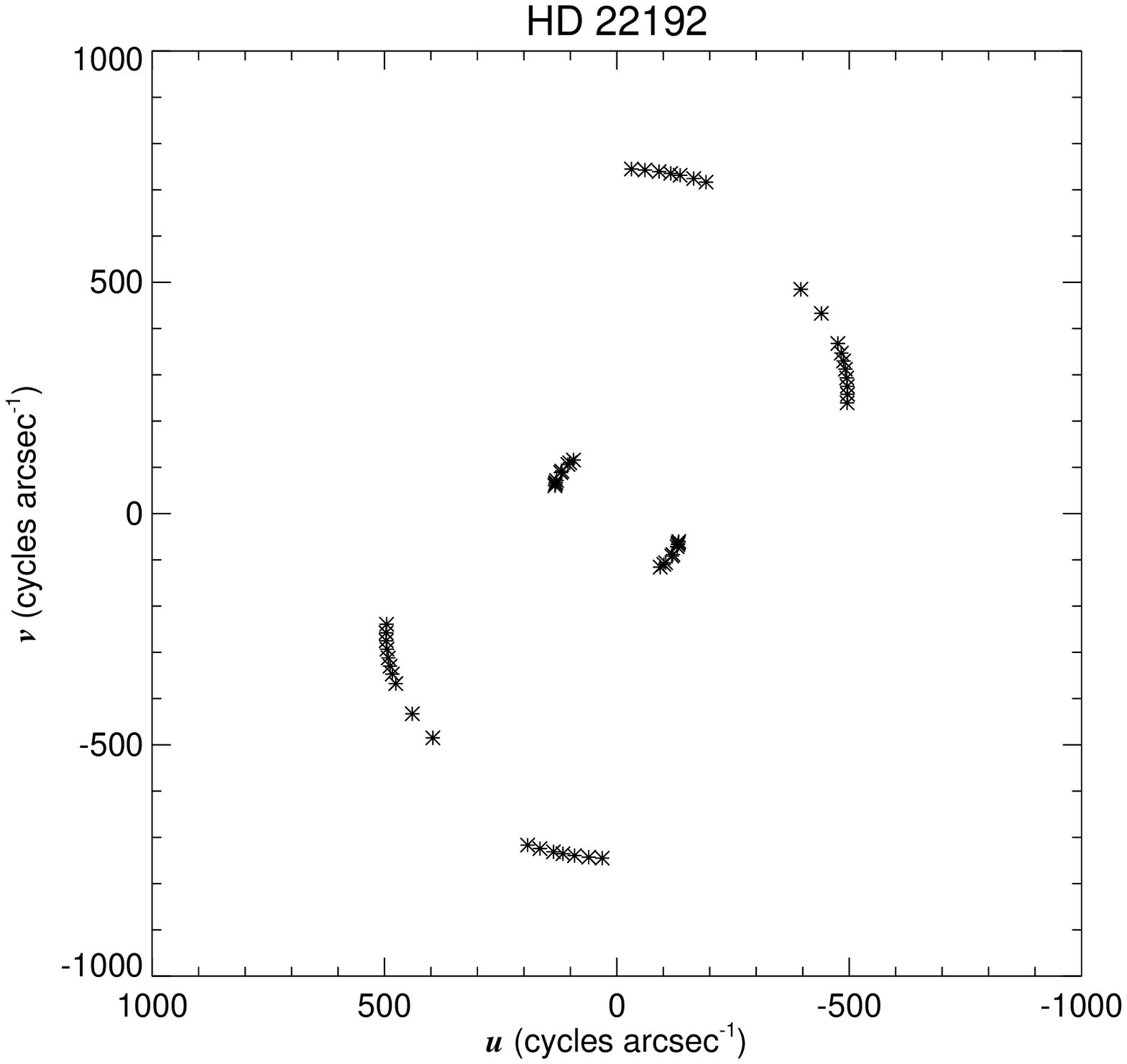}}
{\includegraphics[angle=0,height=5.5cm]{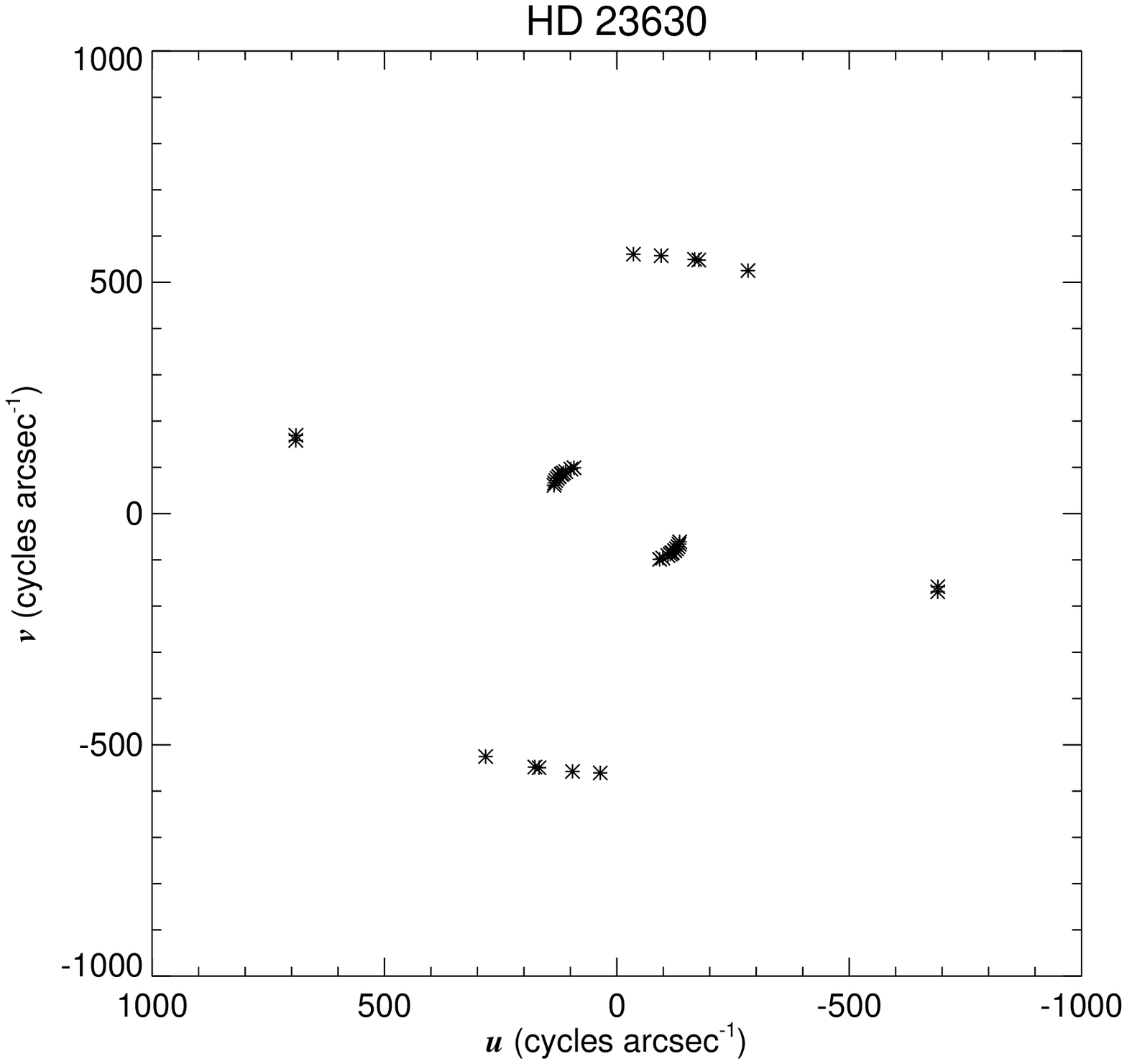}}
{\includegraphics[angle=0,height=5.5cm]{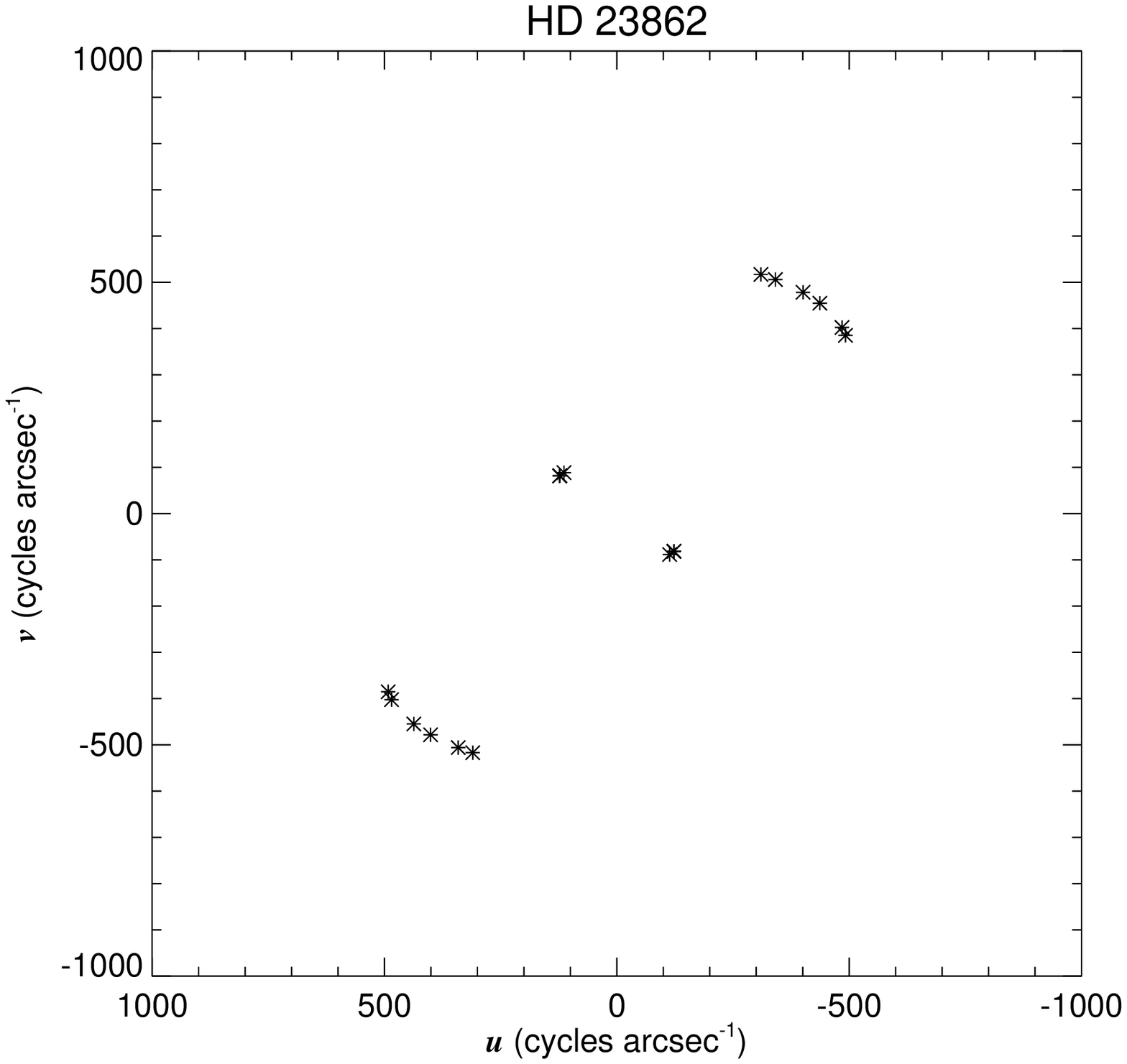}}
{\includegraphics[angle=0,height=5.5cm]{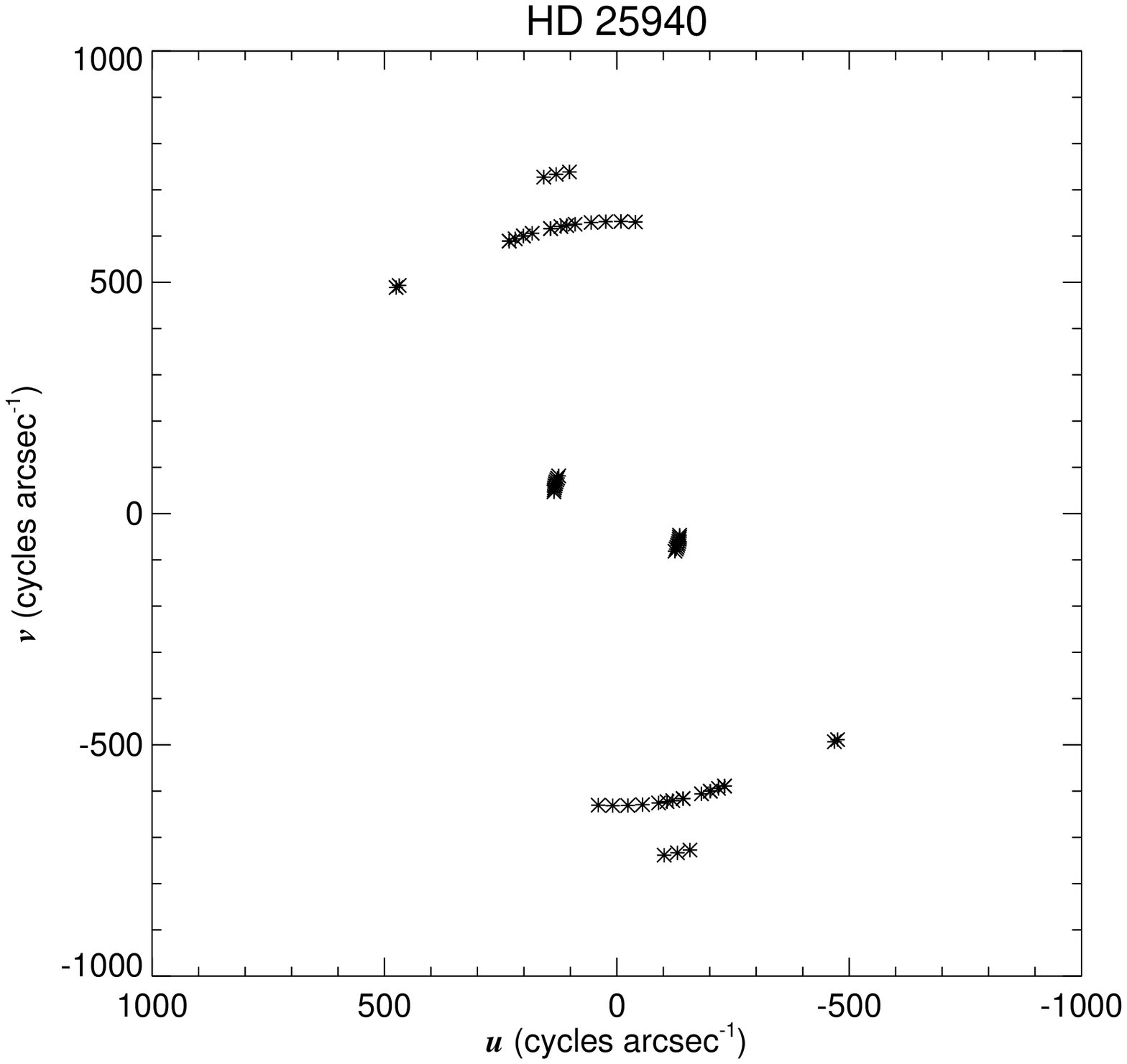}}
{\includegraphics[angle=0,height=5.5cm]{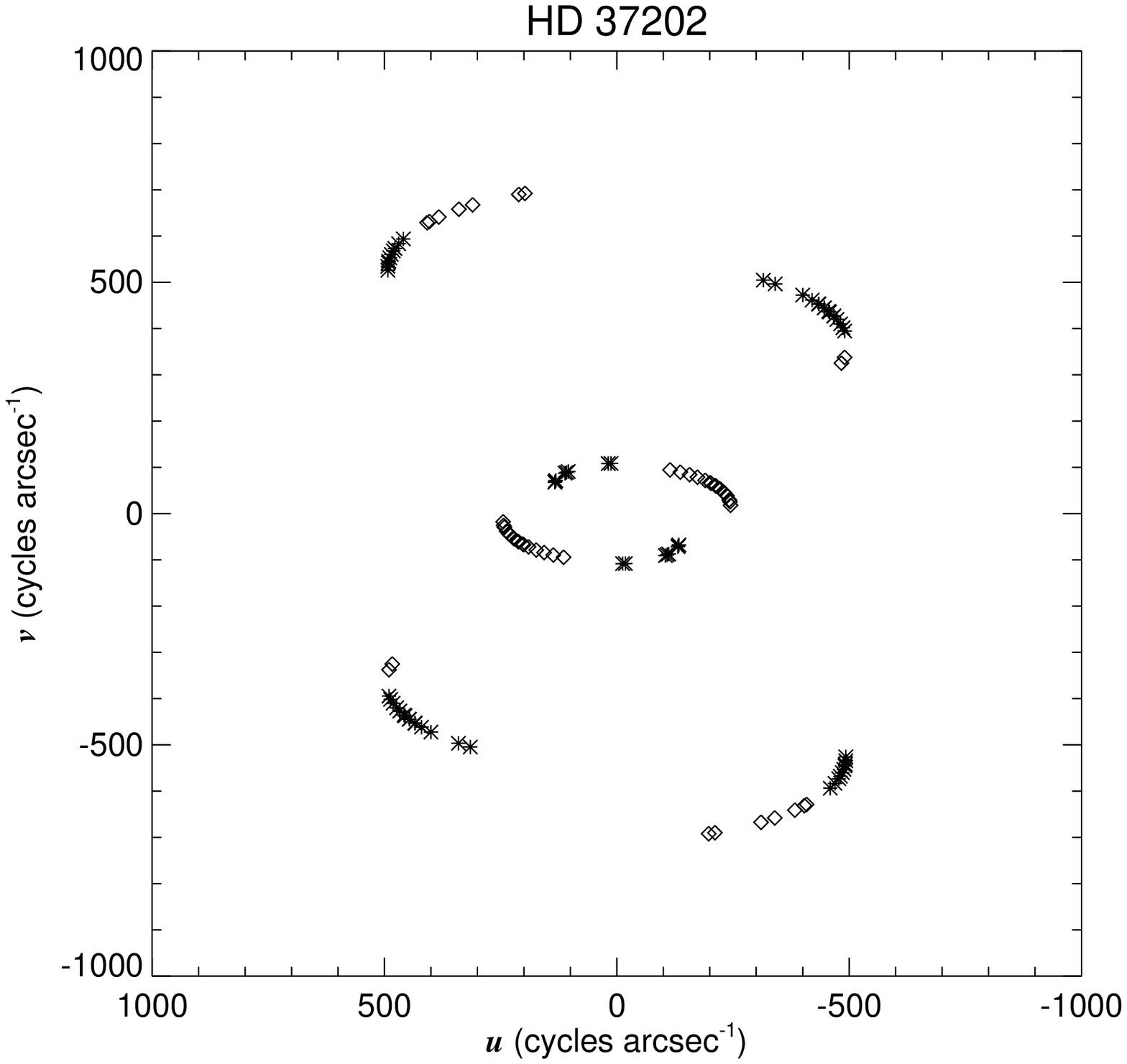}}
\end{center}
\caption[Sampling of the $(u, v)$ plane]
{1.1 -- 1.8. Sampling of the frequency $(u, v)$ plane
for our Be star sample. Observations conducted in this survey are indicated by star symbols while
archived measurements from \citet{gie07} are shown by diamonds.}
\label{uv1}
\end{figure*}

\clearpage

\setcounter{figure}{0}
\begin{figure*}
\begin{center}
{\includegraphics[angle=0,height=5.5cm]{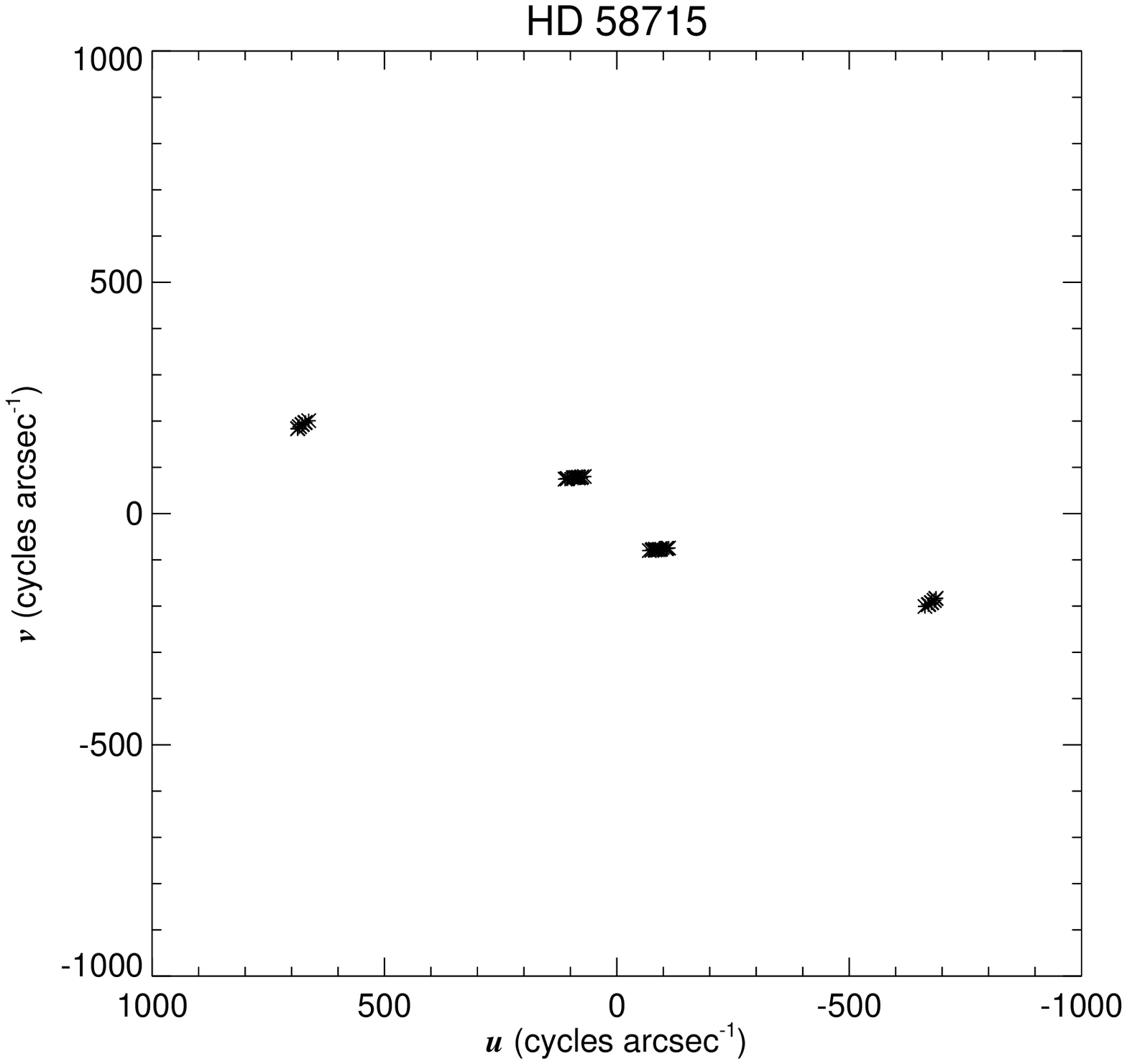}}
{\includegraphics[angle=0,height=5.5cm]{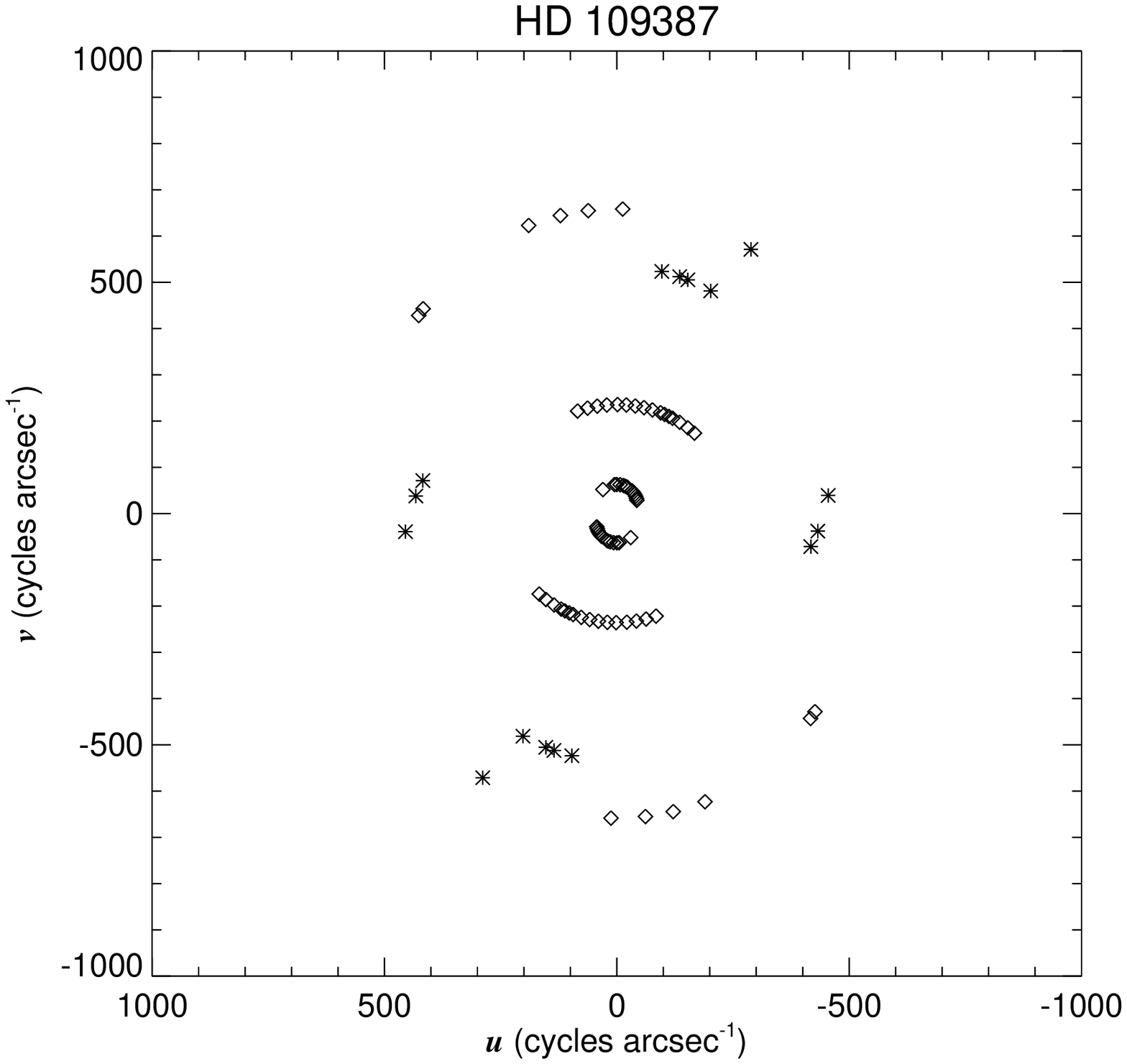}}
{\includegraphics[angle=0,height=5.5cm]{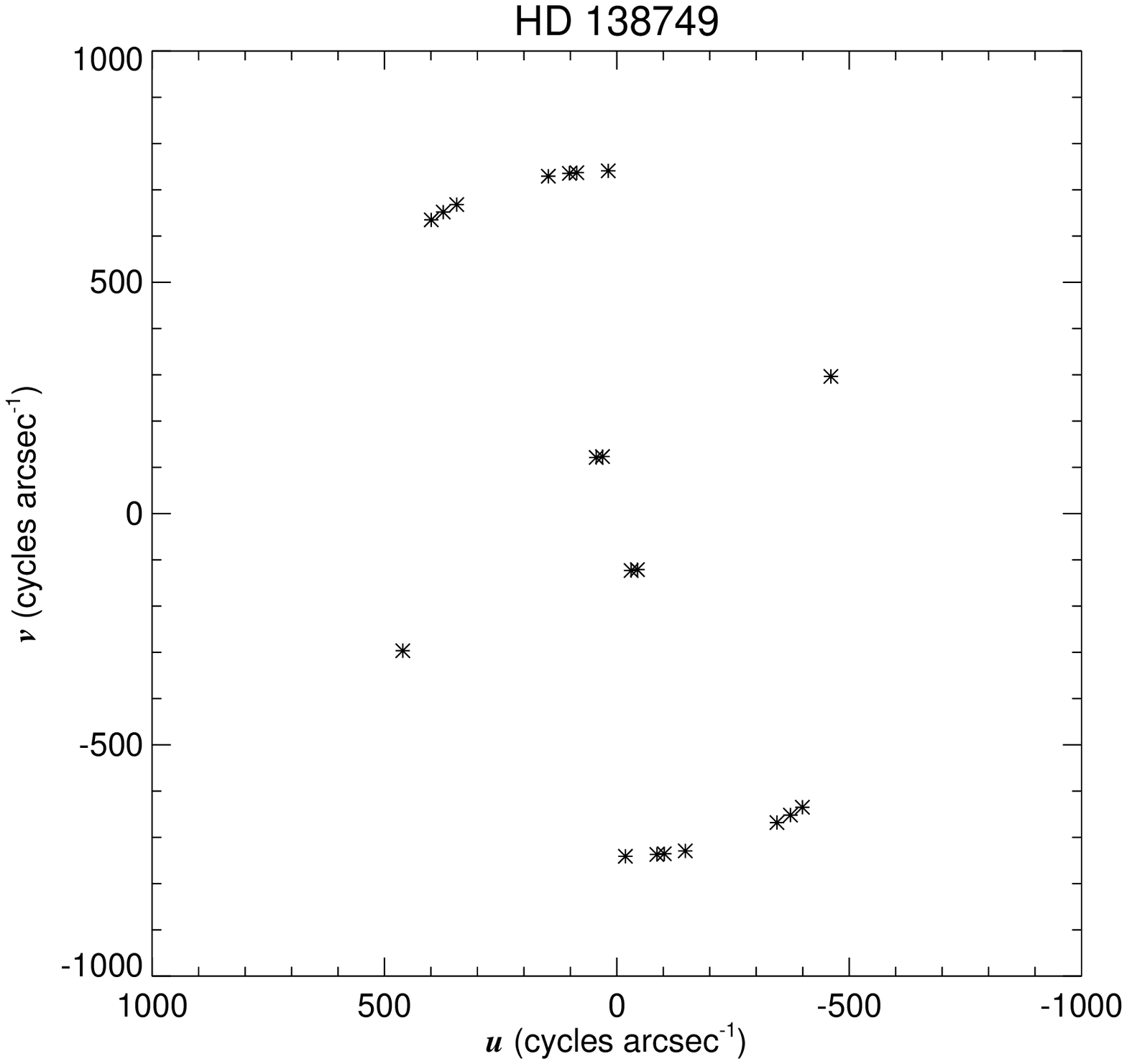}}
{\includegraphics[angle=0,height=5.5cm]{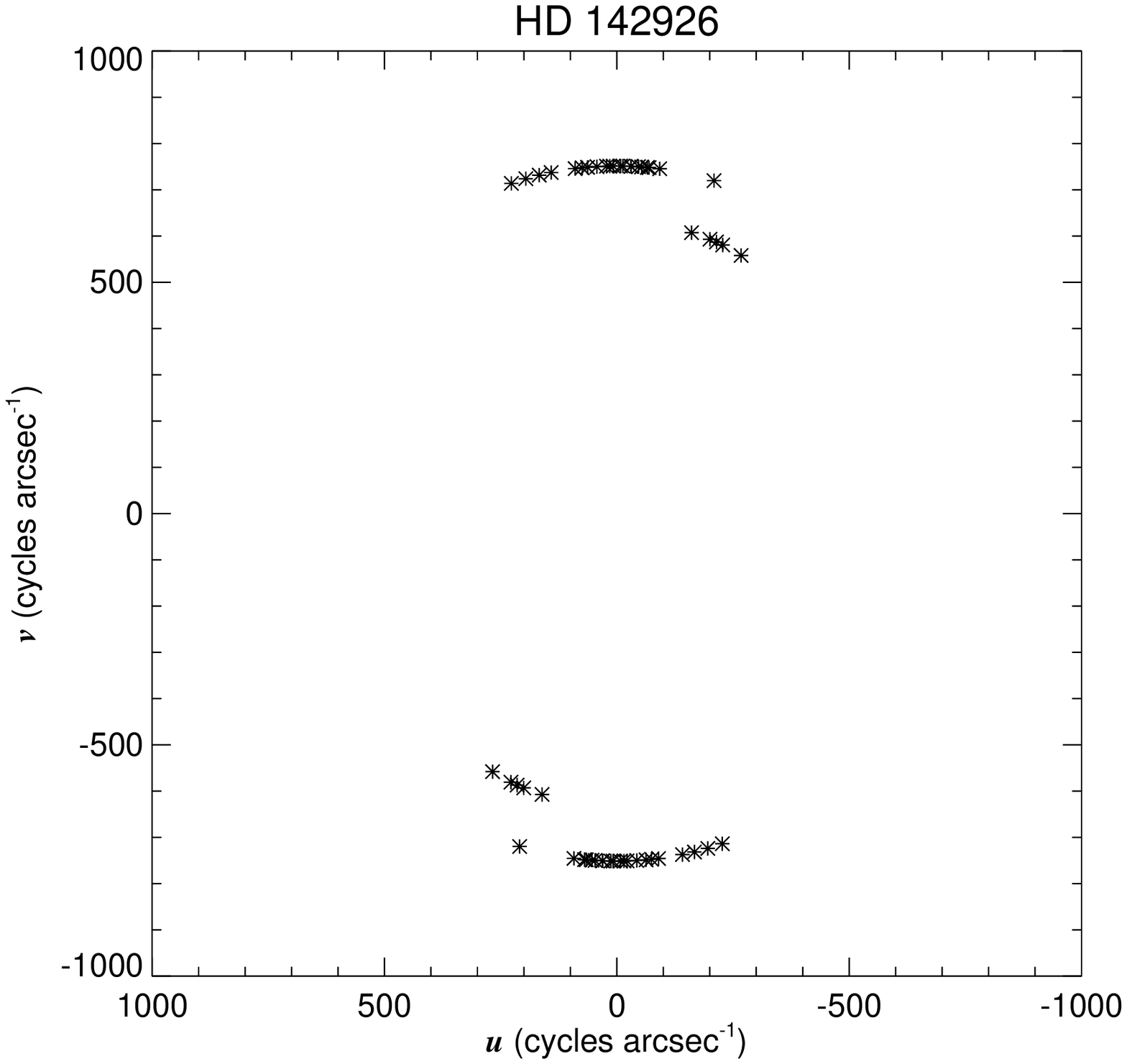}}
{\includegraphics[angle=0,height=5.5cm]{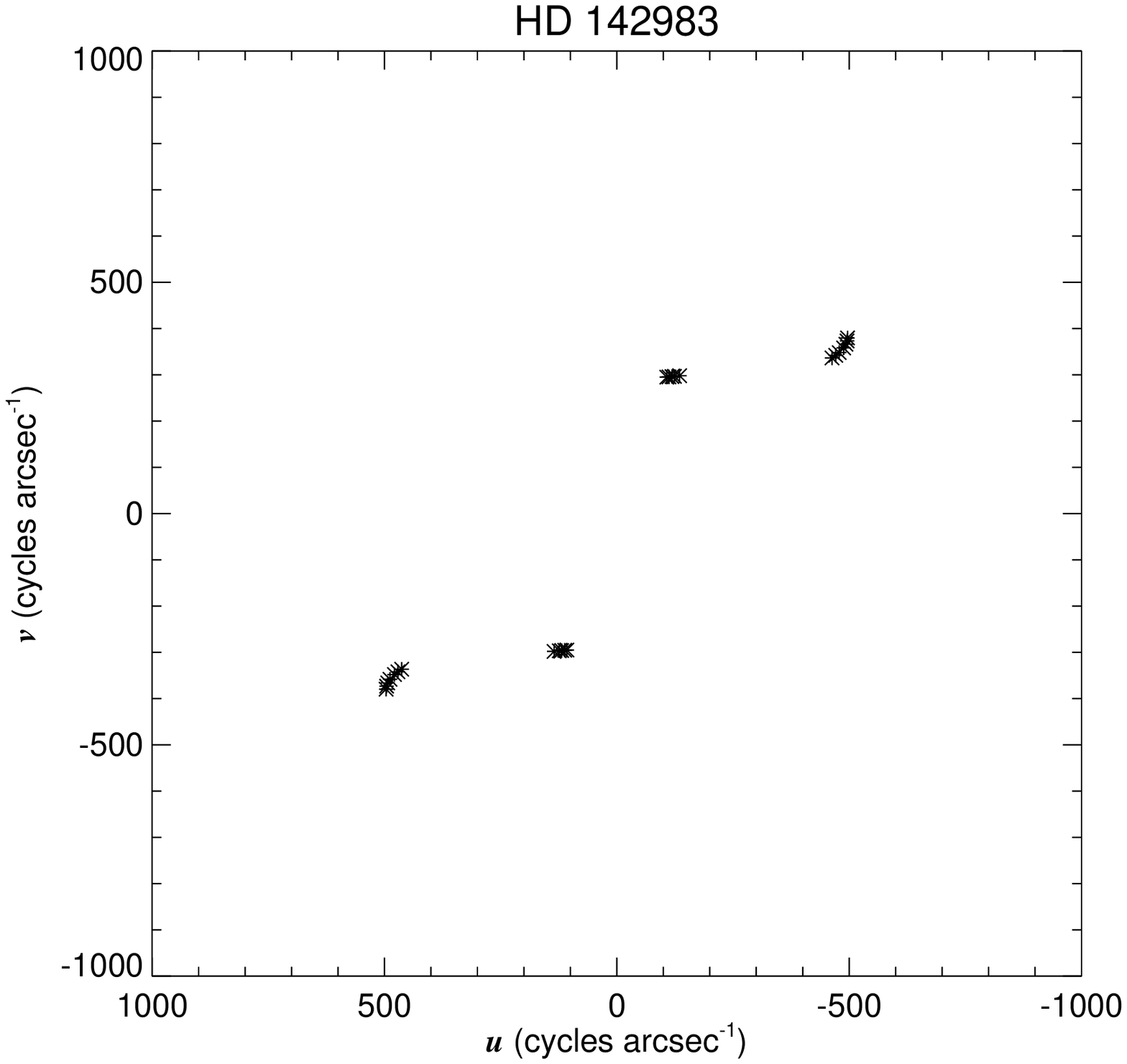}}
{\includegraphics[angle=0,height=5.5cm]{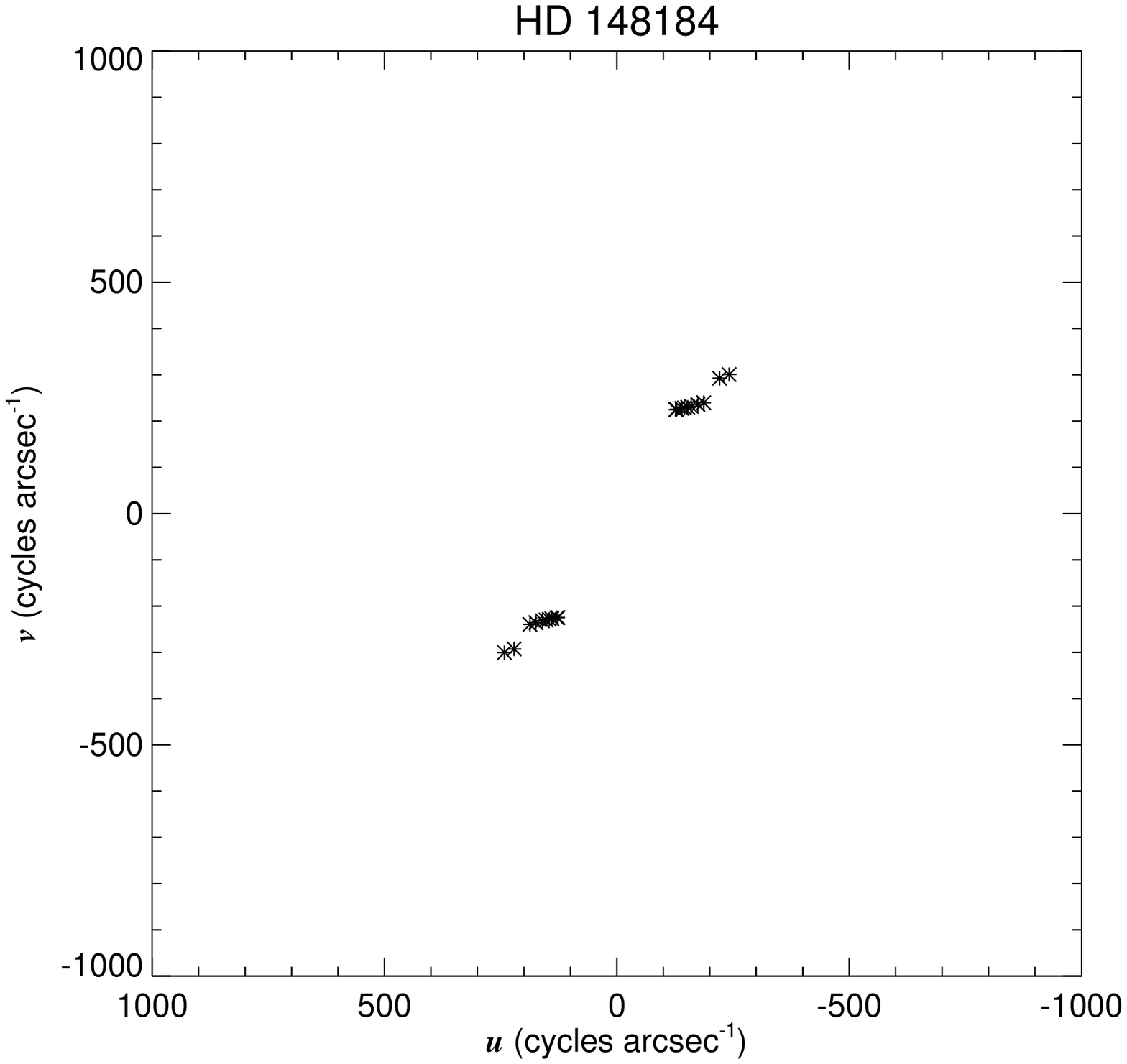}}
{\includegraphics[angle=0,height=5.5cm]{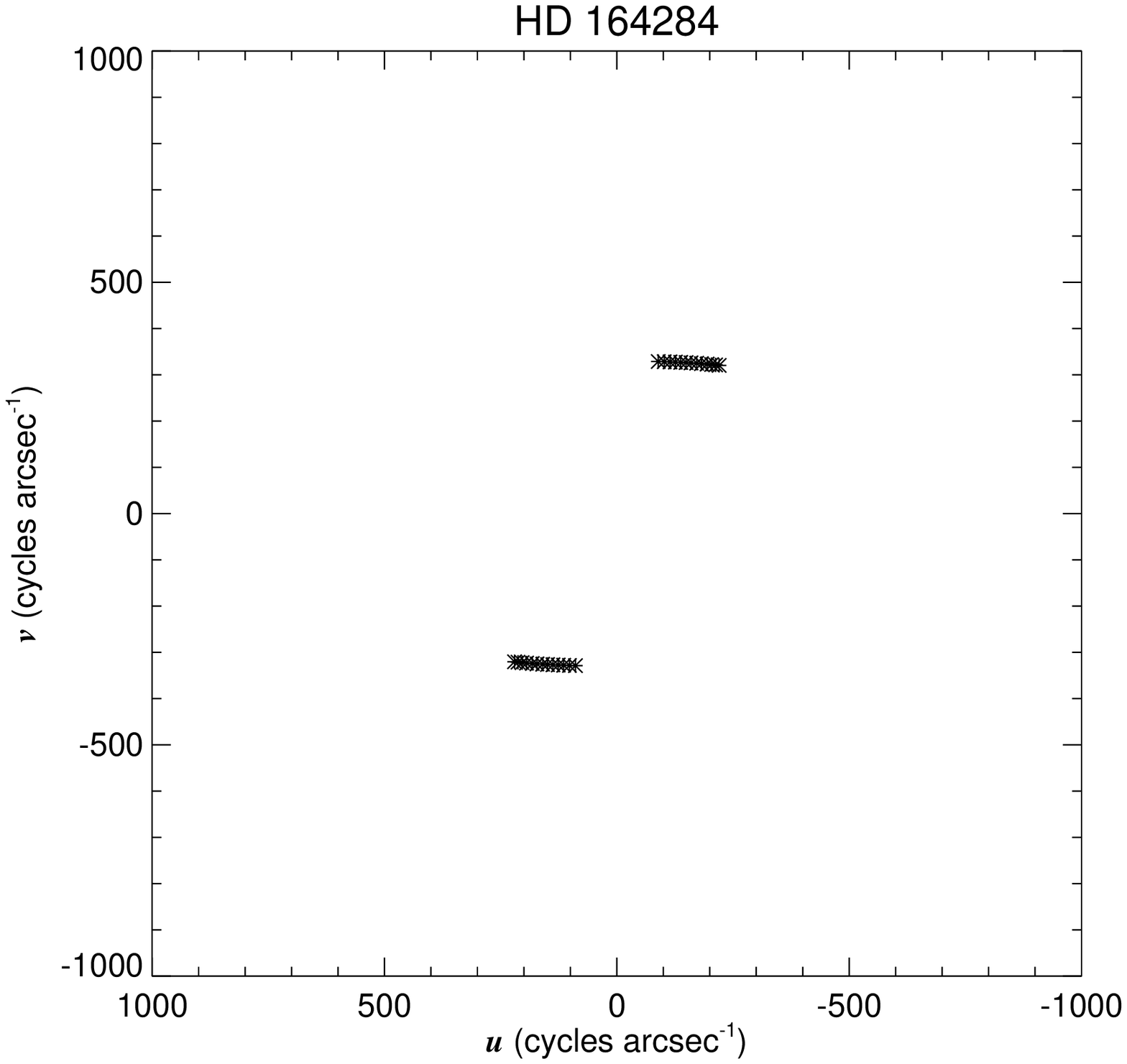}}
{\includegraphics[angle=0,height=5.5cm]{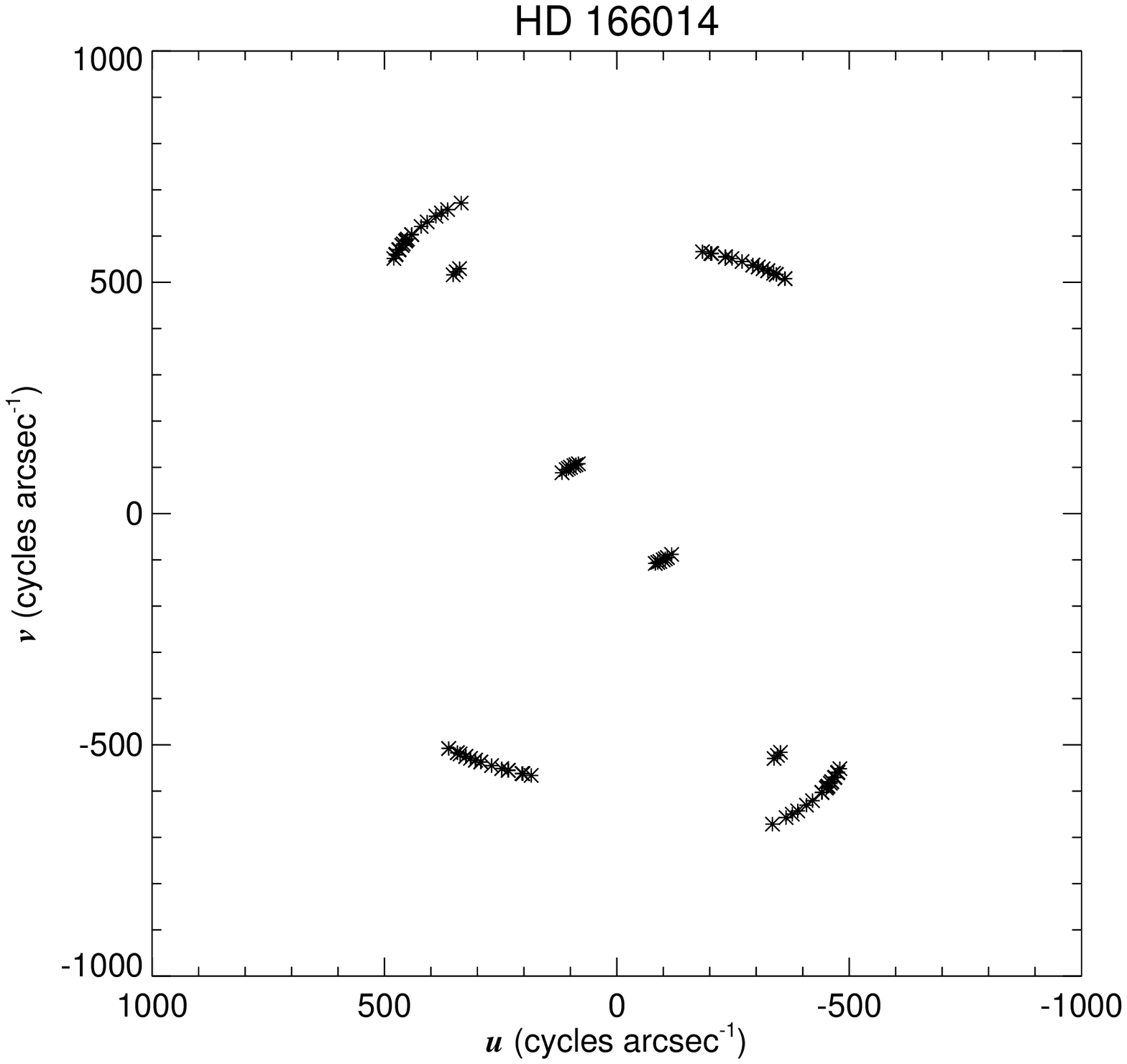}}
\end{center}
\caption[Sampling of the frequency $(u, v)$ plane for the $K$-band observations]
{1.9 -- 1.16. Sampling of the frequency $(u, v)$ plane for our sample stars.}
\label{uv2}
\end{figure*}

\clearpage

\setcounter{figure}{0}
\begin{figure*}
\begin{center}
{\includegraphics[angle=0,height=5.5cm]{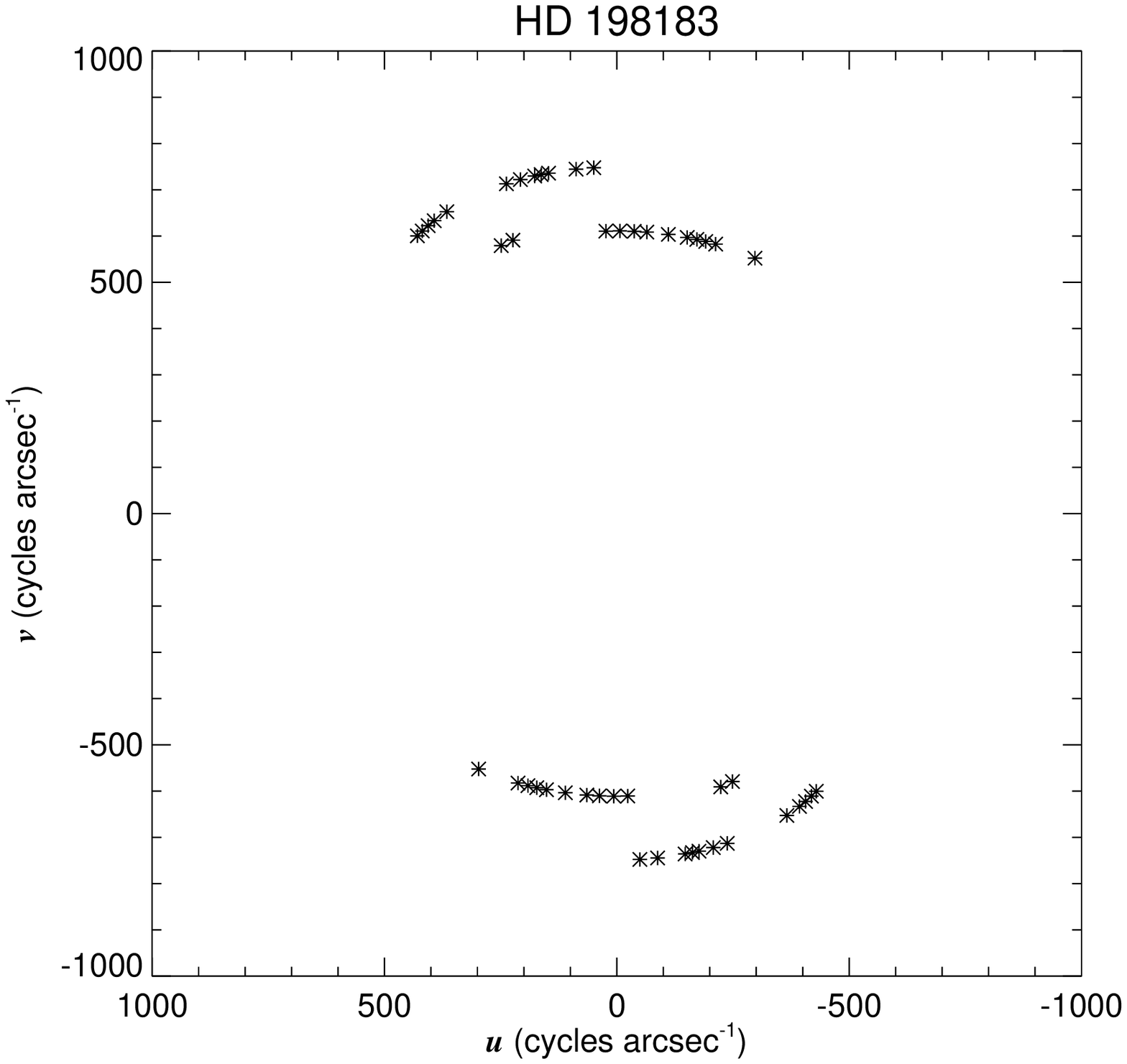}}
{\includegraphics[angle=0,height=5.5cm]{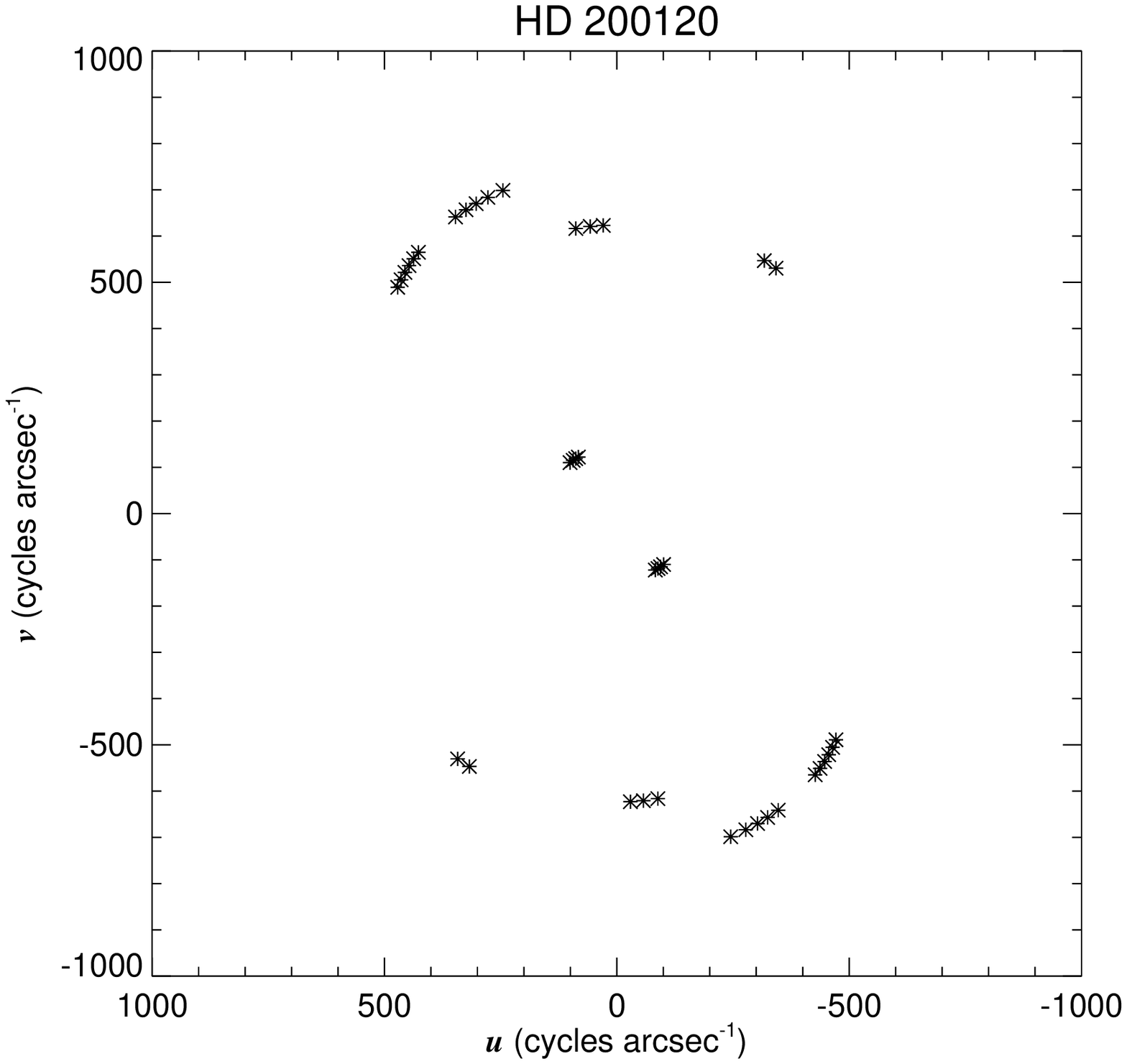}}
{\includegraphics[angle=0,height=5.5cm]{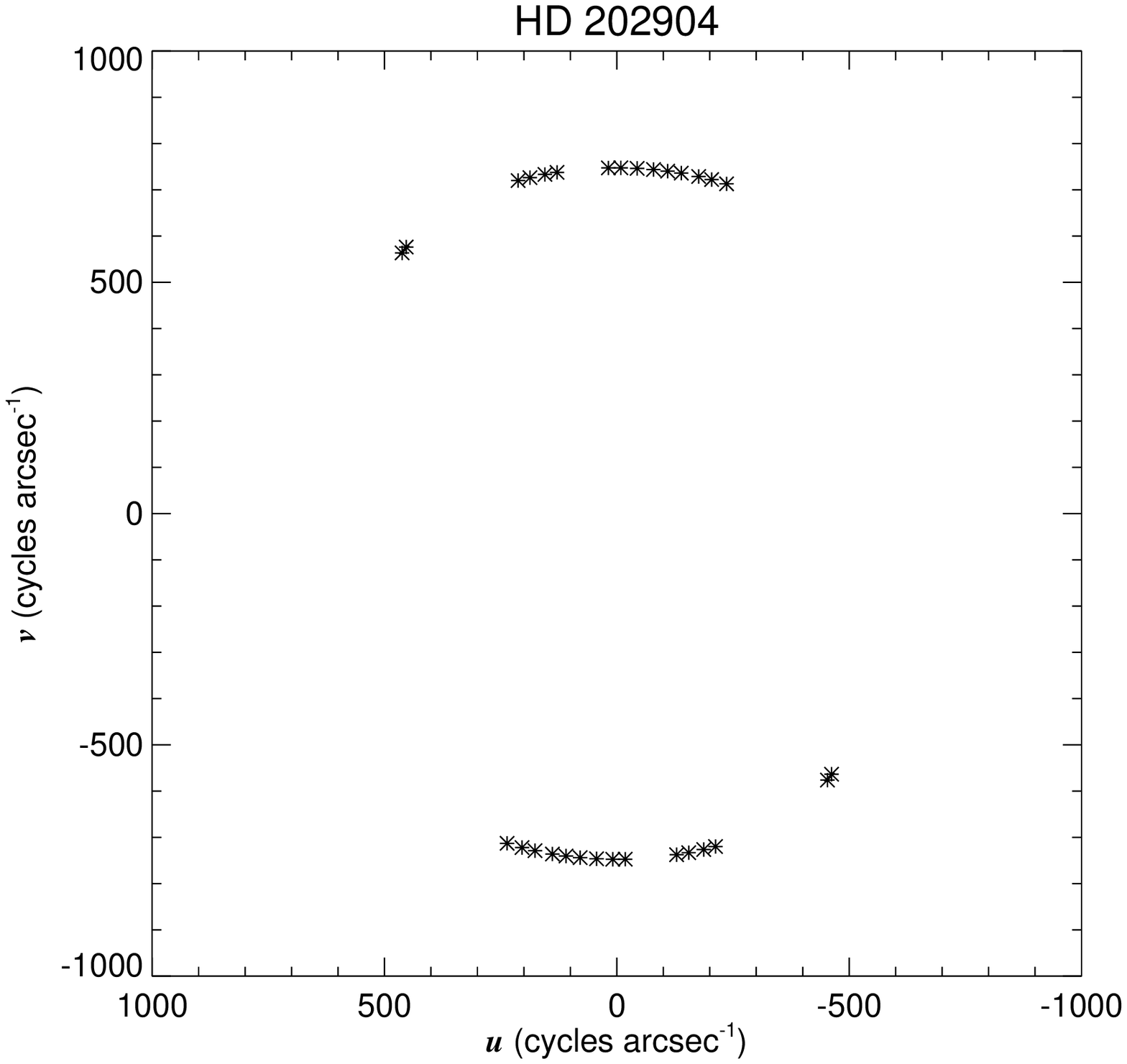}}
{\includegraphics[angle=0,height=5.5cm]{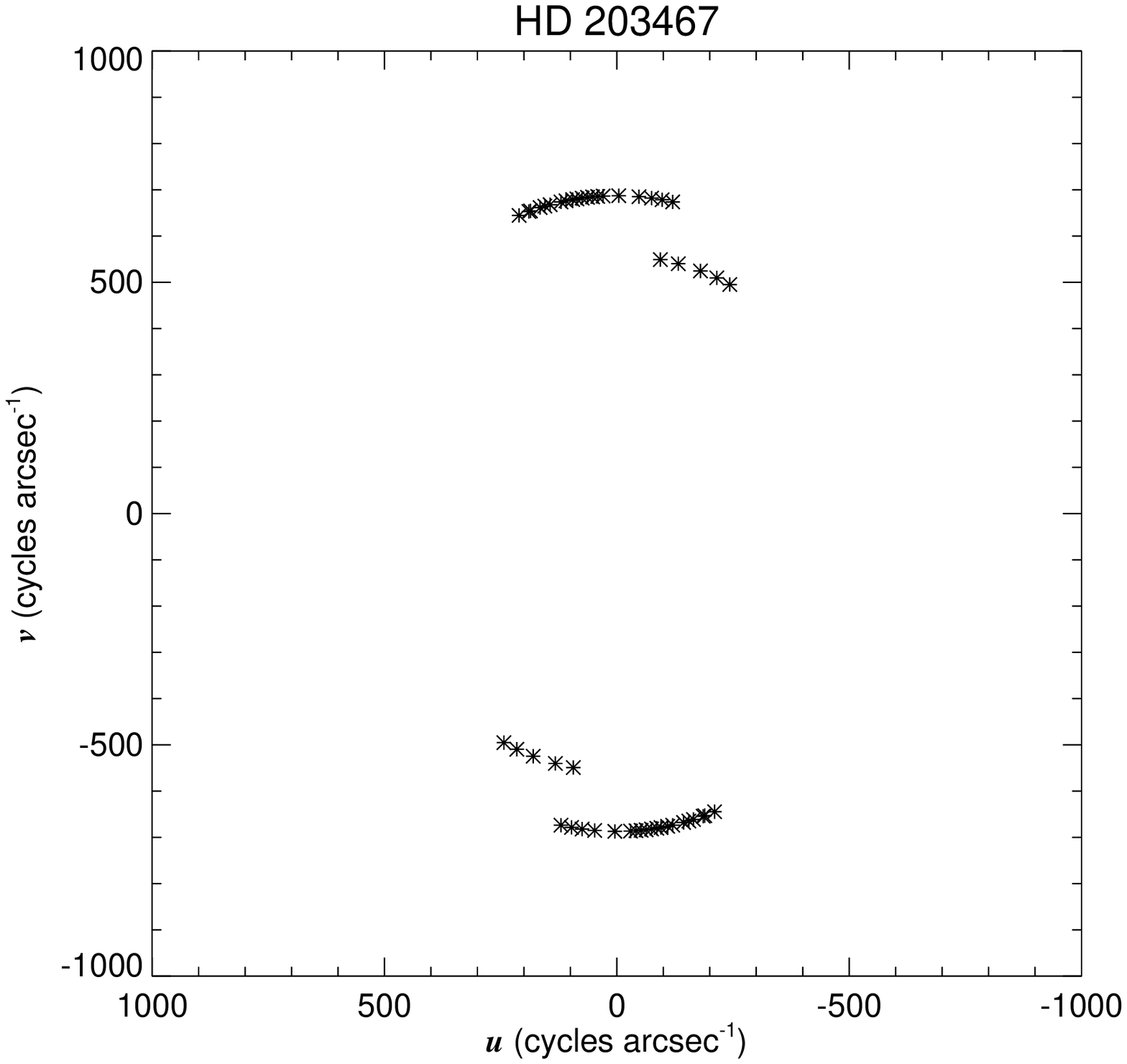}}
{\includegraphics[angle=0,height=5.5cm]{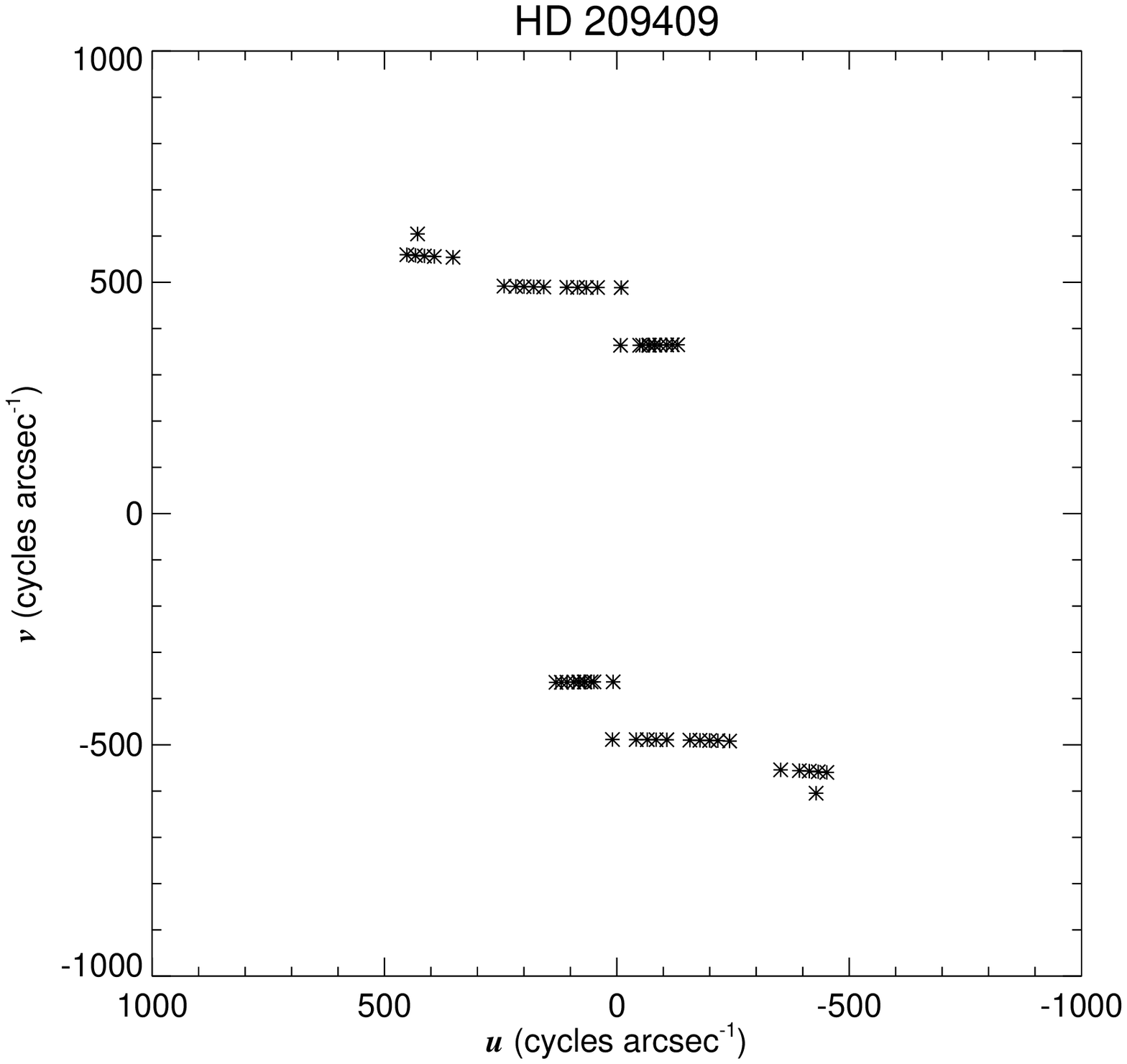}}
{\includegraphics[angle=0,height=5.5cm]{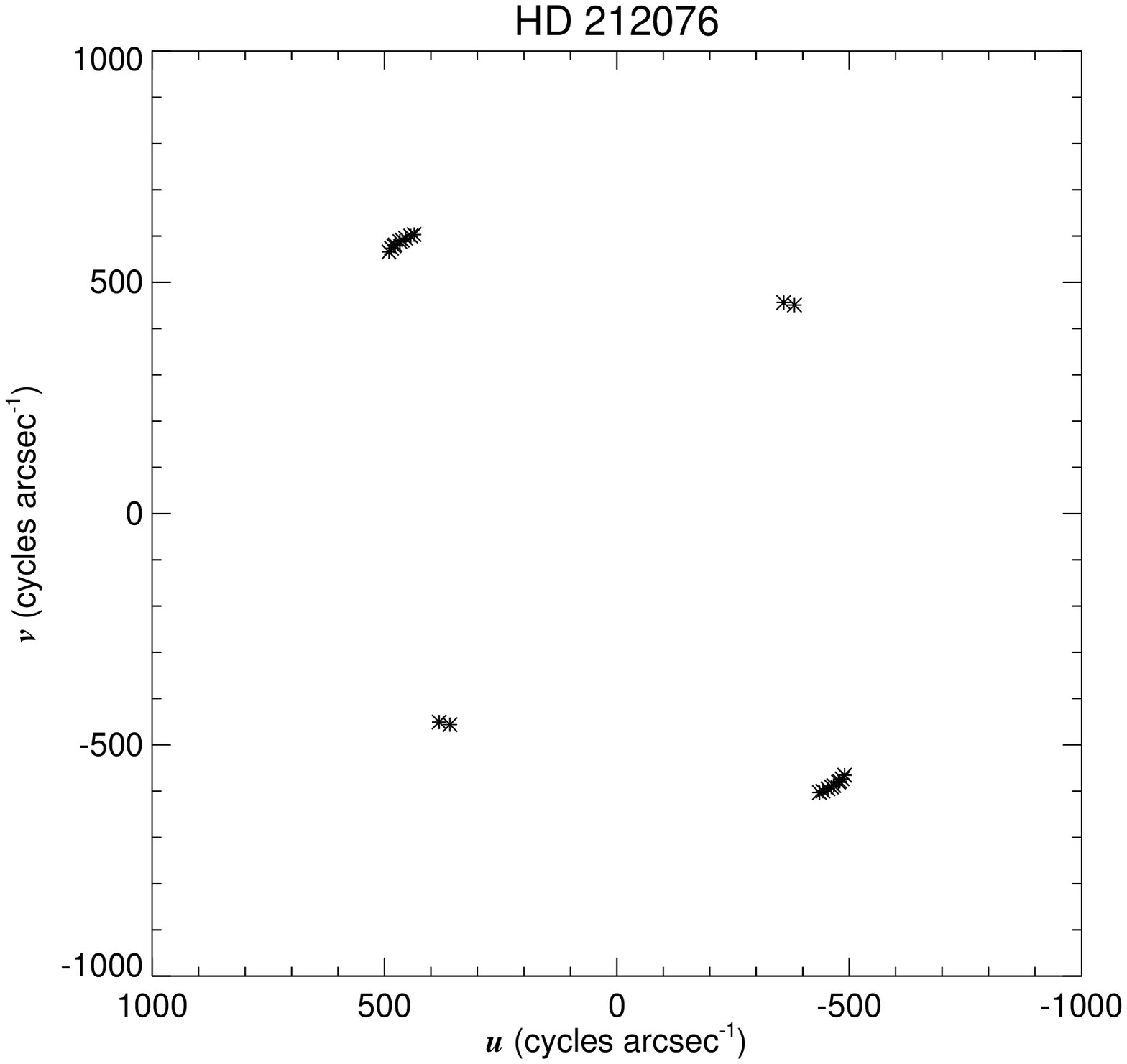}}
{\includegraphics[angle=0,height=5.5cm]{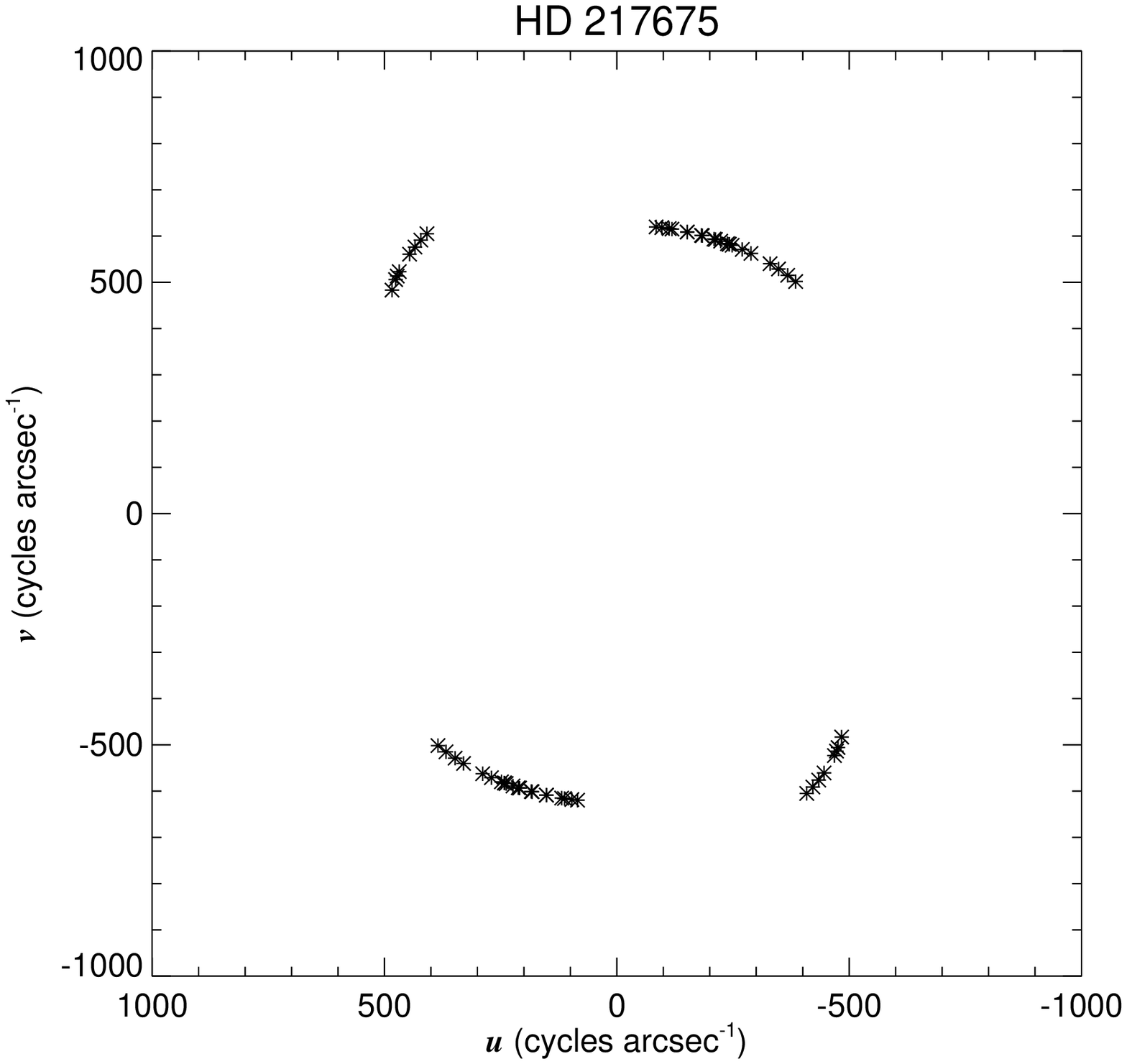}}
{\includegraphics[angle=0,height=5.5cm]{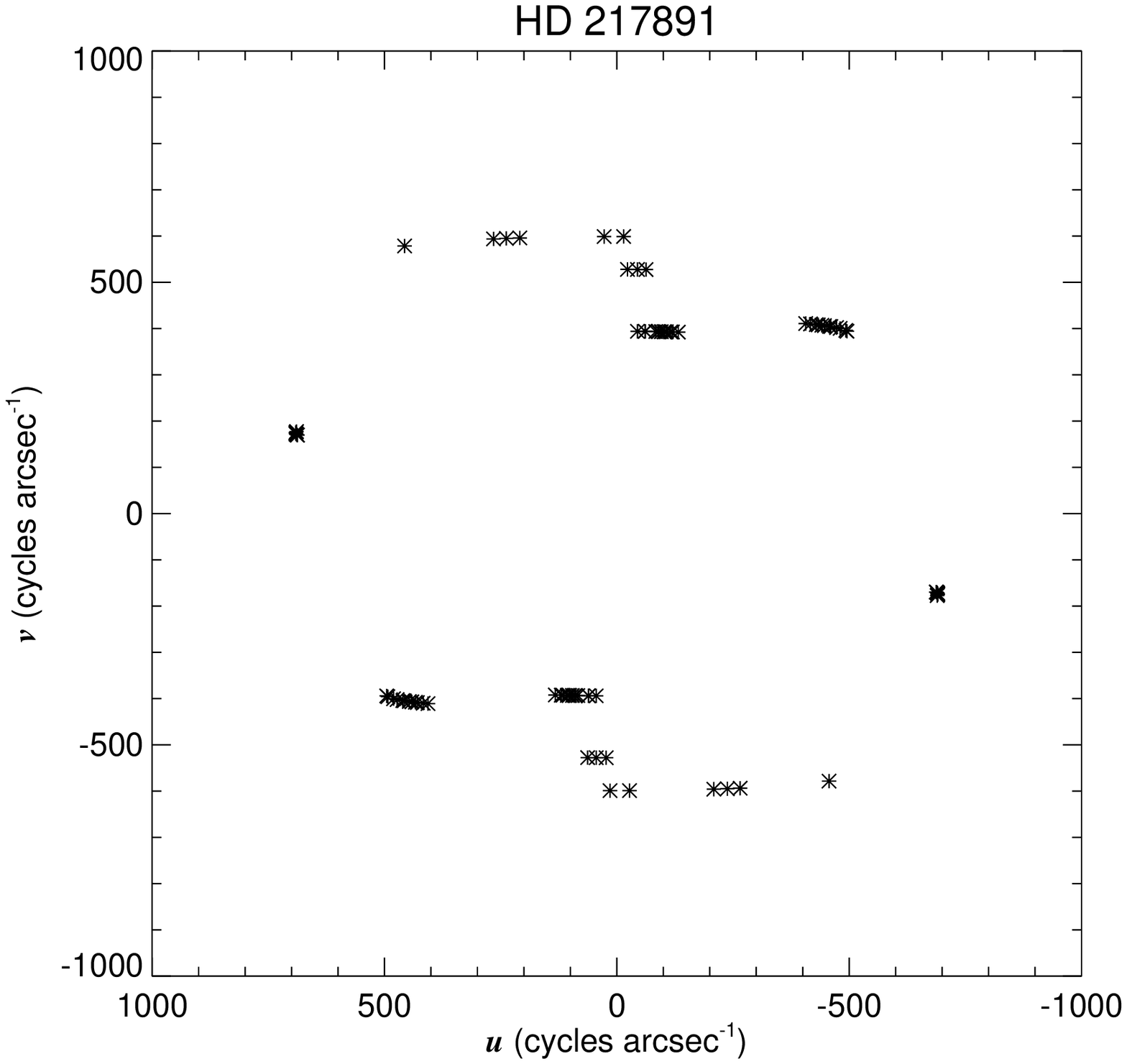}}
\end{center}
\caption[Sampling of the frequency $(u, v)$ plane for the $K$-band observations]
{1.17 -- 1.24. Sampling of the frequency $(u, v)$ plane for our sample stars.}
\label{uv3}
\end{figure*}


\clearpage

\begin{figure*}
\begin{center}
{\includegraphics[angle=0,height=15cm]{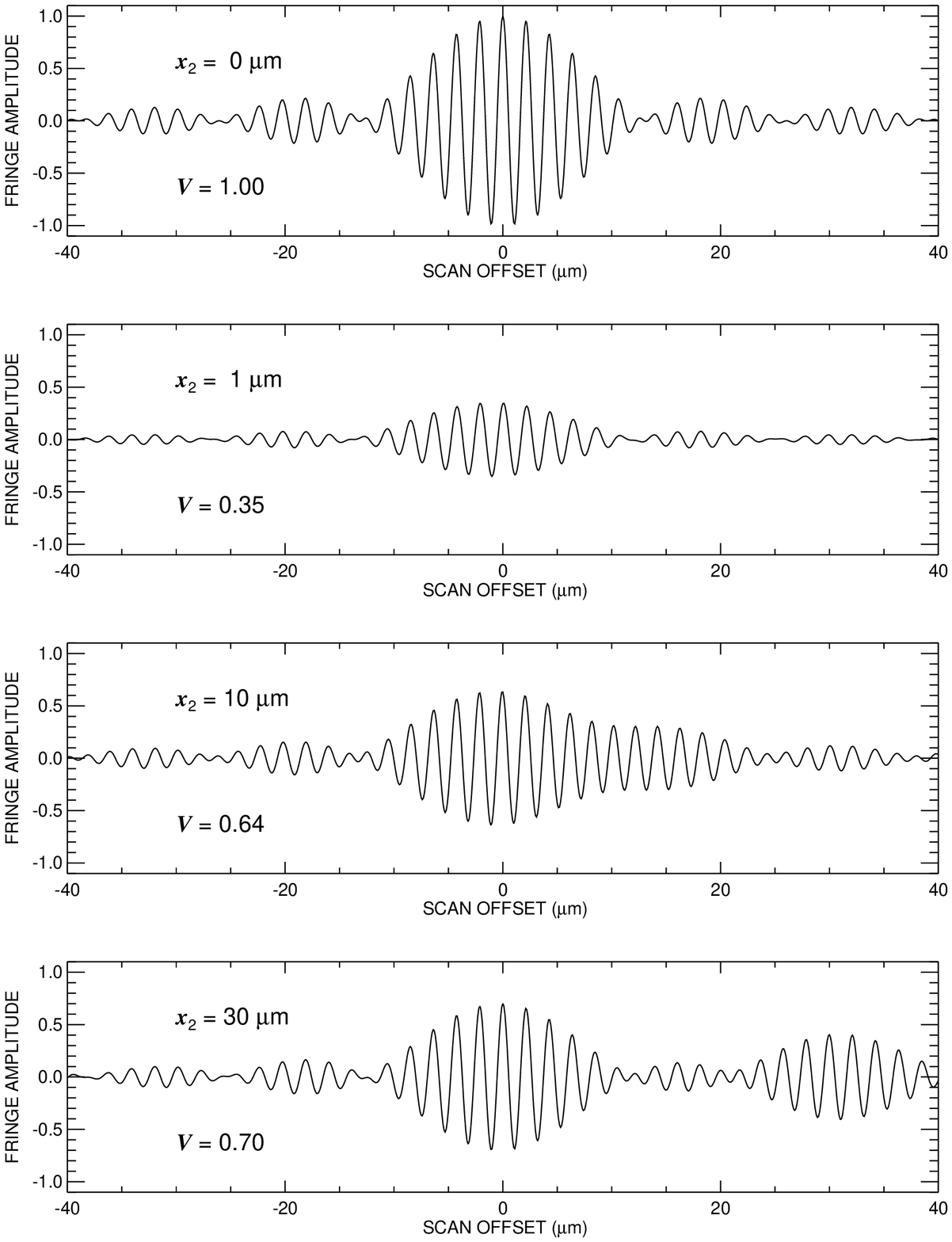}}
\end{center}
\caption[A series of combined fringe patterns of different projected separations]
{A series of combined fringe patterns for an assumed flux ratio of $f_2/f_1 = 0.5$
and a projected separation of zero (upper panel), 1~$\mu$m
(second panel from the top), 10~$\mu$m (third panel from the top),
and 30~$\mu$m (bottom panel).  The 30~$\mu$m separation would correspond to
a projected angular separation of 20.6 mas for a 300~m baseline (see eq.\ 4).}
\label{app1}
\end{figure*}


\clearpage

\begin{figure*}
\begin{center}
{\includegraphics[angle=90,height=10cm]{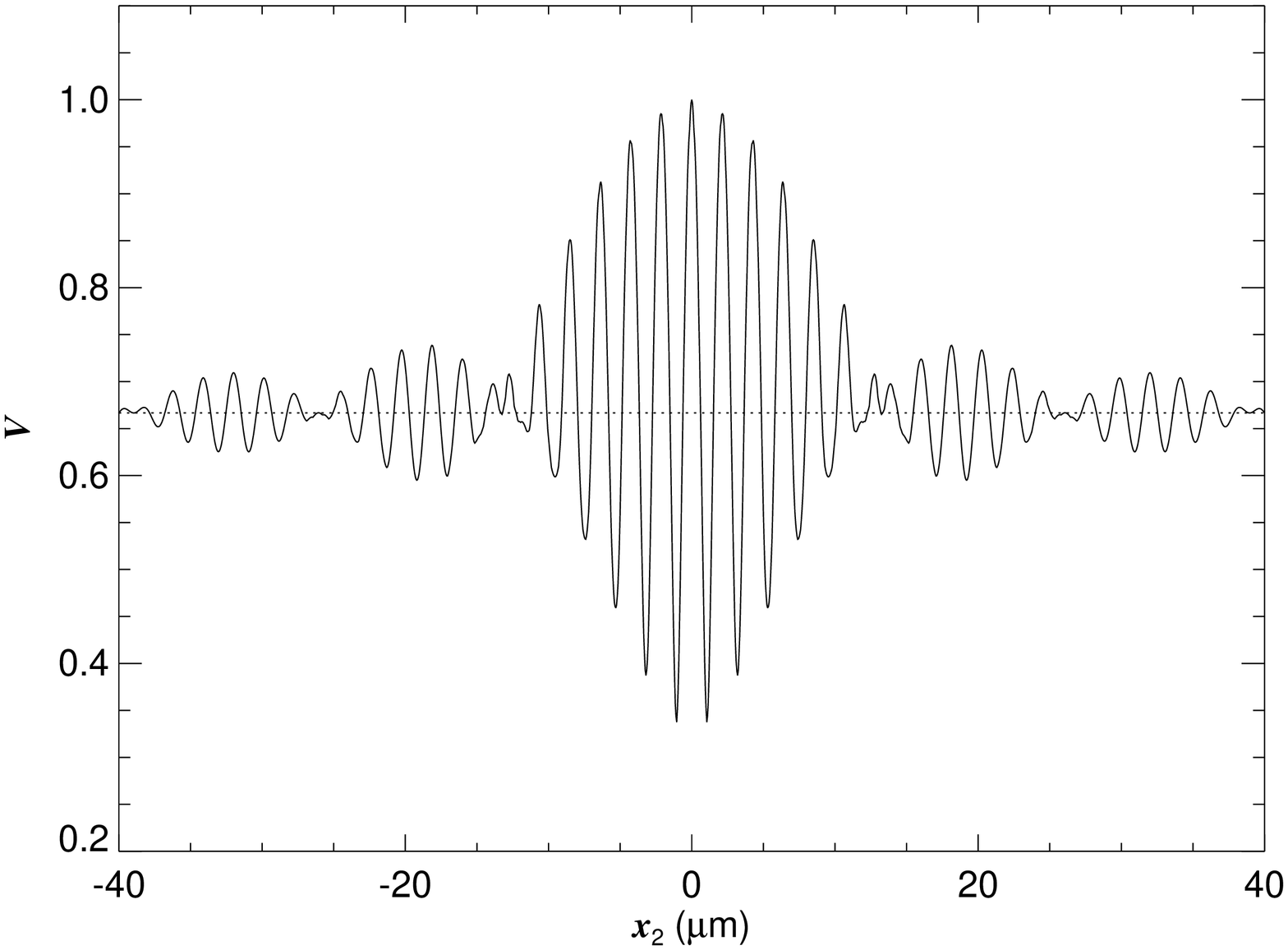}}
\end{center}
\caption[The net visibility as a function of the binary
projected separation]
{The net visibility as a function of the binary
projected separation $x_2$ for the fringe patterns shown in Fig. 2.}
\label{app2}
\end{figure*}


\clearpage

\begin{figure*}
\begin{center}
{\includegraphics[angle=0,height=16cm]{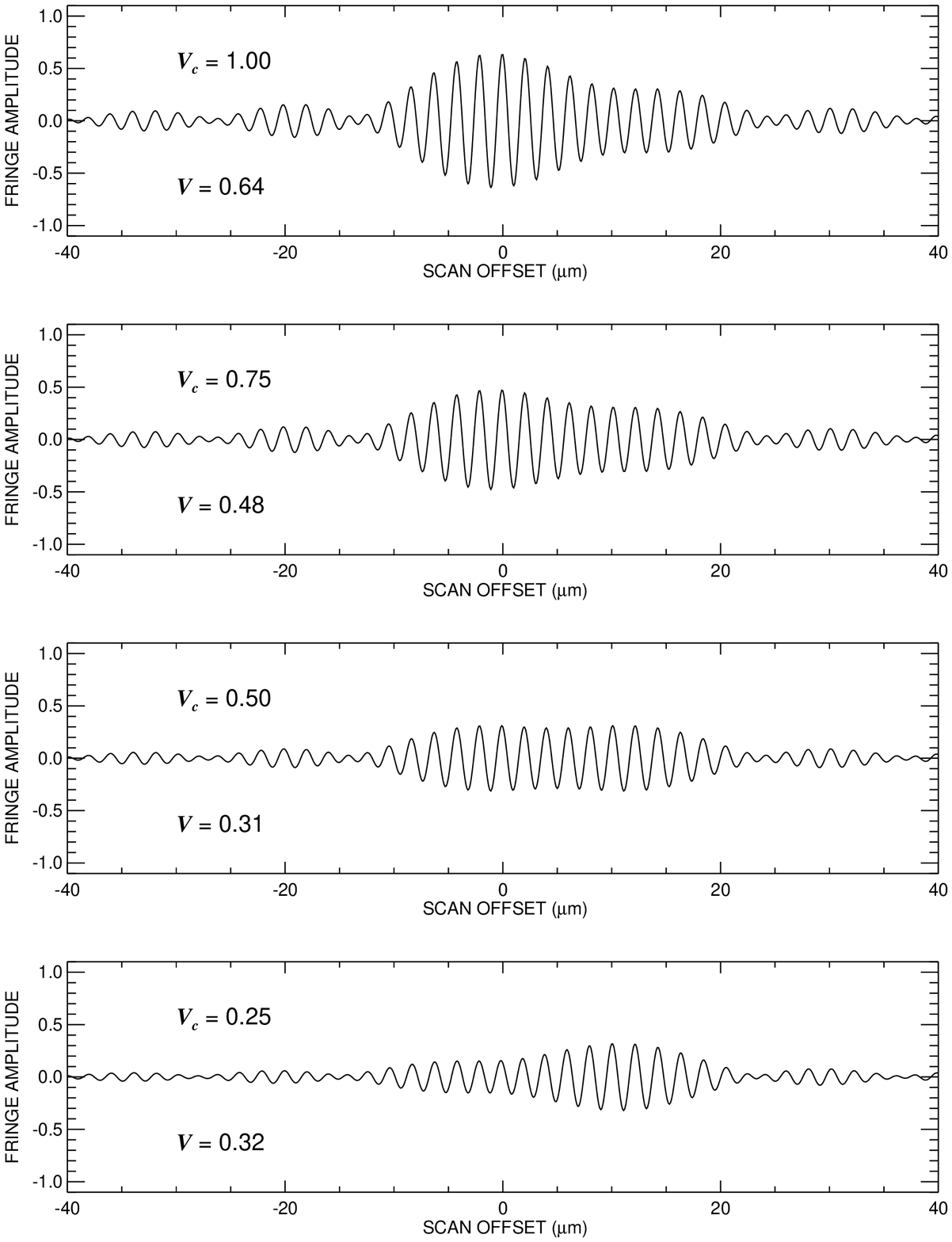}}
\end{center}
\caption[Model binary fringe patterns for $f_2/f_1 = 0.5$ and a separation of
10 $\mu$m]
{Model binary fringe patterns for $f_2/f_1 = 0.5$ and a separation of
10 $\mu$m. From top to bottom, the panels show the progressive
appearance of the combined fringe patterns as the visibility of
a star-plus-disk $V_c$ drops from 1 to 0.25.}
\label{app3}
\end{figure*}


\clearpage

\begin{figure*}
\begin{center}
{\includegraphics[angle=90,height=10cm]{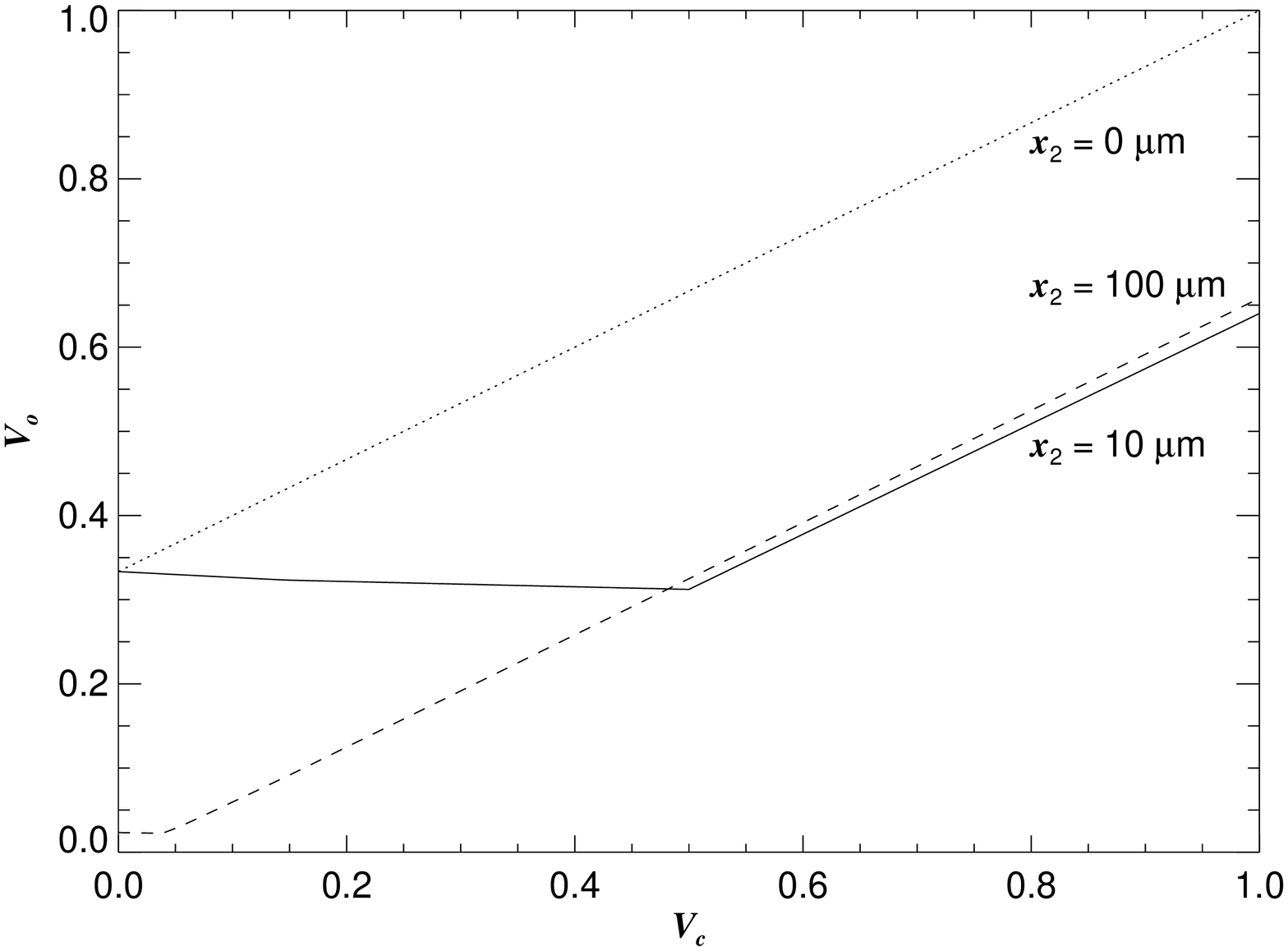}}
\end{center}
\caption[the relationship between the Be star visibility
and the net observed visibility at a  projected separation of zero (dotted line), 10~$\mu$m
(solid line), and 100~$\mu$m (dashed line).]
{The relationship between the Be star visibility $V_c$
and the net observed visibility $V_o$. The various lines show the
predictions for three binary separation values $x_2$
(see Fig.\ 4 for the $x_2 = 10$ $\mu$m case).}
\label{app4}
\end{figure*}


\clearpage

\begin{figure*}
\begin{center}
{\includegraphics[angle=90,height=7cm]{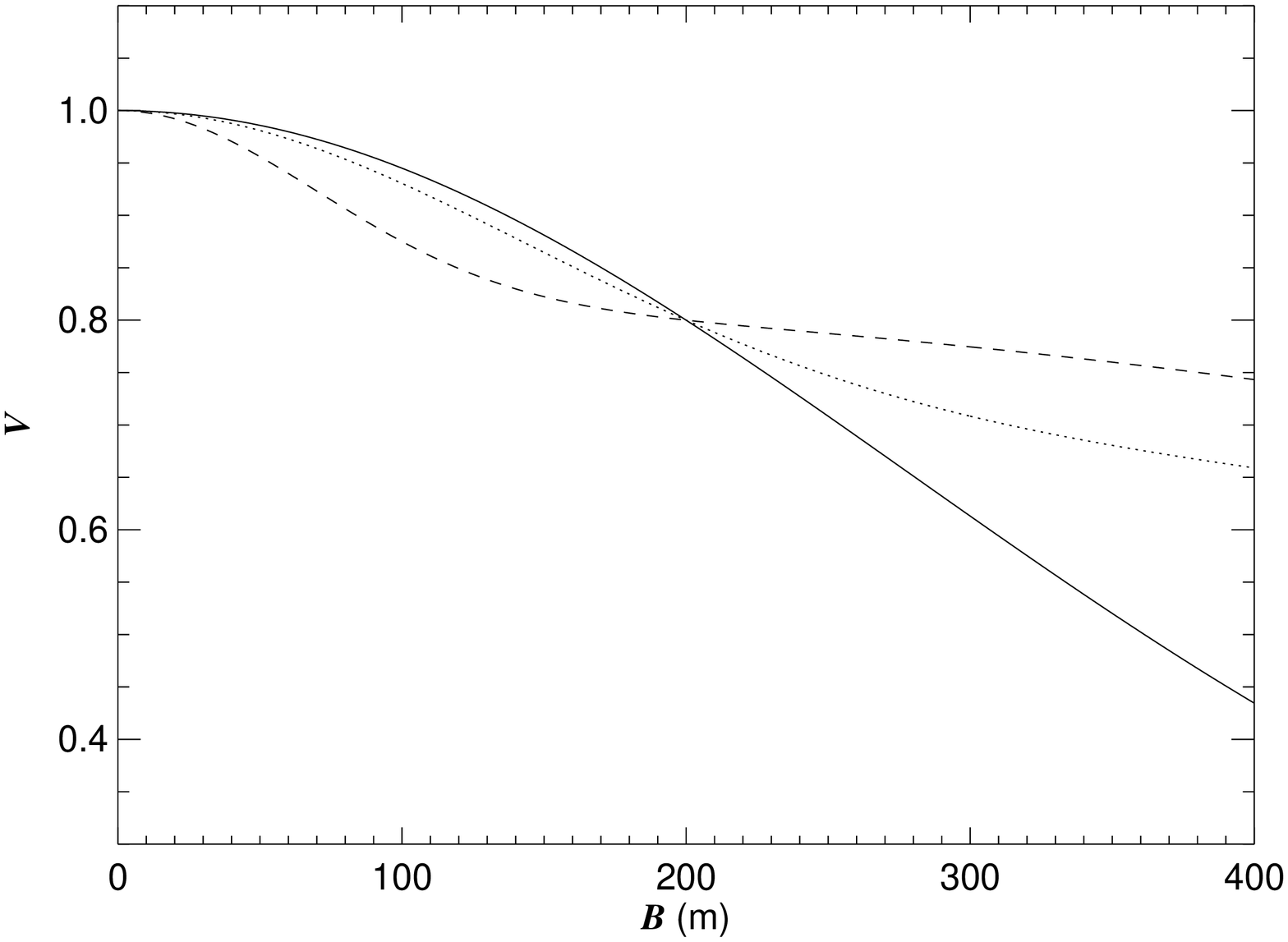}}
\end{center}
\caption[A set of three Gaussian elliptical
models of different ($c_p$, $\theta_{\rm maj}$) that produce a visibility point $V$ = 0.8
at a 200~m projected baseline]{A set of three Gaussian elliptical
models of different ($c_p$, $\theta_{\rm maj}$) that produce a visibility point $V$ = 0.8
at a 200~m projected baseline. The solid curve is for
($c_p$, $\theta_{\rm maj}$) = (0.156, 0.6 mas), the dotted curve is for
($c_p$, $\theta_{\rm maj}$) = (0.719, 1.2 mas), and the dashed curve is for
($c_p$, $\theta_{\rm maj}$) = (0.816, 2.4 mas).}
\label{test1}
\end{figure*}


\begin{figure*}
\begin{center}
{\includegraphics[angle=90,height=7cm]{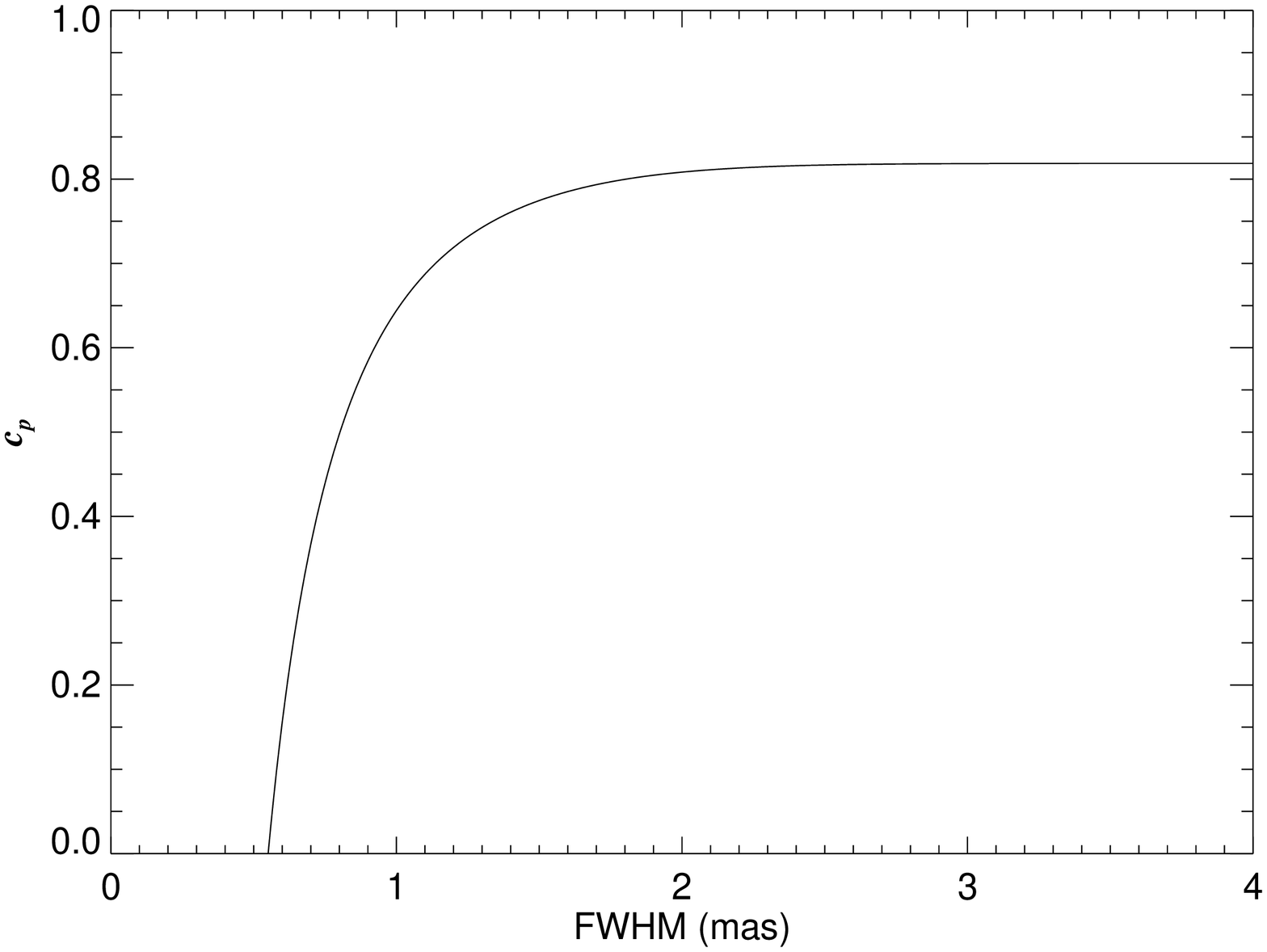}}
\end{center}
\caption[The relation between $c_p$ and $\theta_{\rm maj}$ = FWHM for Gaussian elliptical visibility
models]{The relation between $c_p$ and $\theta_{\rm maj}$ = FWHM for Gaussian elliptical visibility
models that go through the same observed point of $V = 0.8$ at a 200 m baseline.}
\label{test2}
\end{figure*}


\clearpage

\setcounter{figure}{7}
\begin{figure*}
\begin{center}
{\includegraphics[angle=90,height=5cm]{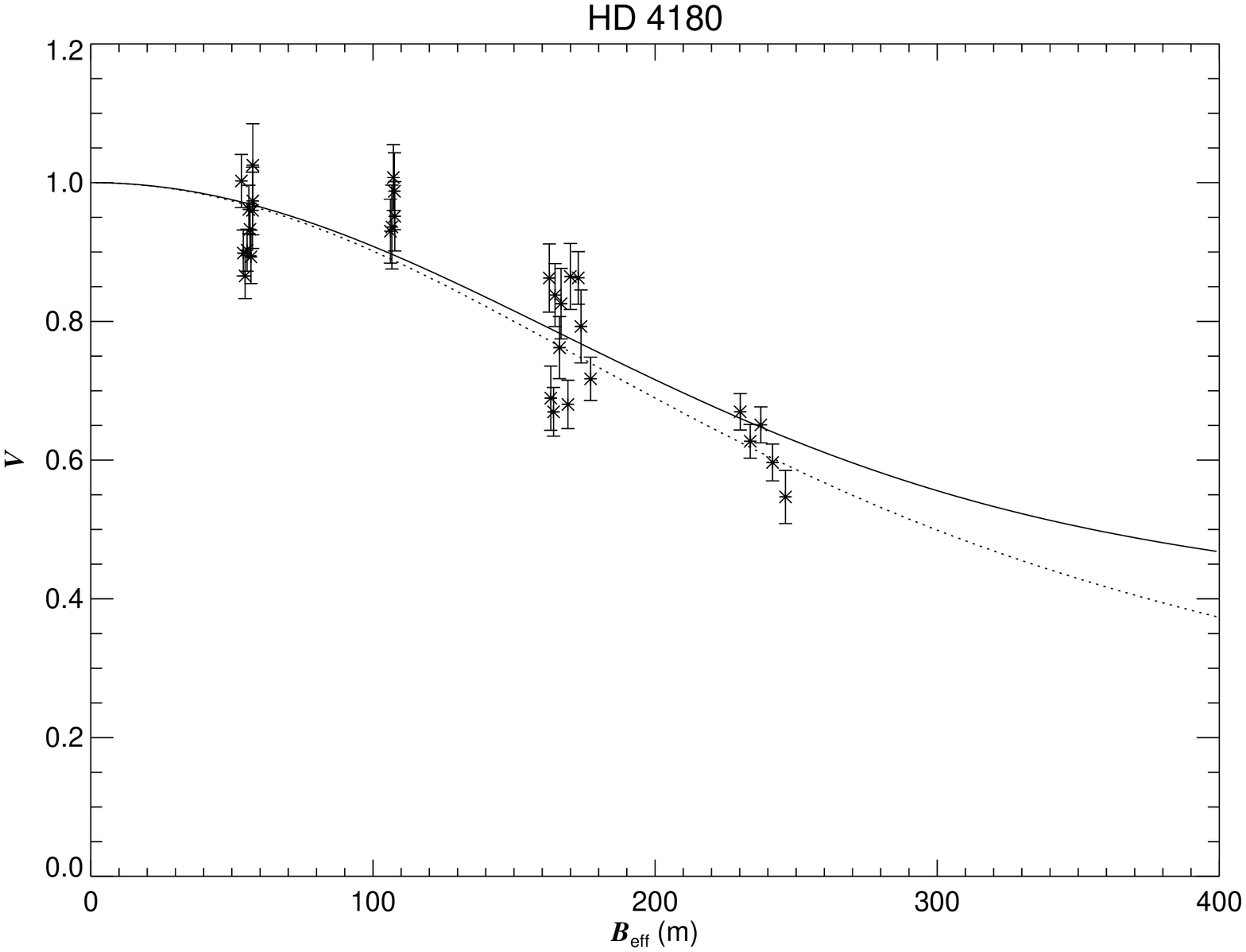}}
{\includegraphics[angle=90,height=5cm]{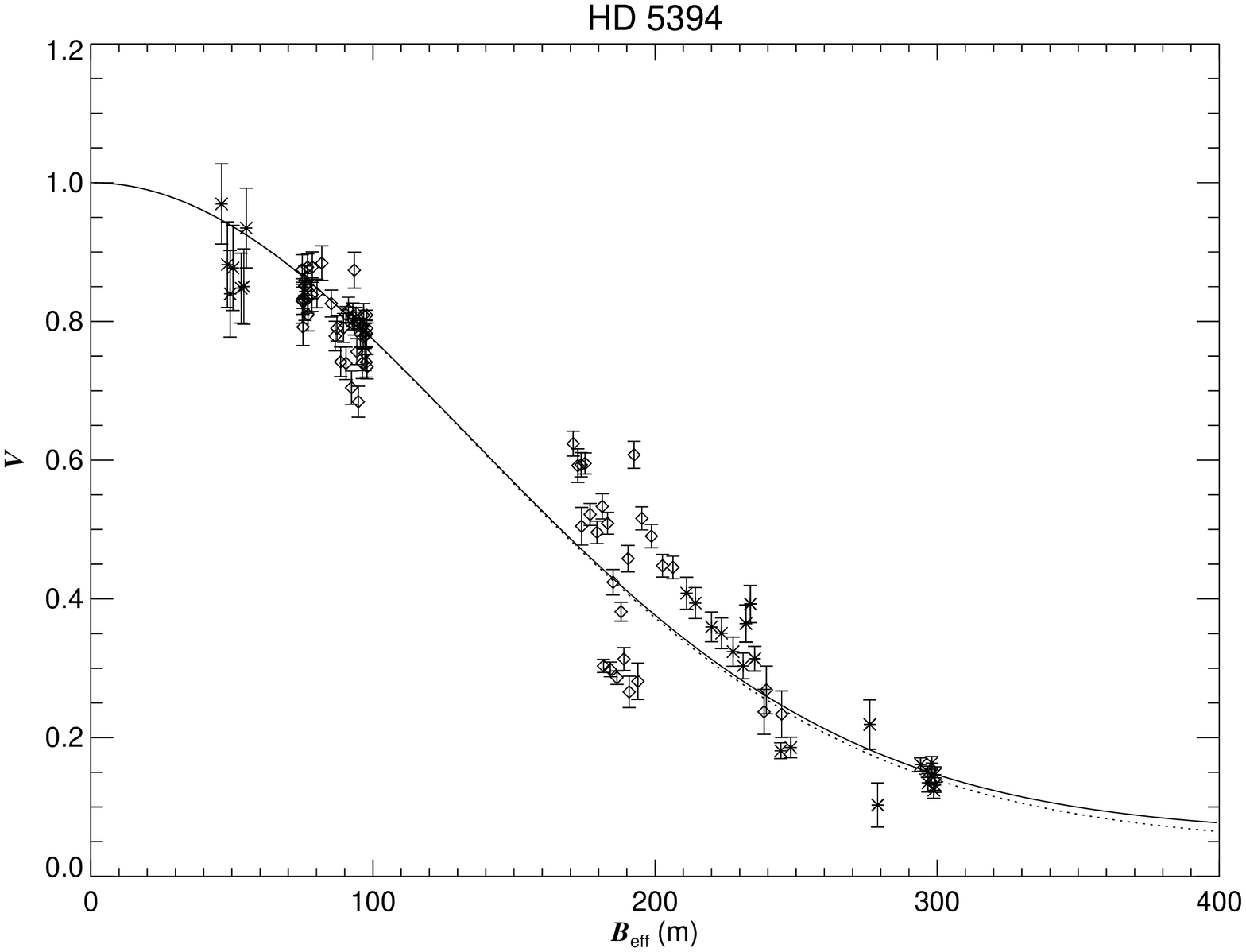}}
{\includegraphics[angle=90,height=5cm]{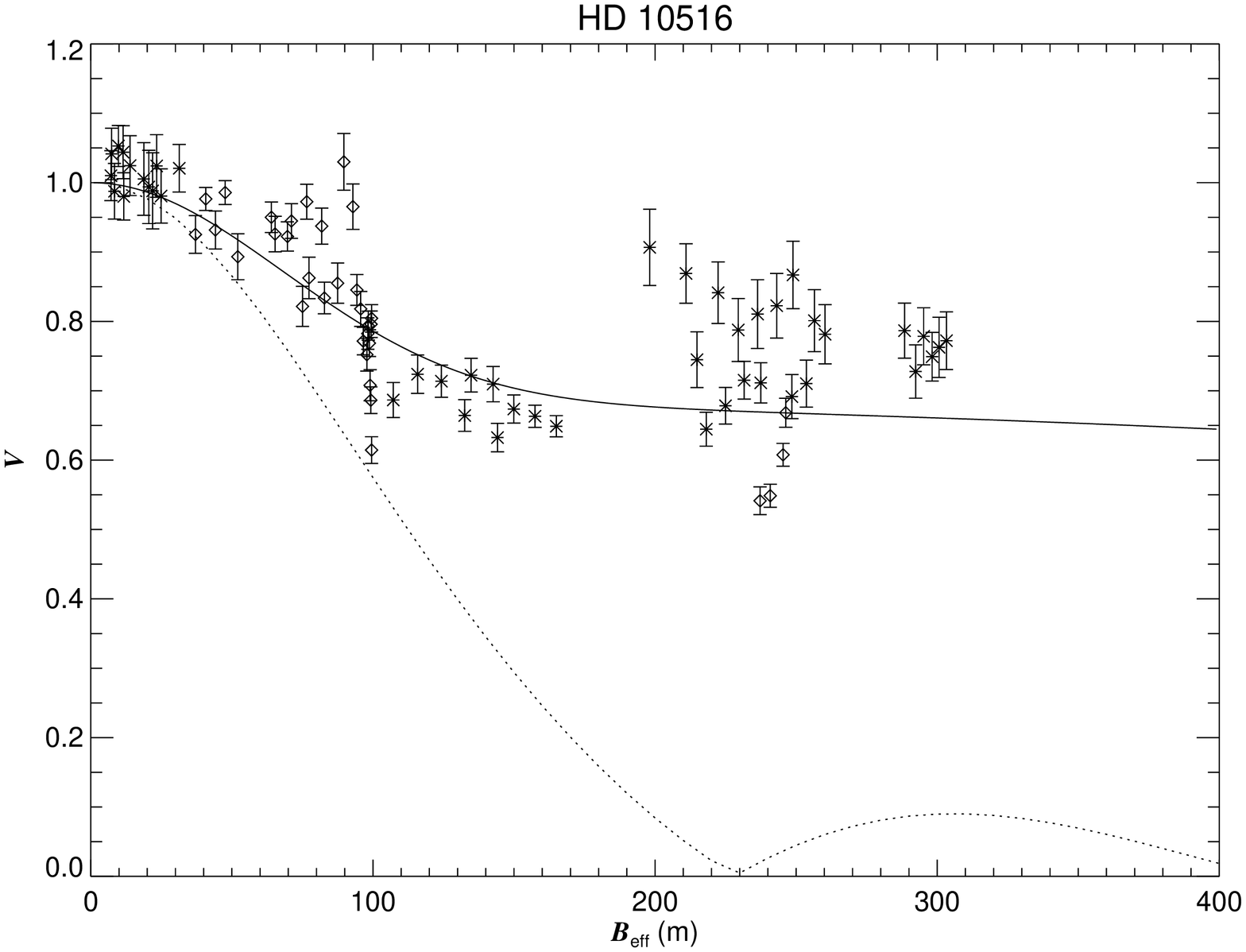}}
{\includegraphics[angle=90,height=5cm]{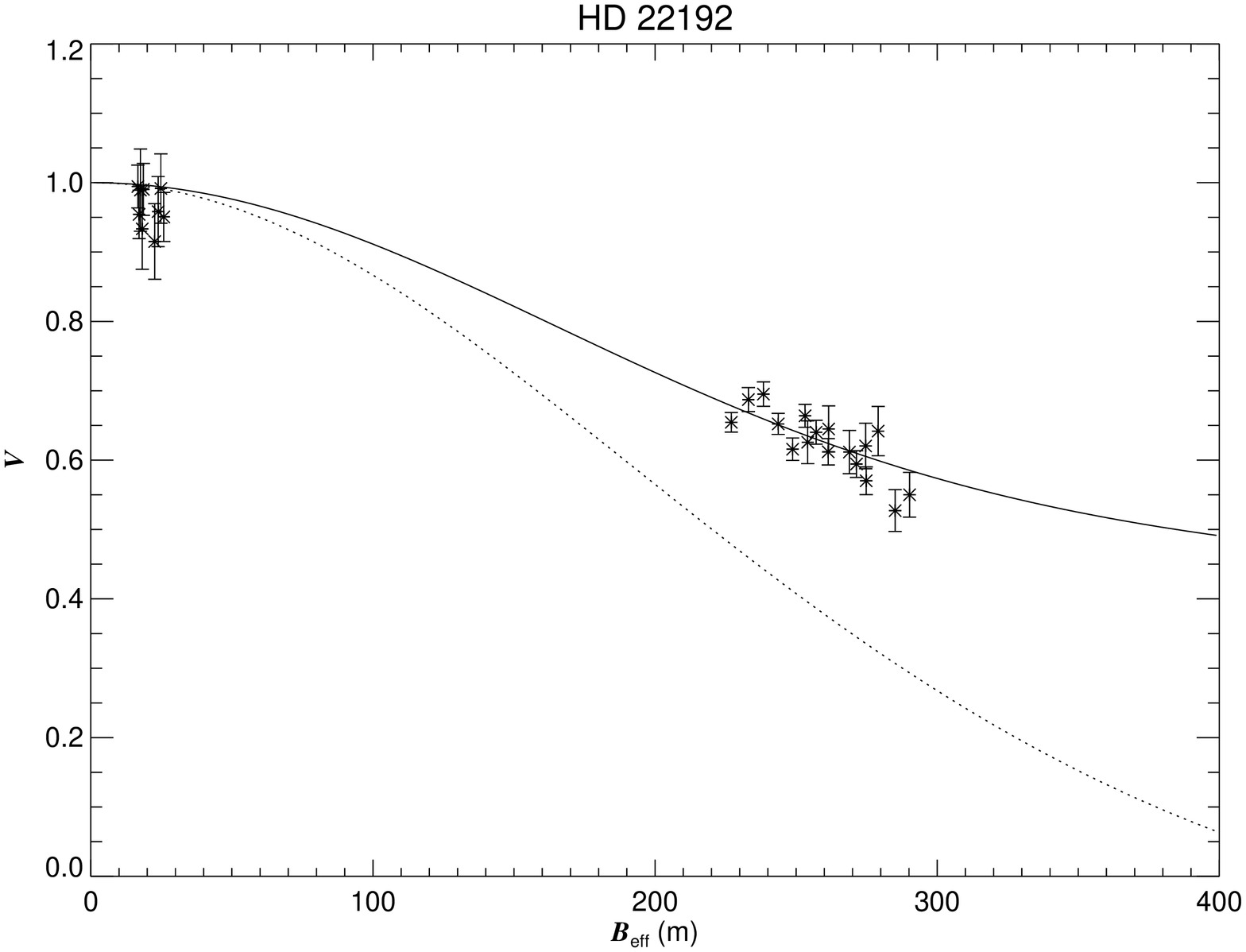}}
{\includegraphics[angle=90,height=5cm]{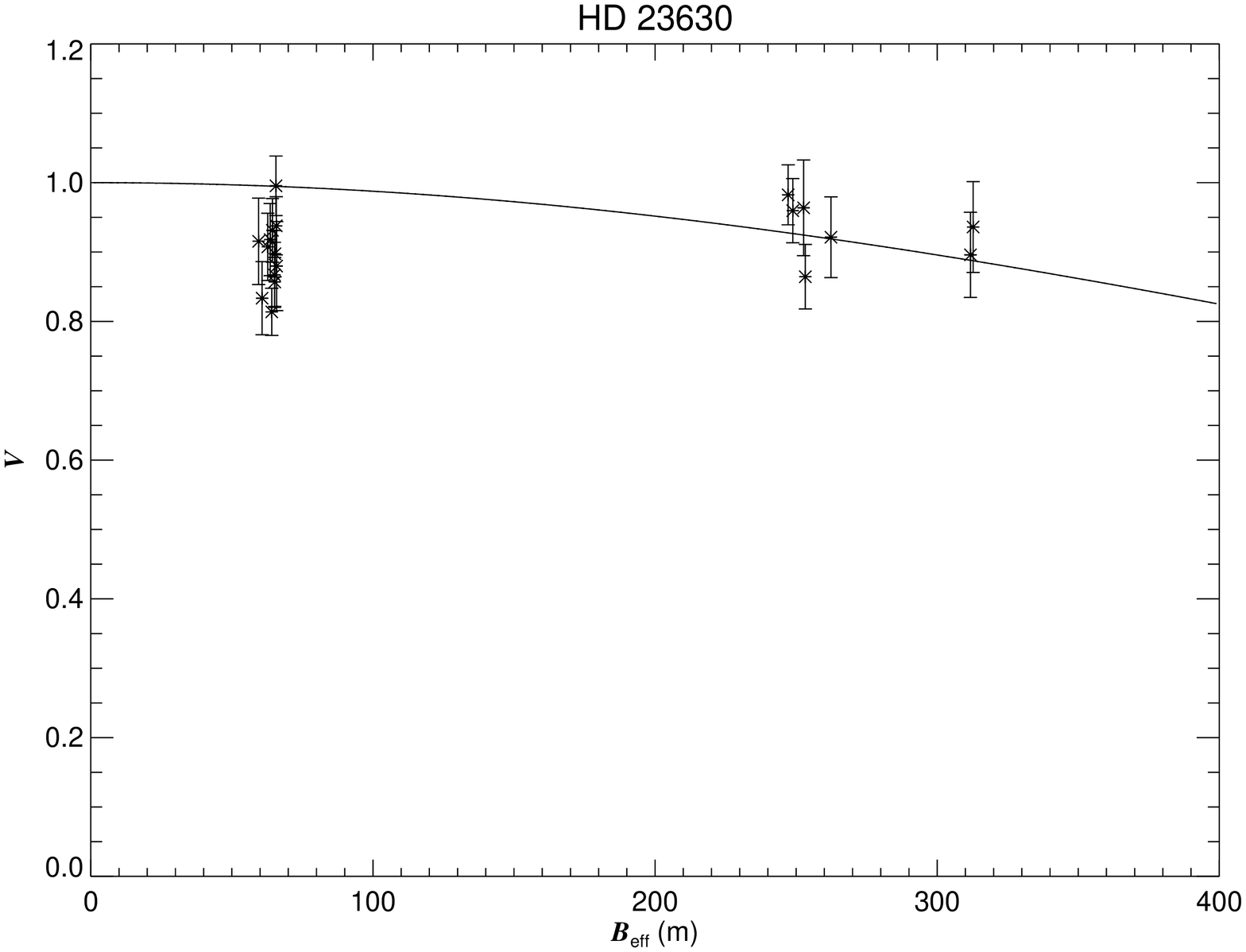}}
{\includegraphics[angle=90,height=5cm]{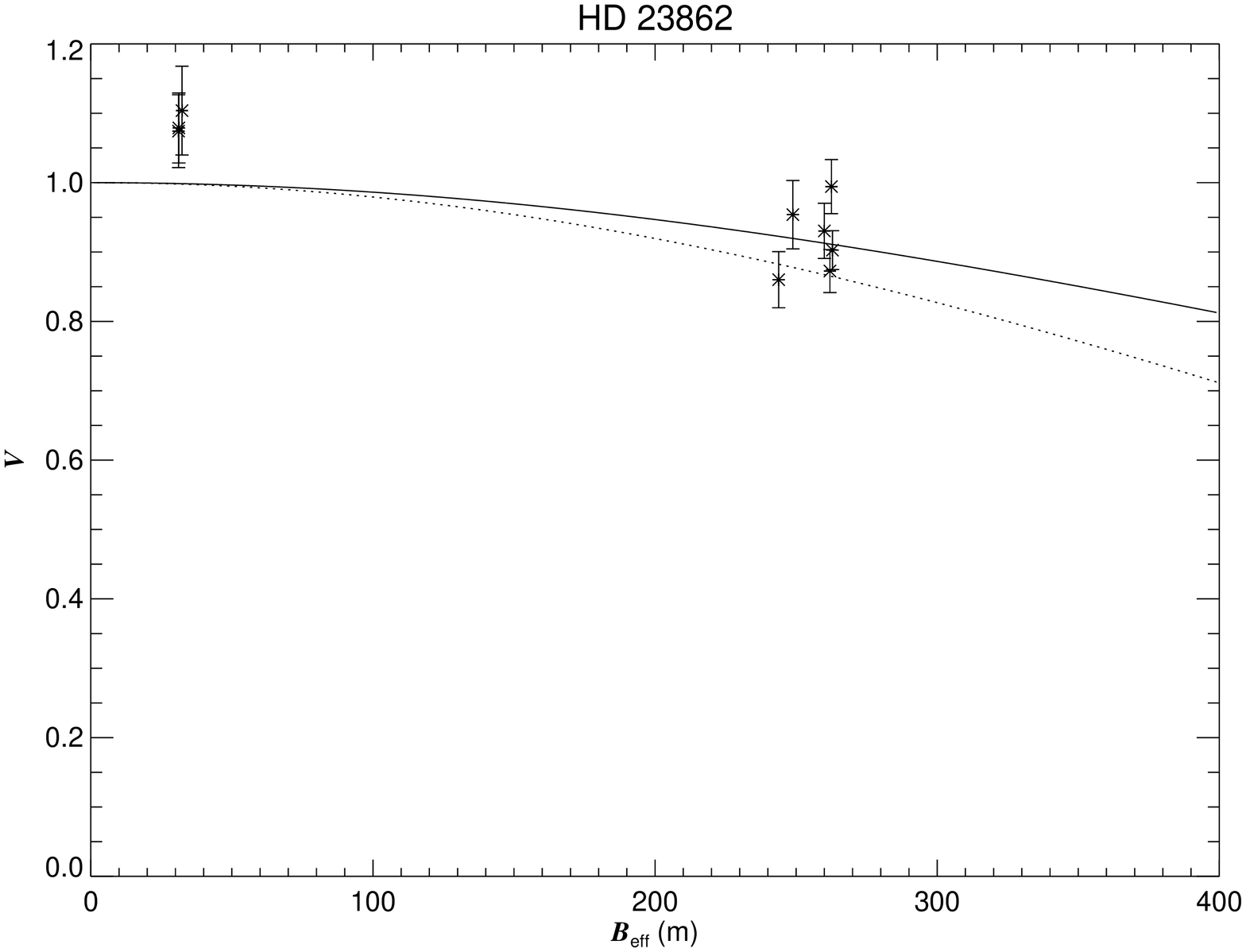}}
{\includegraphics[angle=90,height=5cm]{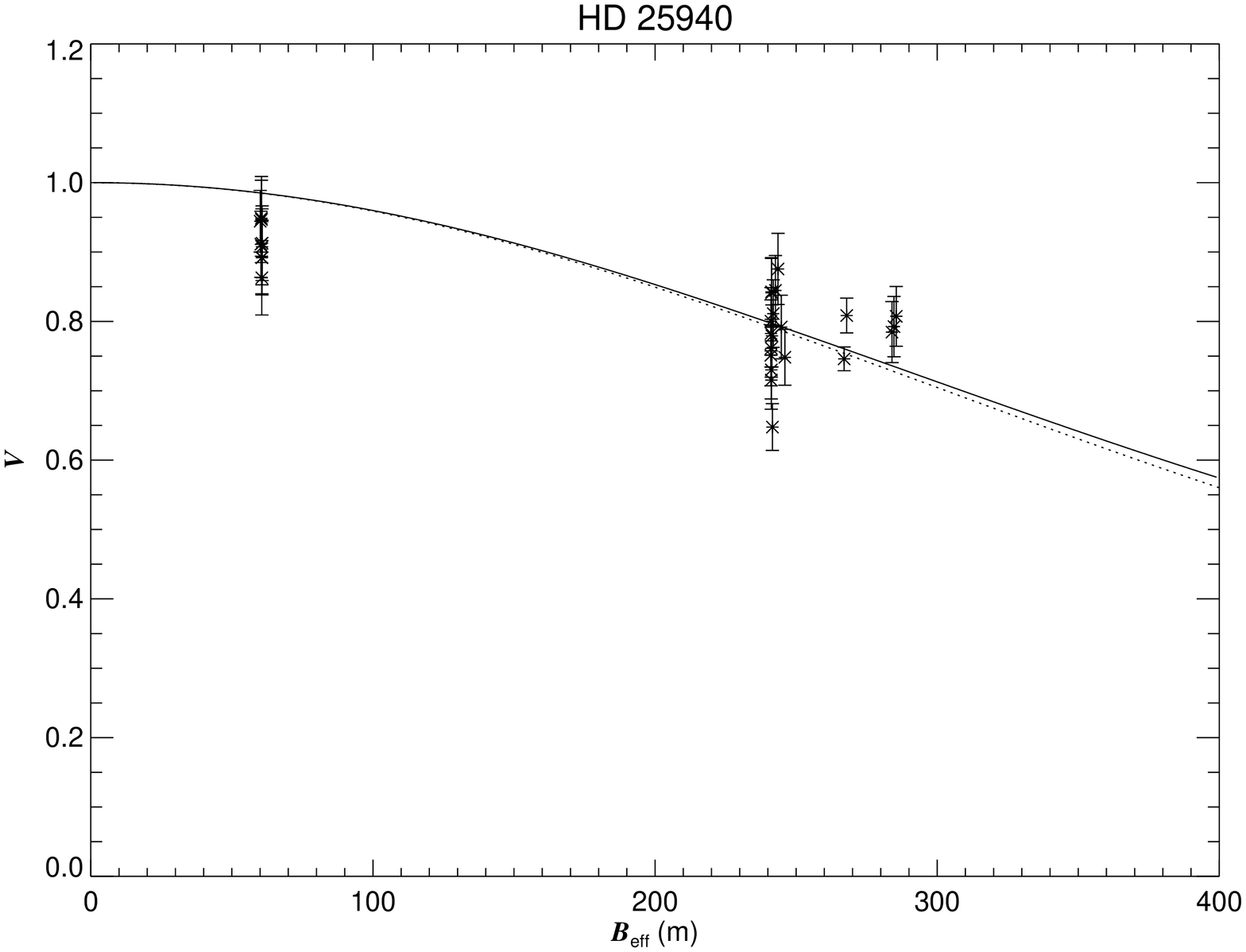}}
{\includegraphics[angle=90,height=5cm]{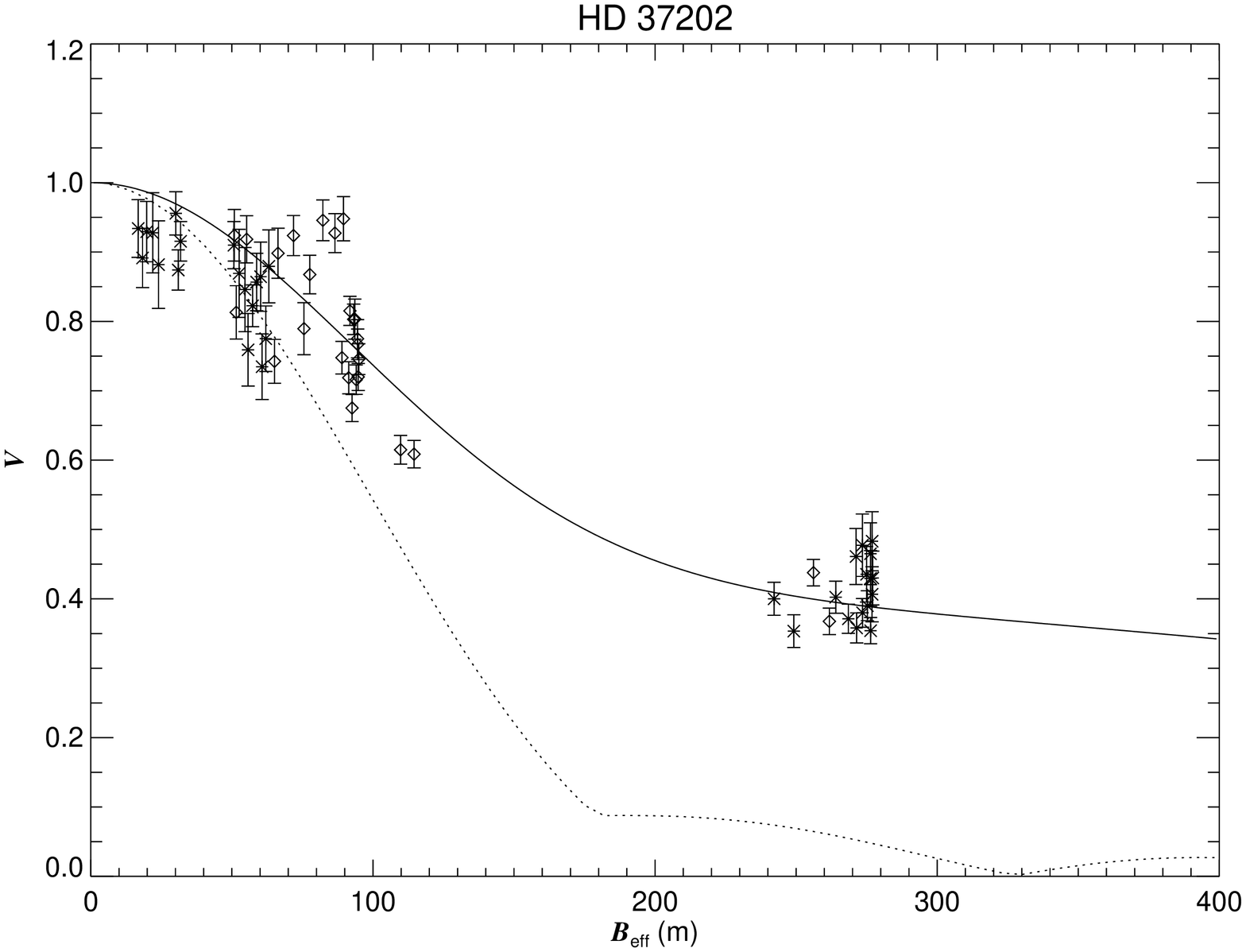}}
\end{center}
\caption[Calibrated visibilities versus baseline]
{8.1 -- 8.8.  Calibrated visibilities versus the effective baseline. The solid line and the dotted
lines represent the Gaussian elliptical model along the major and minor axes, respectively,
and the star signs represent the interferometric data.}
\label{vis1}
\end{figure*}


\clearpage

\setcounter{figure}{7}
\begin{figure*}
\begin{center}
{\includegraphics[angle=90,height=5cm]{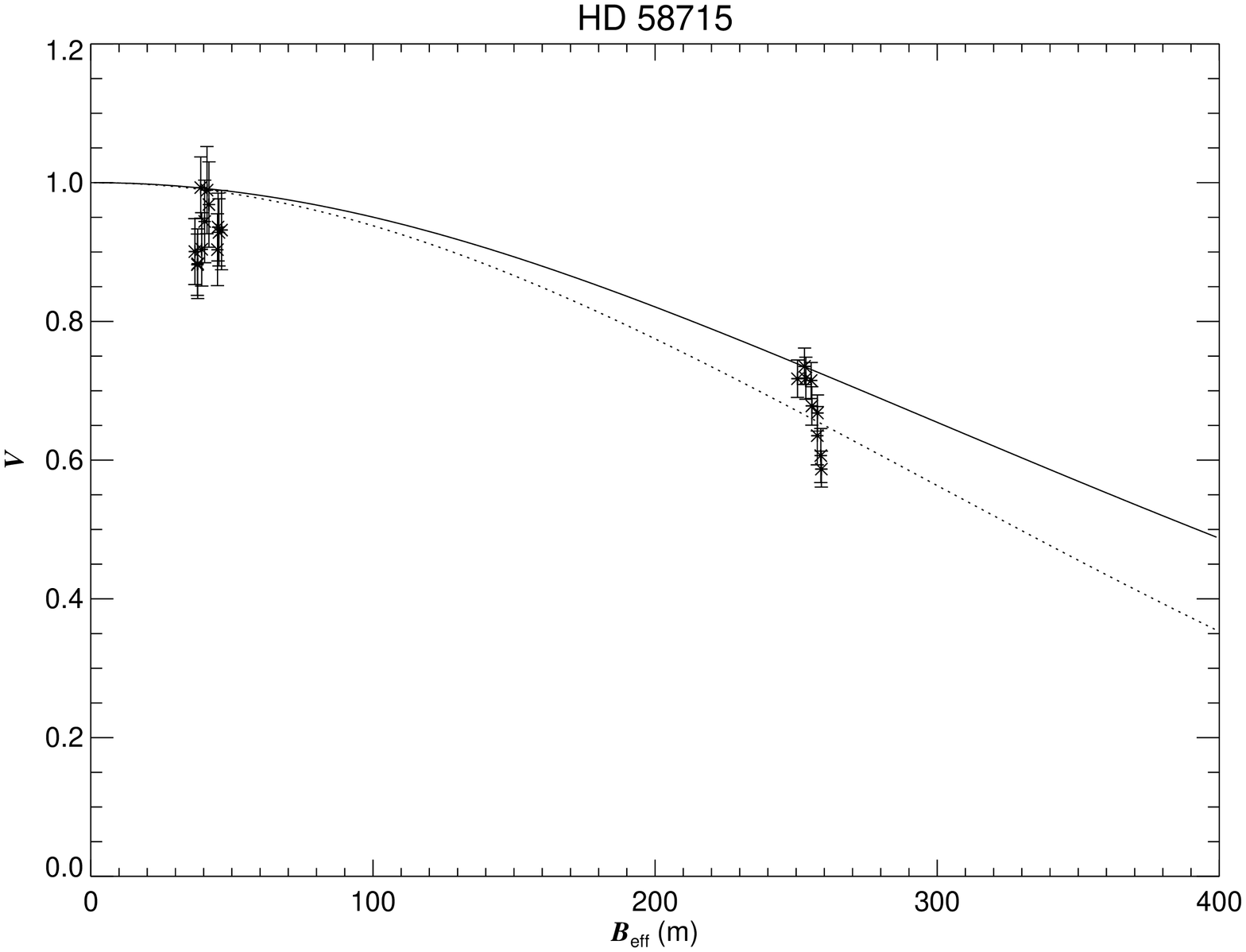}}
{\includegraphics[angle=90,height=5cm]{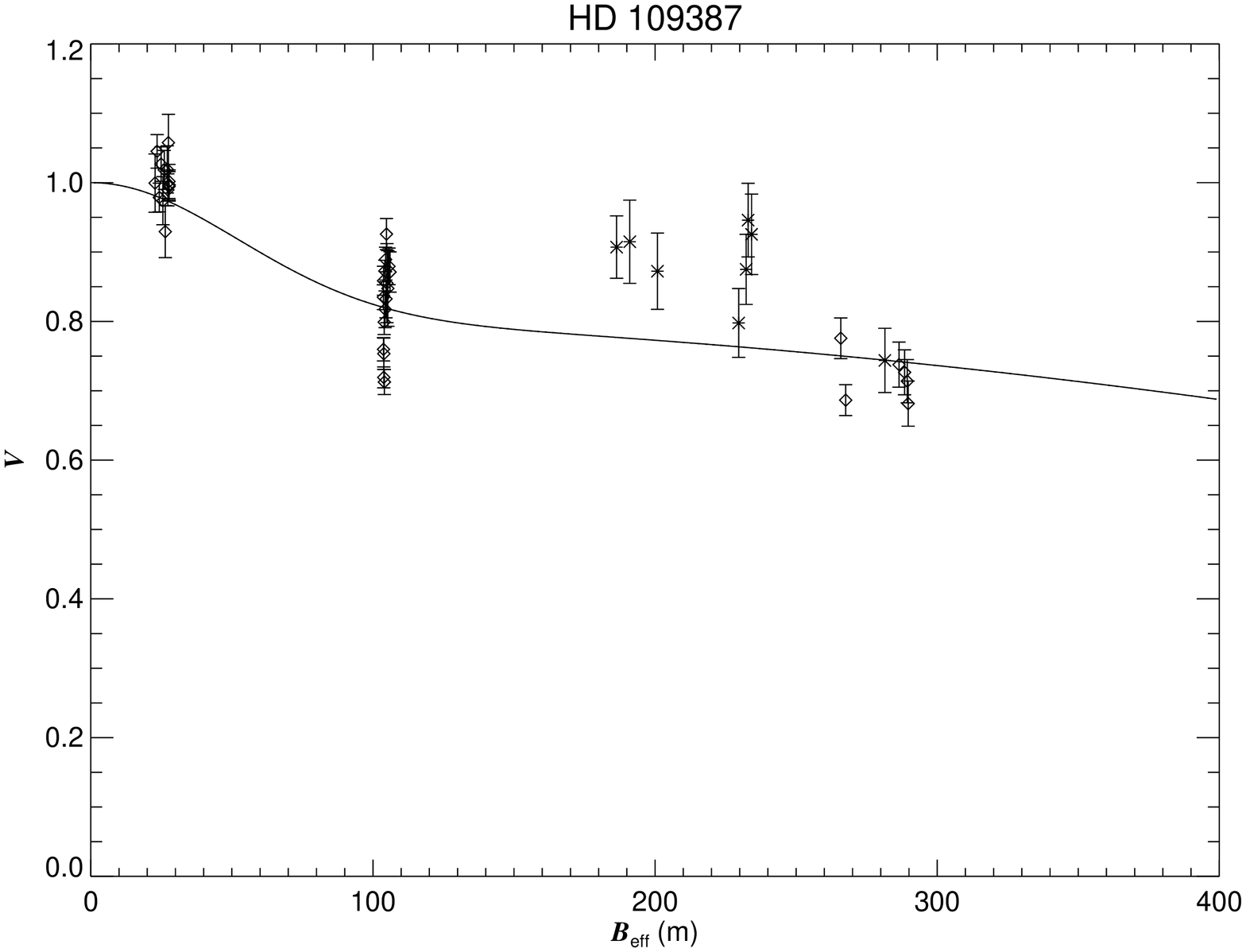}}
{\includegraphics[angle=90,height=5cm]{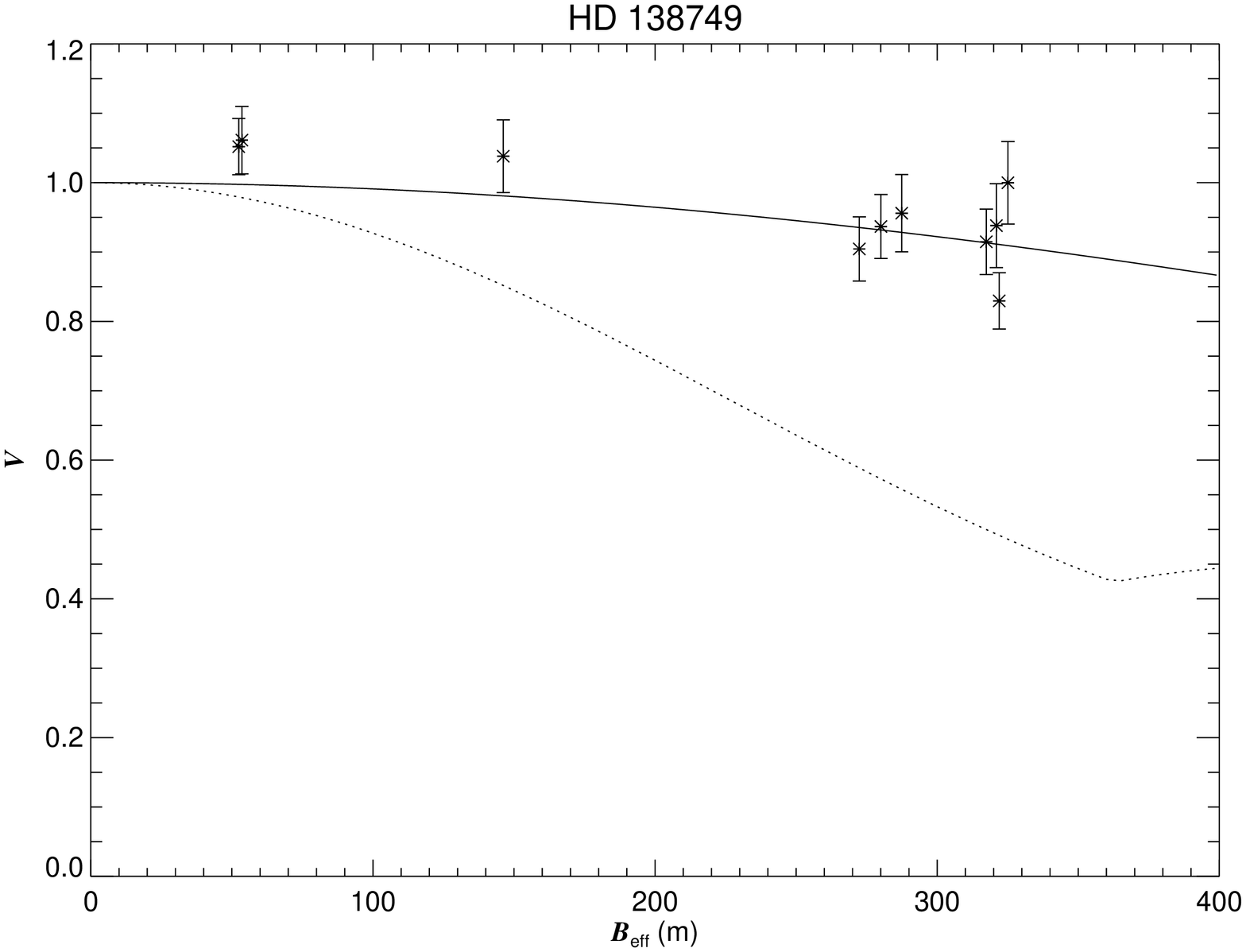}}
{\includegraphics[angle=90,height=5cm]{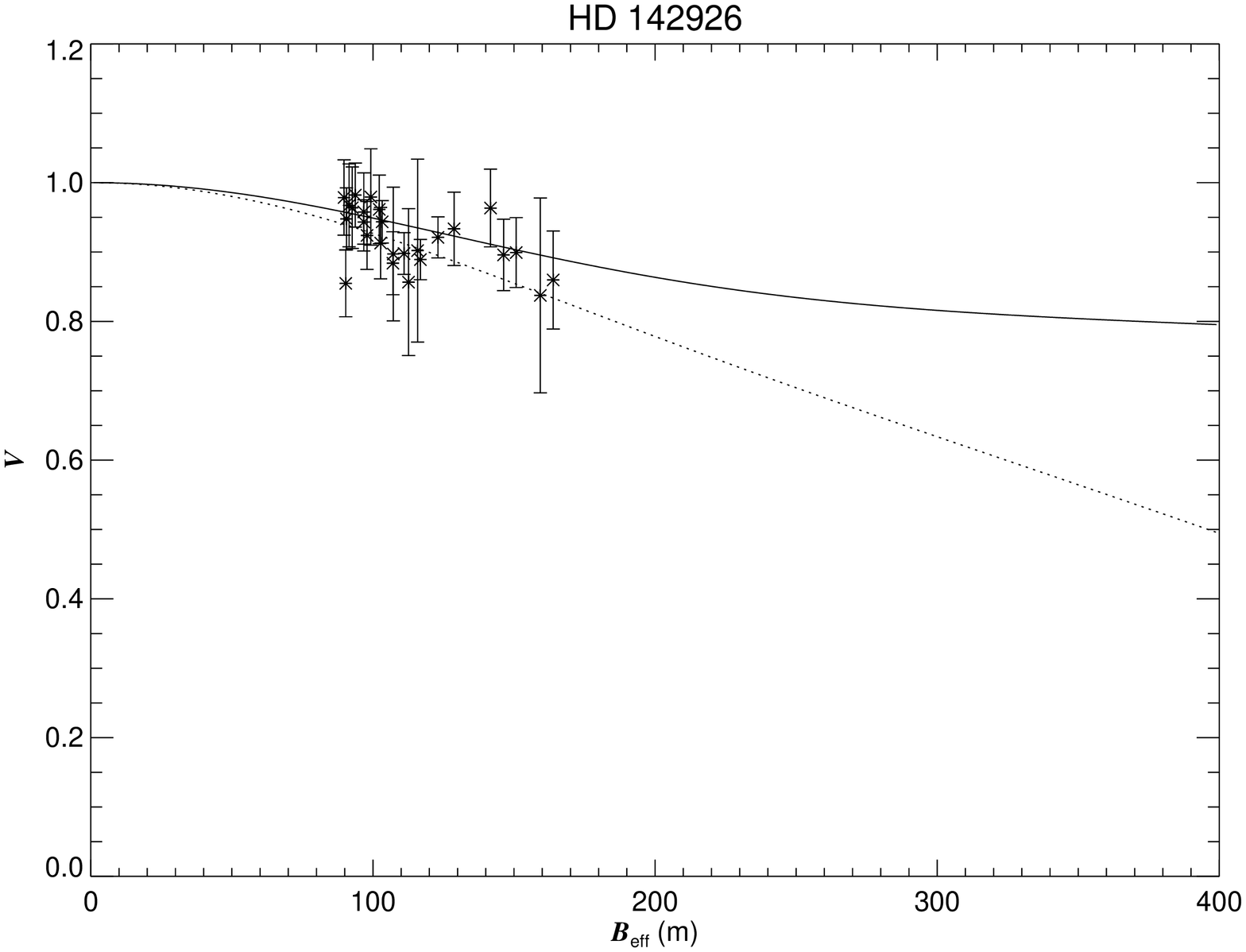}}
{\includegraphics[angle=90,height=5cm]{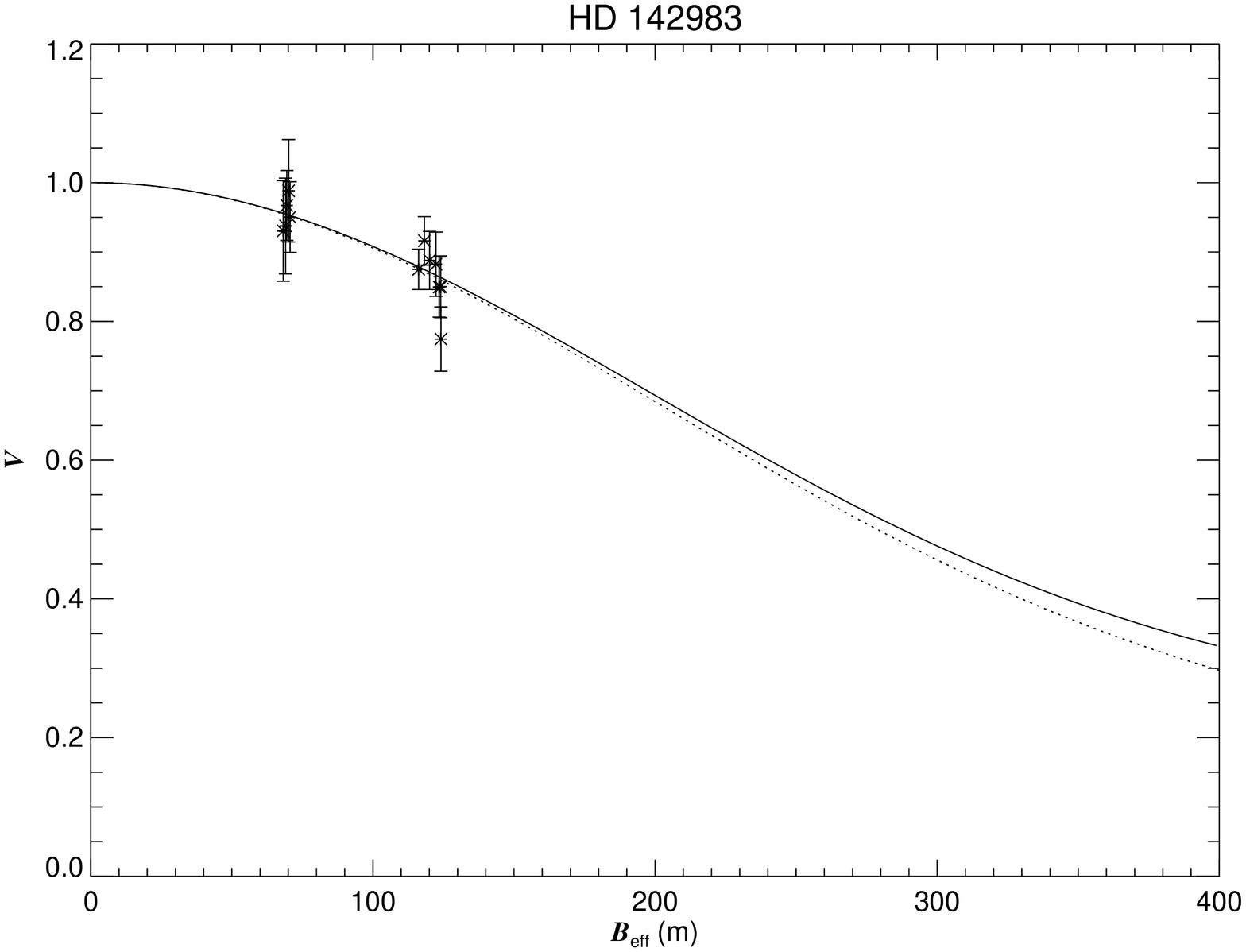}}
{\includegraphics[angle=90,height=5cm]{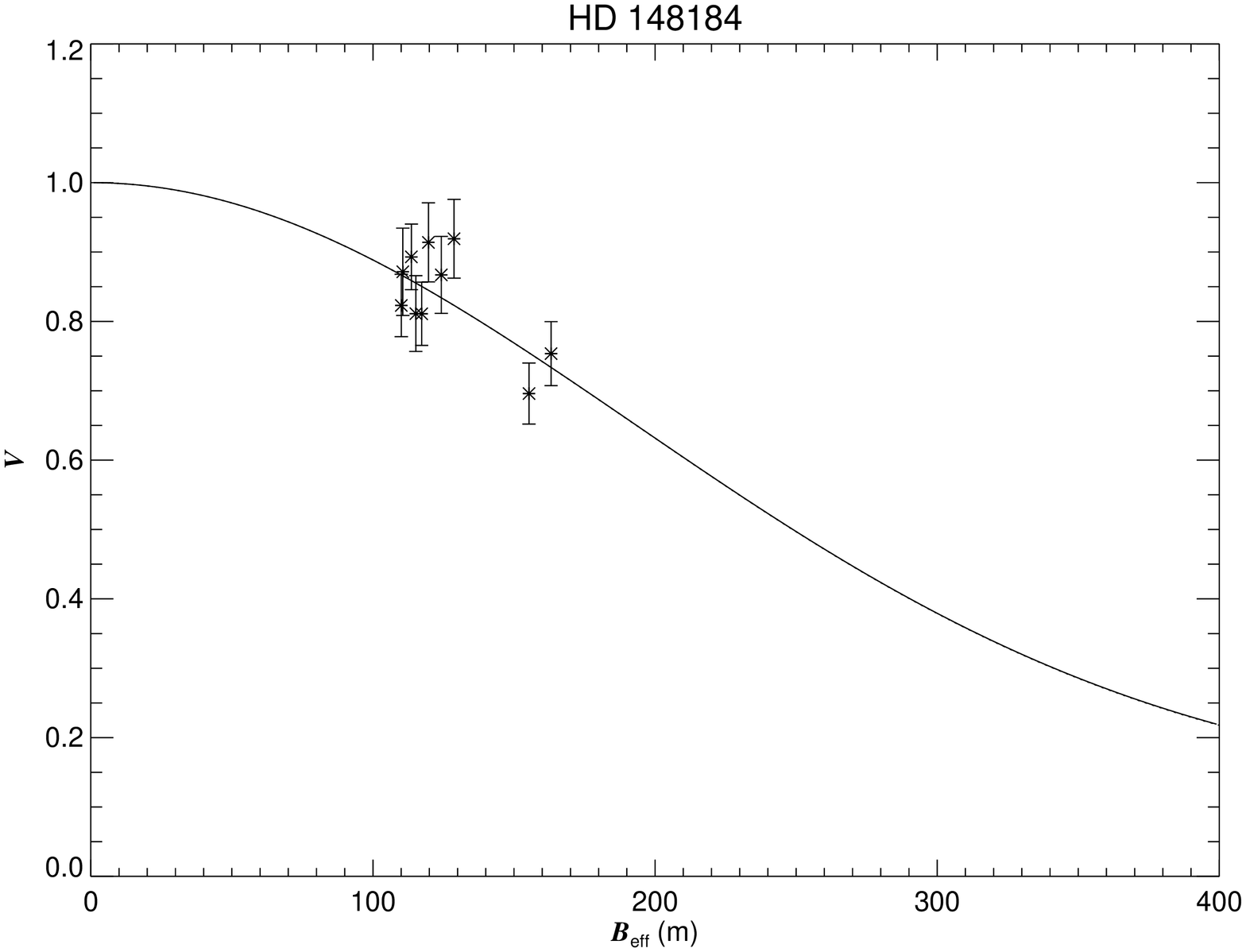}}
{\includegraphics[angle=90,height=5cm]{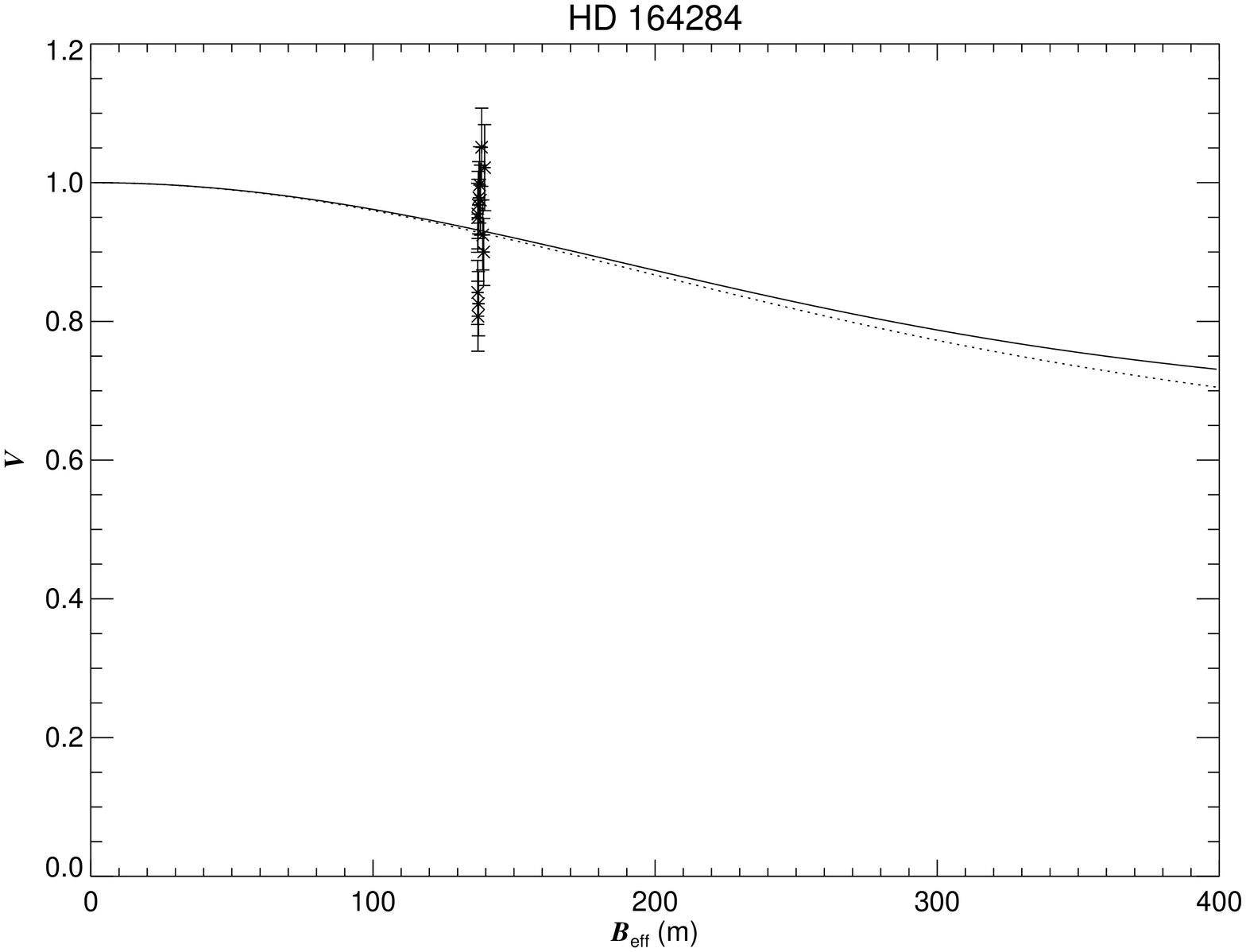}}
{\includegraphics[angle=90,height=5cm]{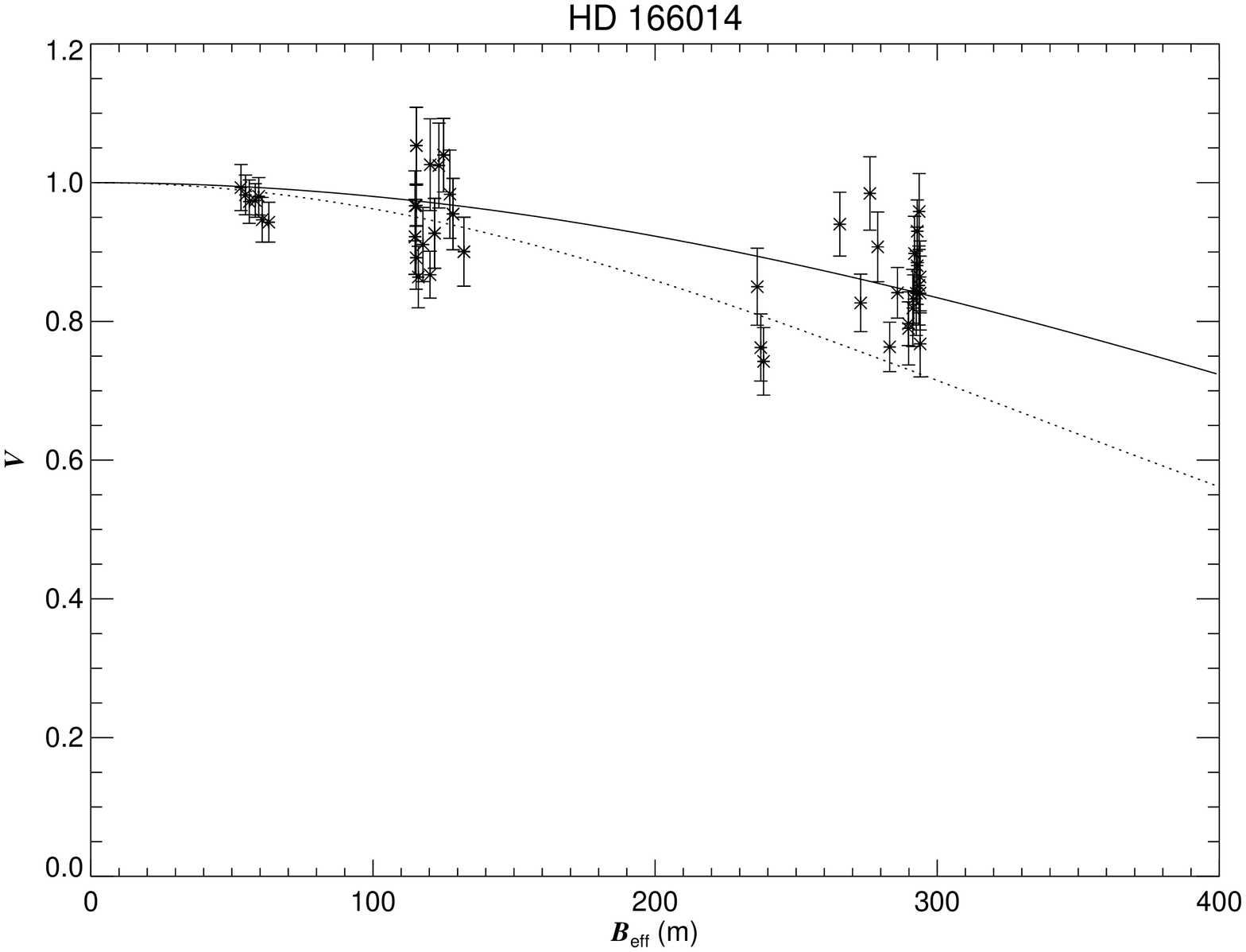}}
\end{center}
\caption[Calibrated visibilities versus baseline]
{8.9 -- 8.16. Calibrated visibilities versus the effective baseline. The solid and the dotted
lines represent the Gaussian elliptical model along the major and minor axes, respectively,
and the star signs represent the interferometric data.}
\label{vis2}
\end{figure*}


\clearpage

\setcounter{figure}{7}
\begin{figure*}
\begin{center}
{\includegraphics[angle=90,height=5cm]{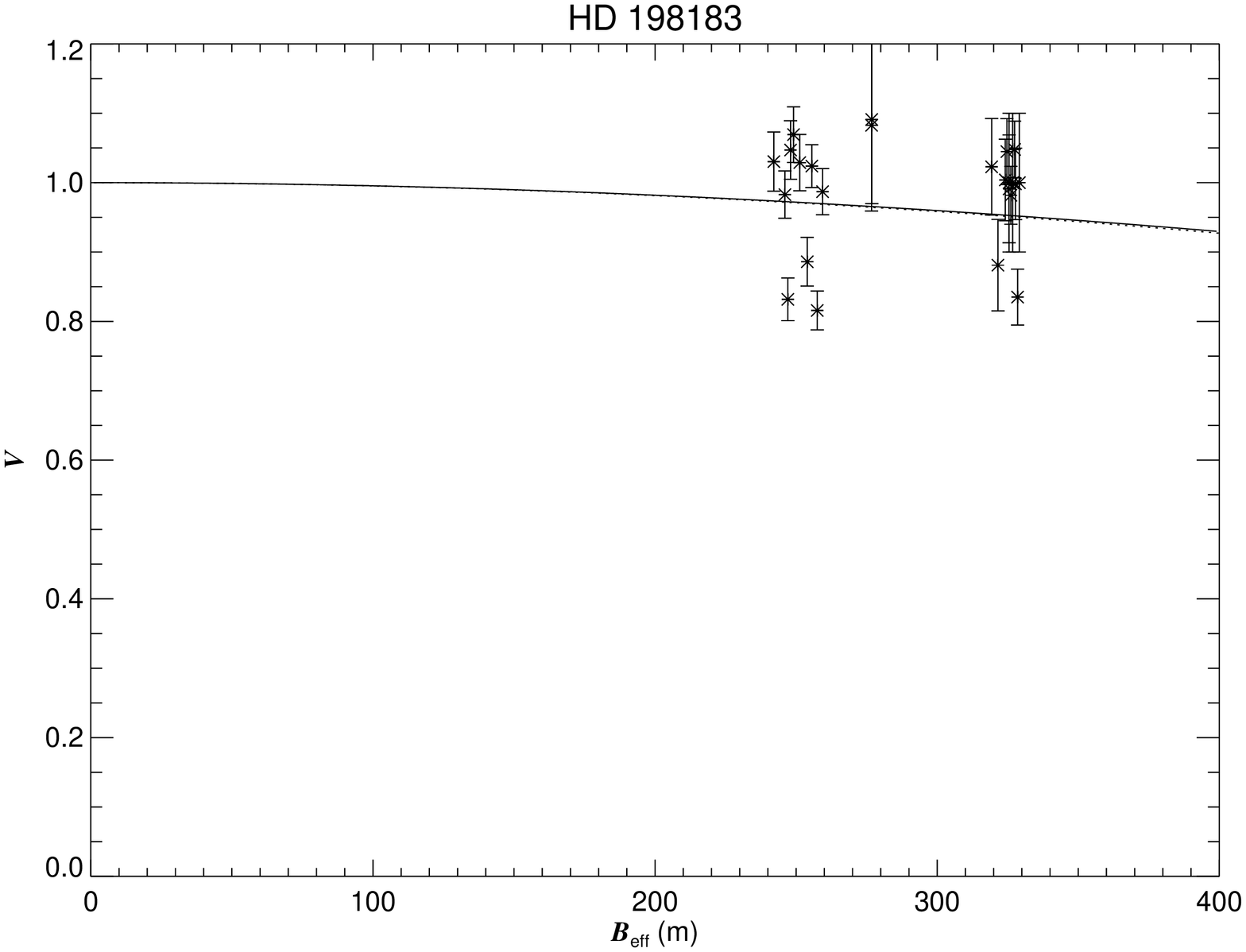}}
{\includegraphics[angle=90,height=5cm]{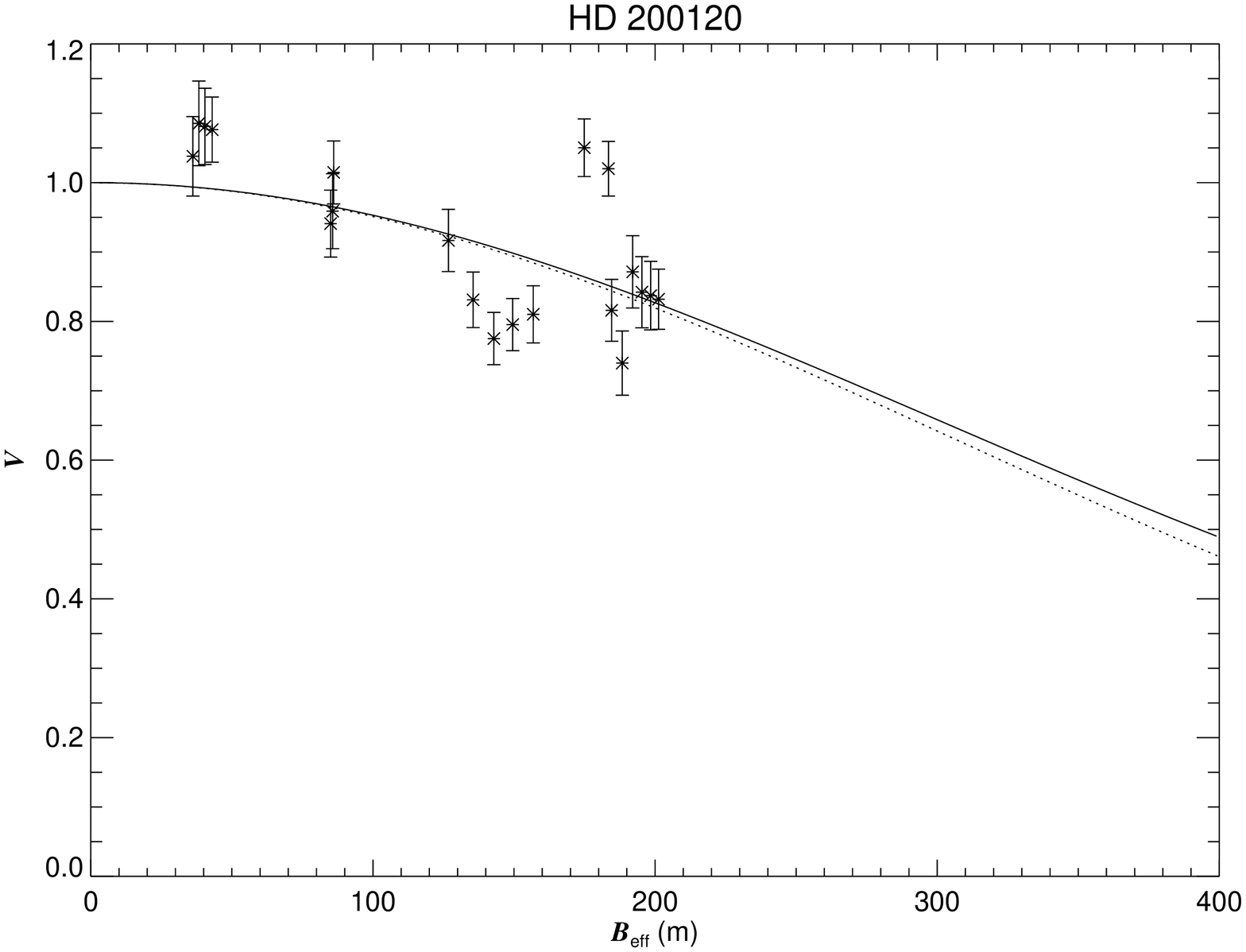}}
{\includegraphics[angle=90,height=5cm]{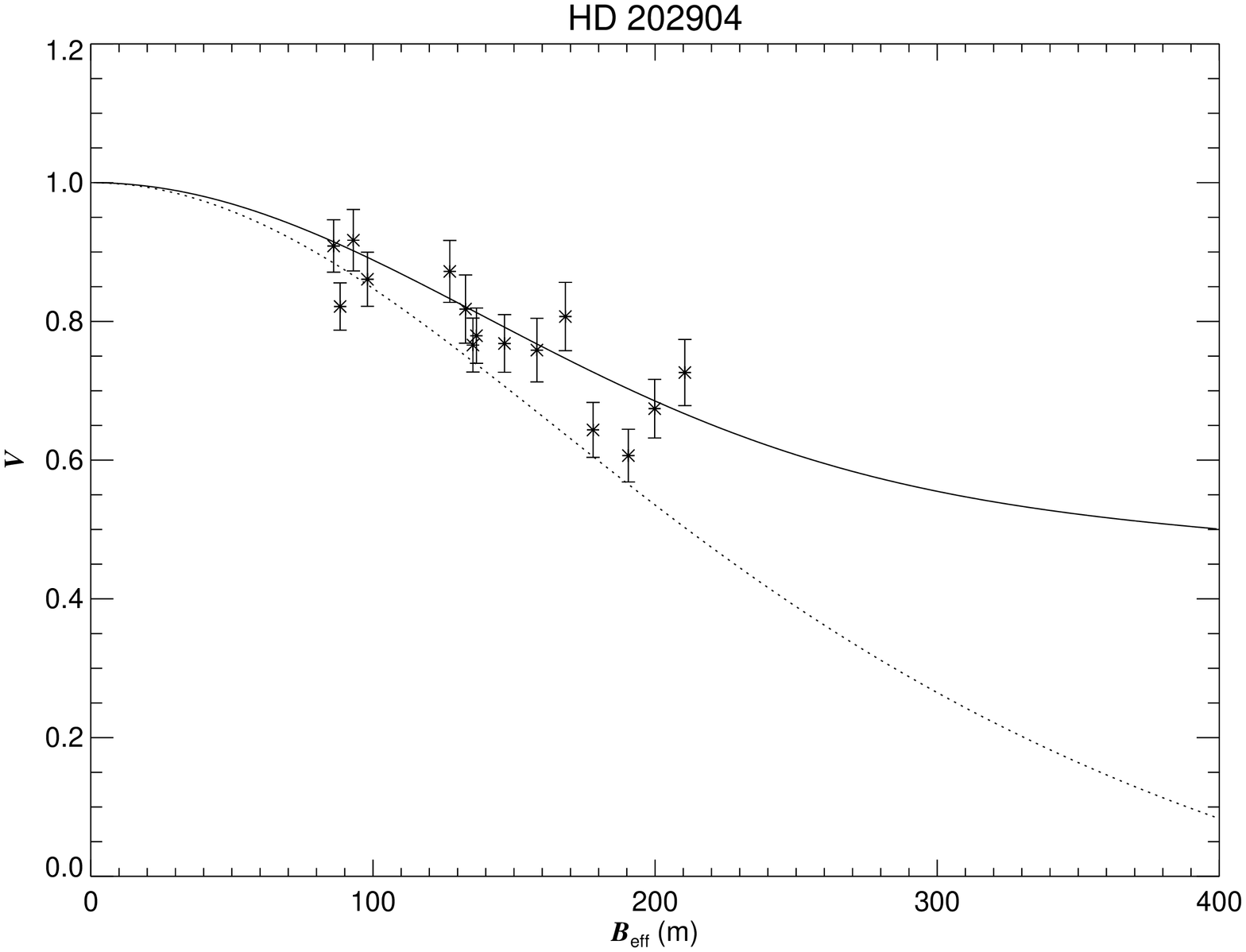}}
{\includegraphics[angle=90,height=5cm]{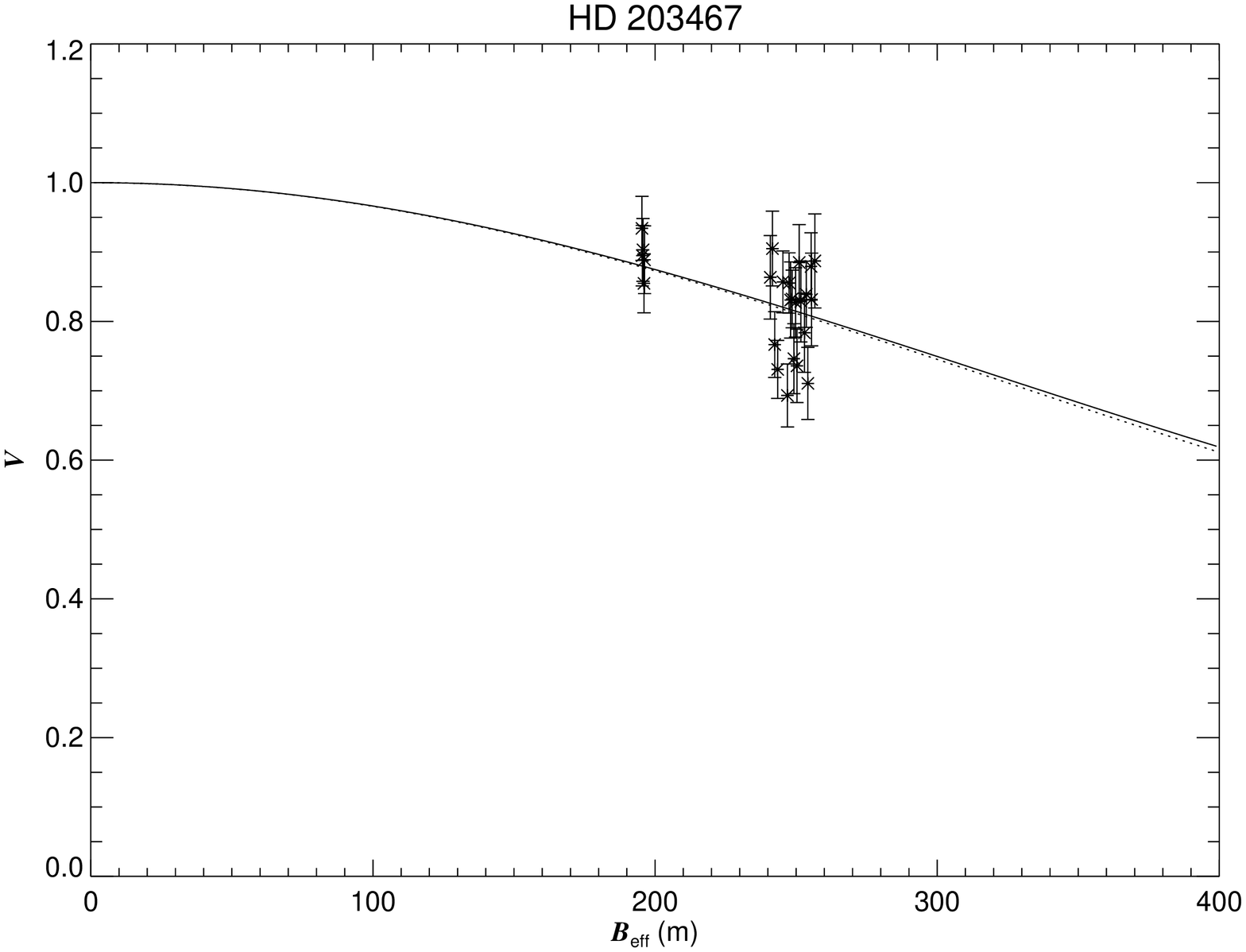}}
{\includegraphics[angle=90,height=5cm]{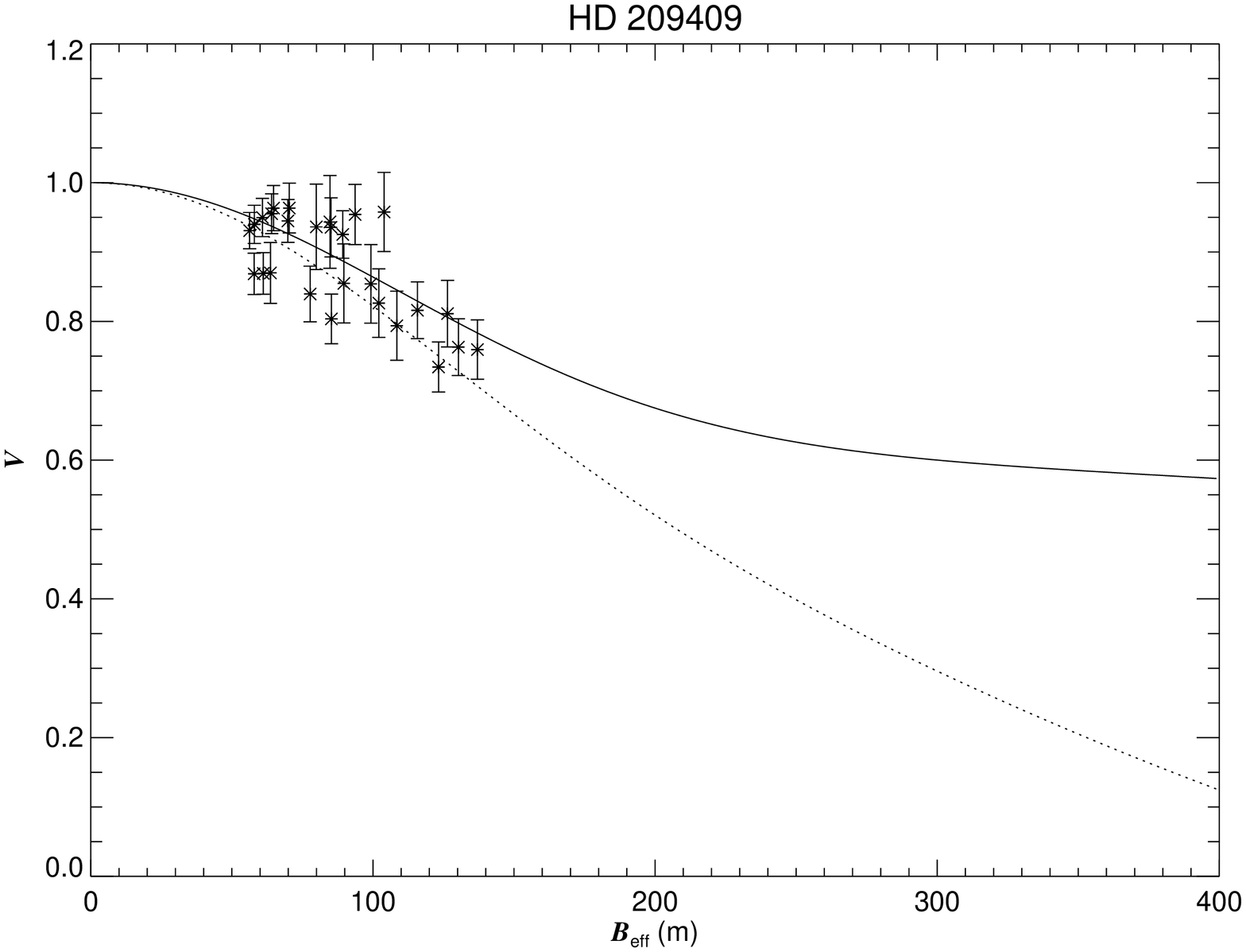}}
{\includegraphics[angle=90,height=5cm]{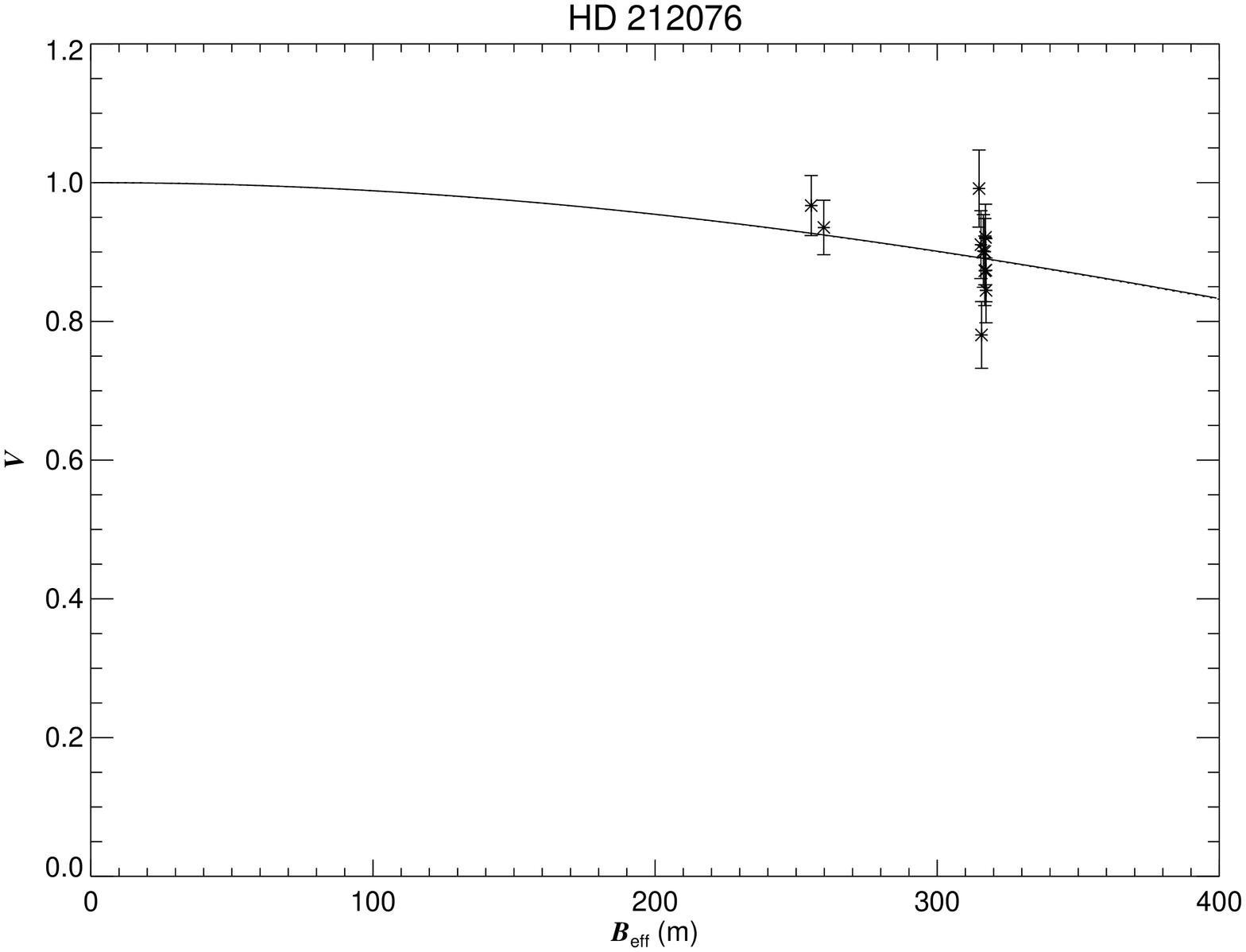}}
{\includegraphics[angle=90,height=5cm]{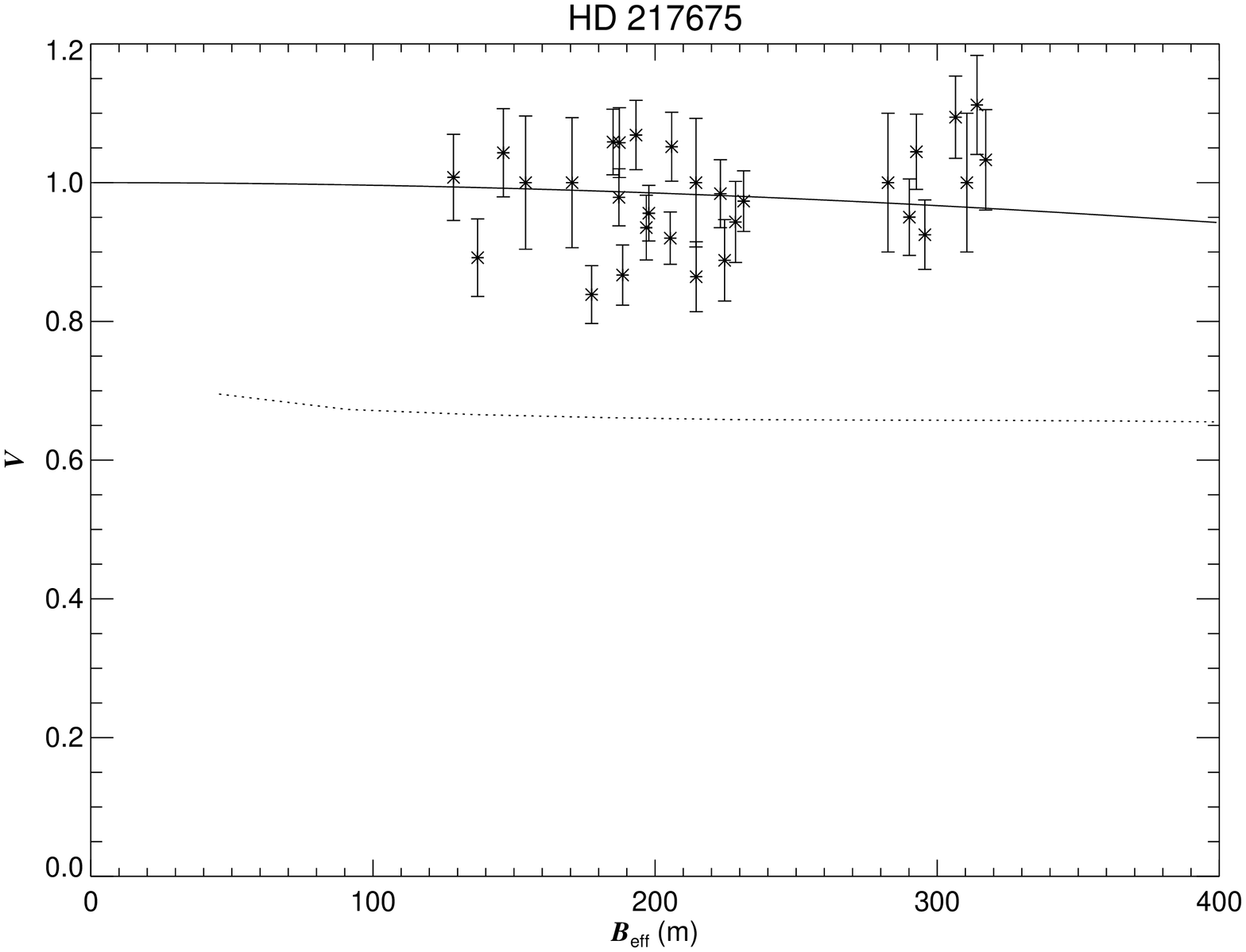}}
{\includegraphics[angle=90,height=5cm]{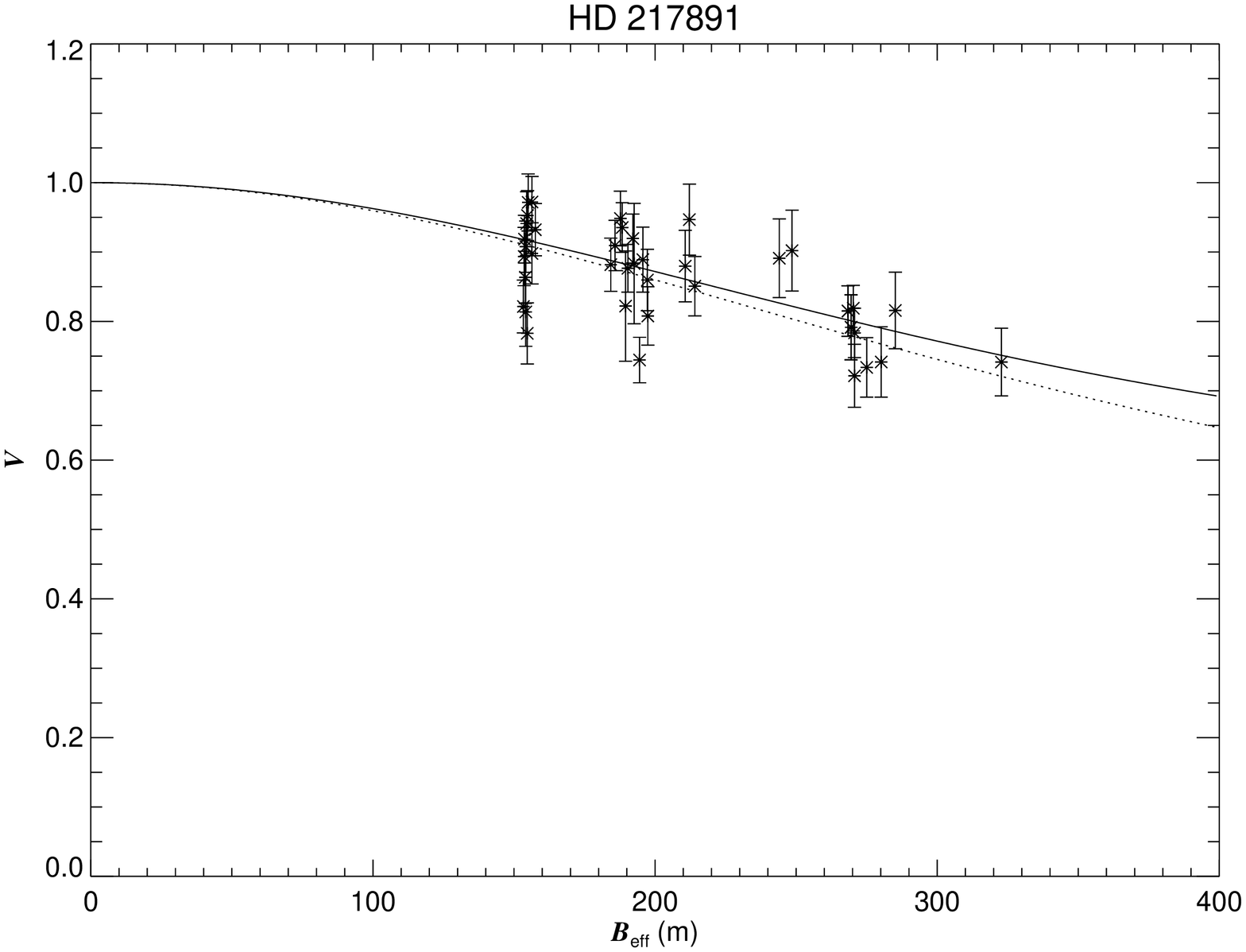}}
\end{center}
\caption[Calibrated visibilities versus baseline]
{8.17 -- 8.24. Calibrated visibilities versus the effective baseline. The solid line and the dotted
lines represent the Gaussian elliptical model along the major and minor axes, respectively,
and the star signs represent the interferometric data.}
\label{vis3}
\end{figure*}

\clearpage

\begin{figure*}
\begin{center}
{\includegraphics[angle=90,height=10cm]{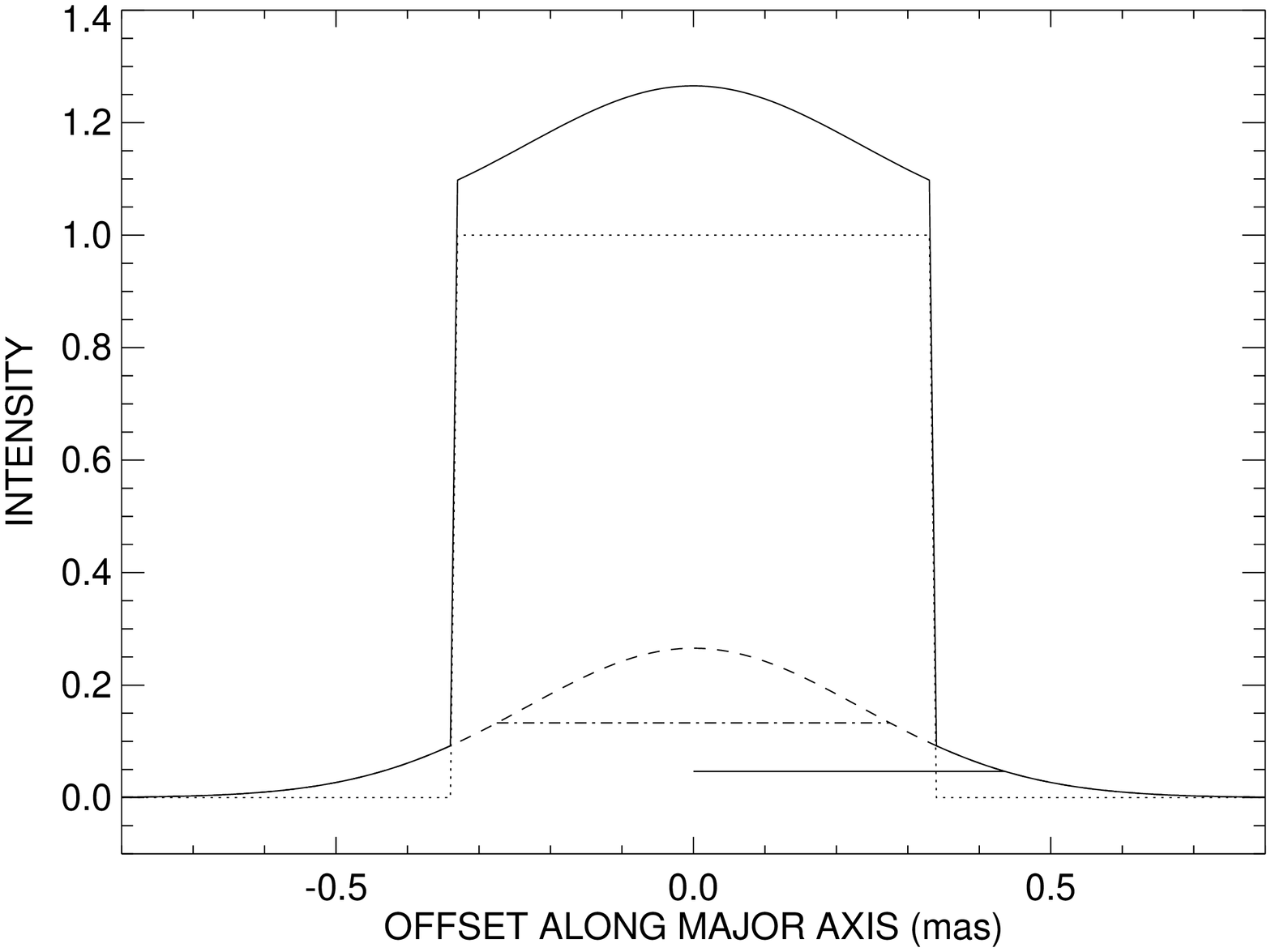}}
\end{center}
\caption[Model intensity profile for a faint disk case]{A set of intensity
profiles for a faint disk case. The diagram shows the stellar
(uniform disk) flux (dotted line), the Gaussian circumstellar disk
flux (dashed line), and their sum (upper solid line).  In this case
the Gaussian FWHM (indicated by the dashed-dotted line) is smaller
than the stellar diameter, and the revised circumstellar disk
radius (from eq.\ 21) is shown as lower solid line on the right side.}
\label{gaucorr}
\end{figure*}


\begin{figure*}
\begin{center}
{\includegraphics[angle=90,height=10cm]{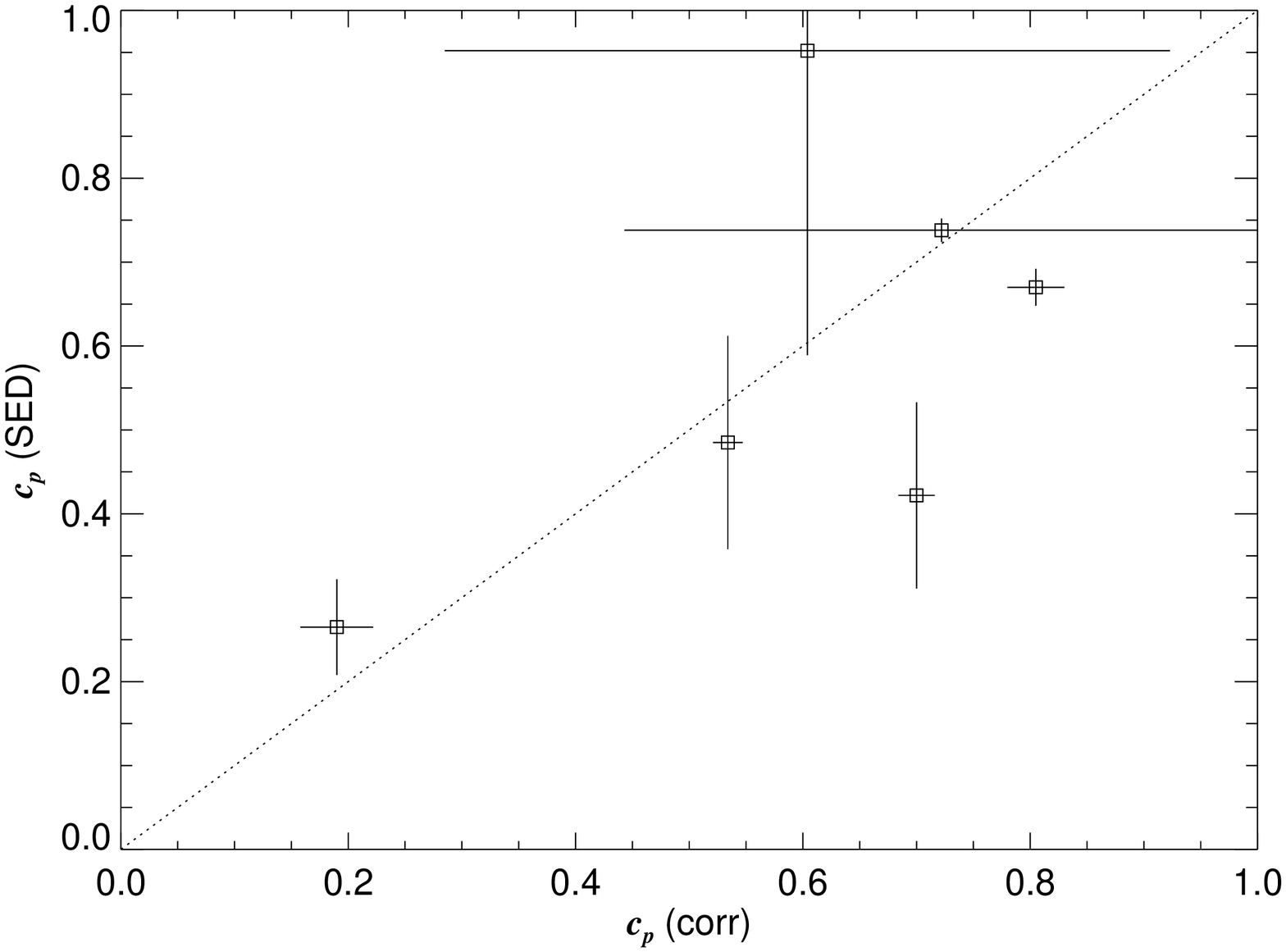}}
\end{center}
\caption[A comparison between the values of the photospheric contribution
$c_p$ derived from the SEDs and interferometry]
{A comparison between the values of the $K$-band photospheric contribution
$c_p$(SED) derived from the SED fits and those from the Gaussian
elliptical fits $c_p$(corr).}
\label{cpcomp}
\end{figure*}


\clearpage
\begin{figure*}
\begin{center}
{\includegraphics[angle=90,height=10cm]{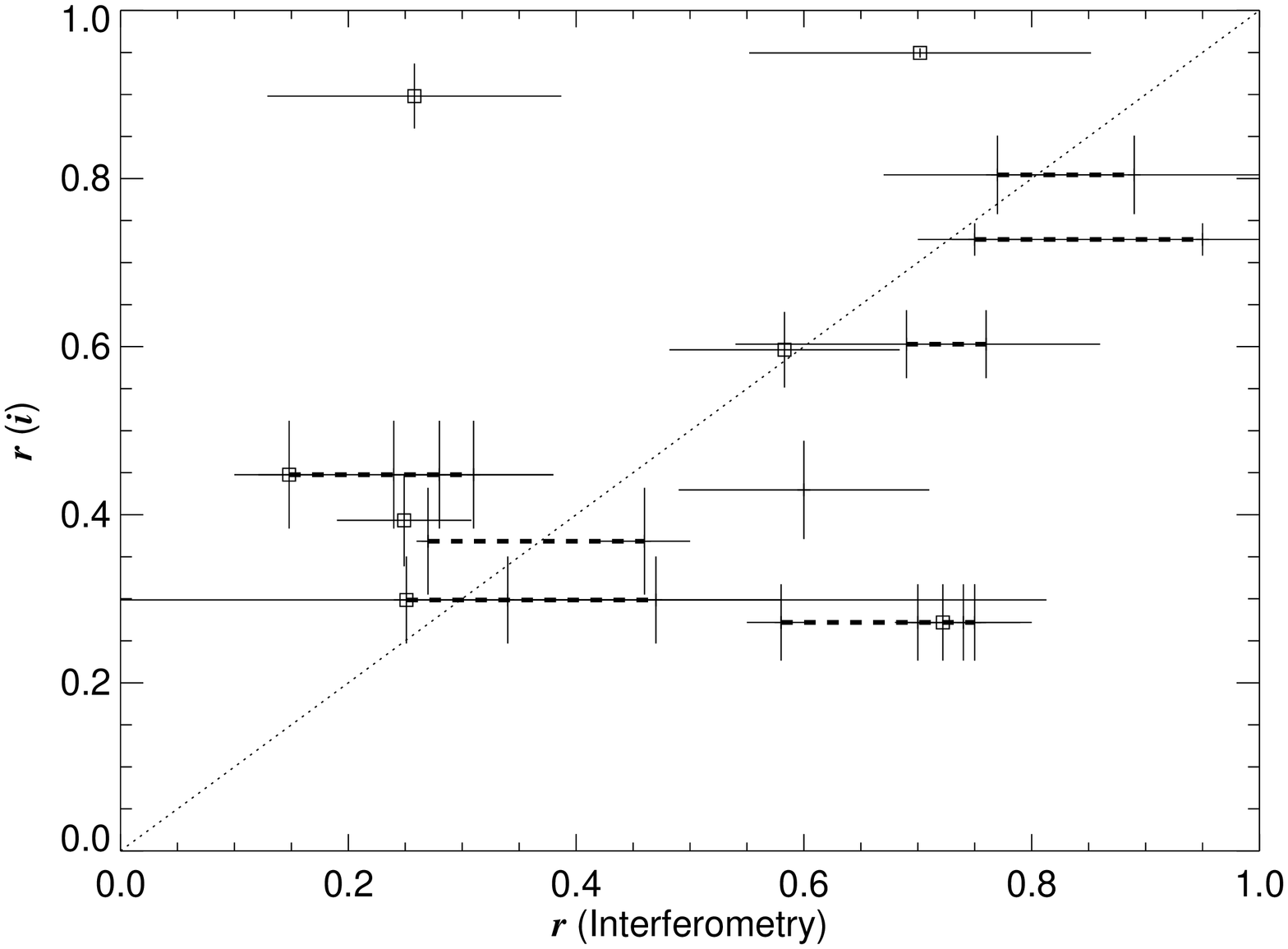}}
\end{center}
\caption[A comparison between $r$ fixed values and derived from interferometry]
{A comparison between the values of the disk axial ratio $r(i)$ adopted from
the stellar inclinations given by \citet{fre05} and those derived from
interferometry (our estimates are indicated by square symbols while the
rest are from prior work listed in Table~8).  Thick dashed lines connect
the various estimates from interferometry for the same star.}
\label{rcomp}
\end{figure*}


\begin{figure*}
\begin{center}
{\includegraphics[angle=90,height=10cm]{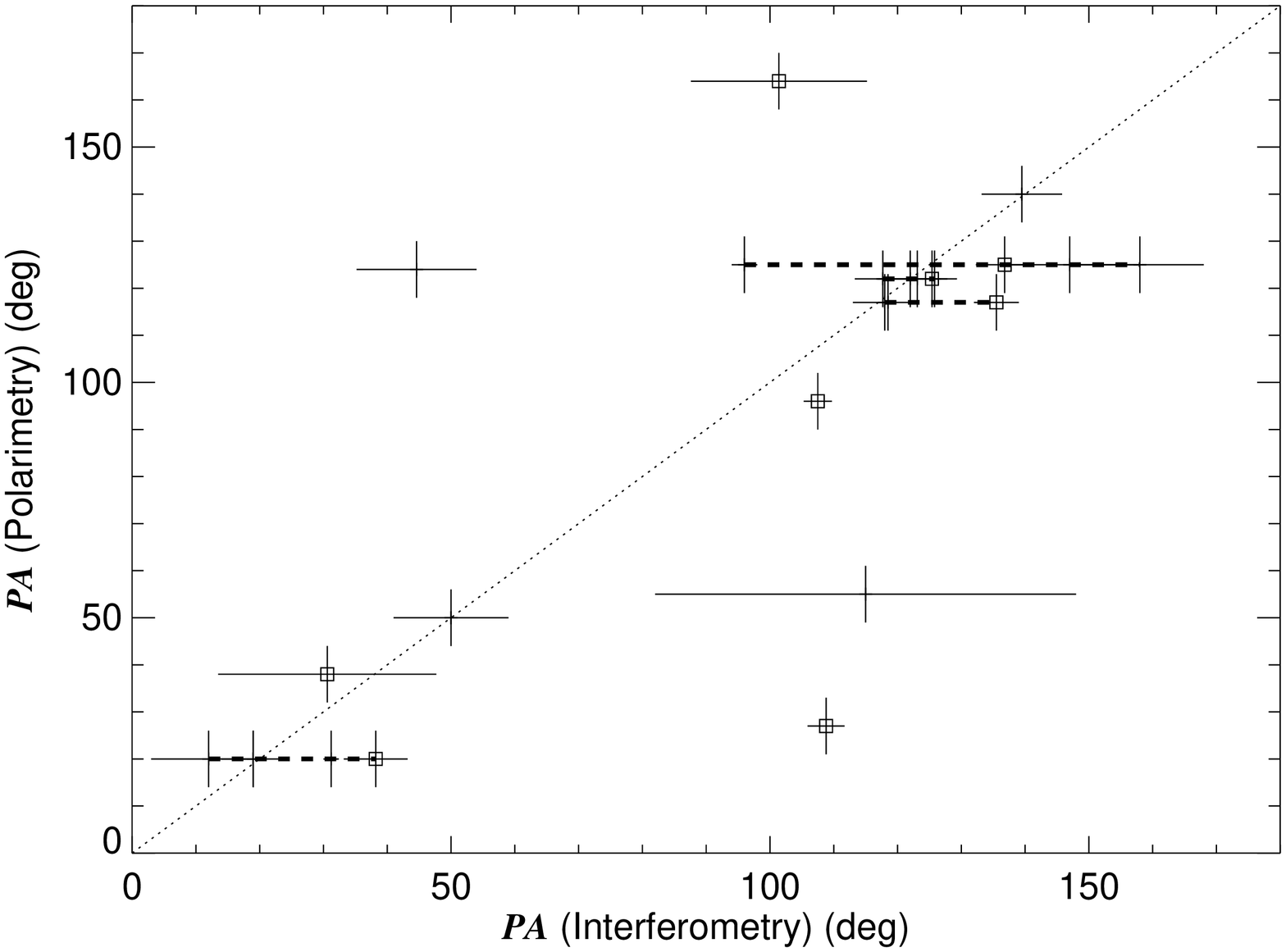}}
\end{center}
\caption[A comparison between the values of the disk position angle]
{A comparison between the values of the disk long axis position angle
$PA$ adopted from the intrinsic polarization angle plus 90$^\circ$
(\citealt{mcd99}; Yudin et al.\ 2001) and those derived from
interferometry  (our estimates are indicated by square symbols while the
rest are from prior work listed in Table~8).  Thick dashed lines connect
the various estimates from interferometry for the same star.}
\label{pacomp}
\end{figure*}


\clearpage
\begin{figure*}
\begin{center}
{\includegraphics[angle=90,height=10cm]{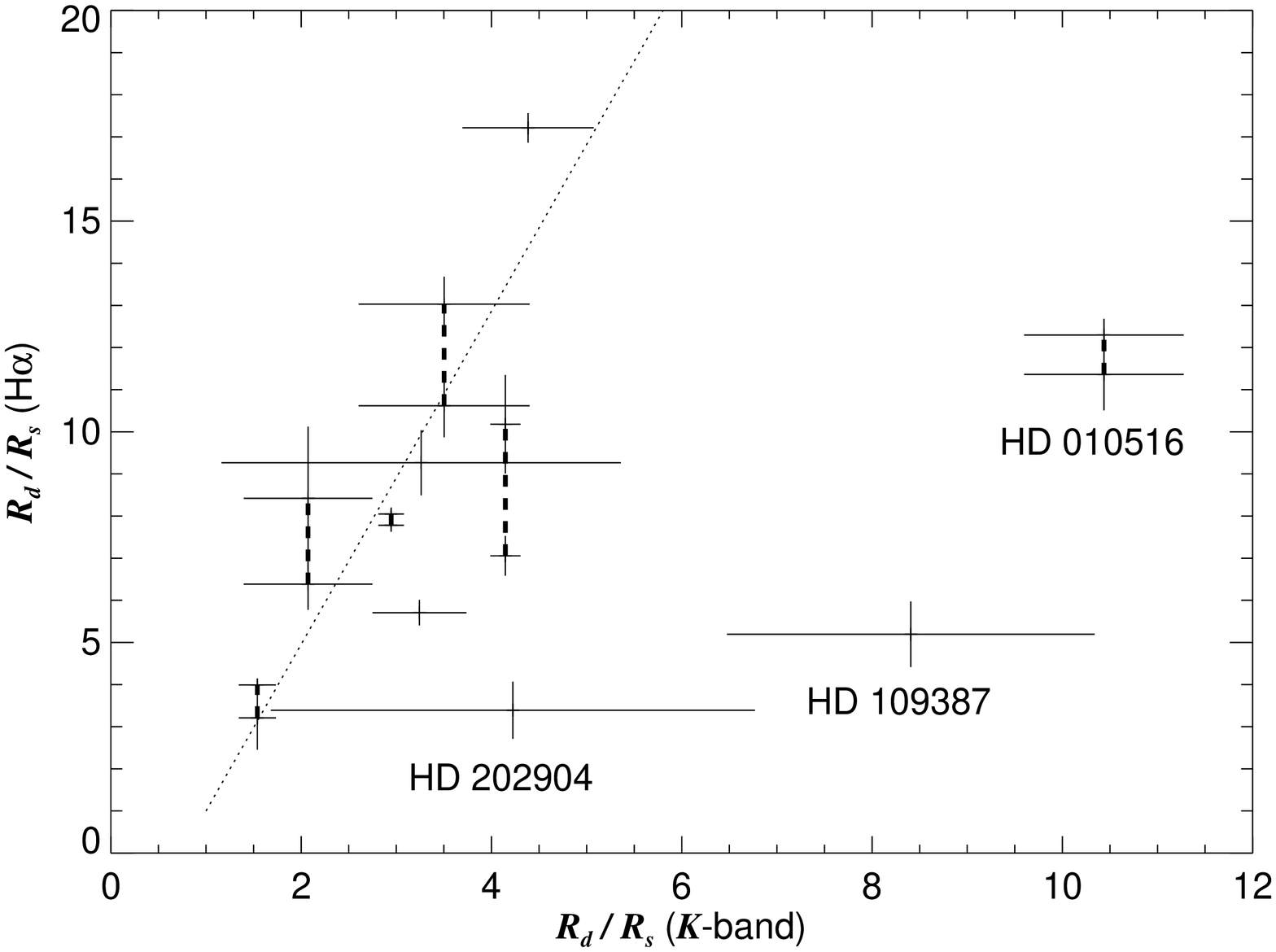}}
\end{center}
\caption[A comparison of the $K$-band disk angular diameters with
the H$\alpha$ angular diameters]{A comparison of the $K$-band disk
sizes with the H$\alpha$ disk sizes ($R_d / R_s = \theta_{\rm maj} / \theta_s$
with $\theta_{\rm maj}$ from Table~8 and $\theta_s$ from Table~7).
The dotted line represents a linear fit to the data with a slope of 4.0.
Thick dashed lines connect multiple H$\alpha$ measurements
for the same star.}
\label{Kha}
\end{figure*}


\clearpage
\begin{figure*}
\begin{center}
{\includegraphics[angle=90,height=10cm]{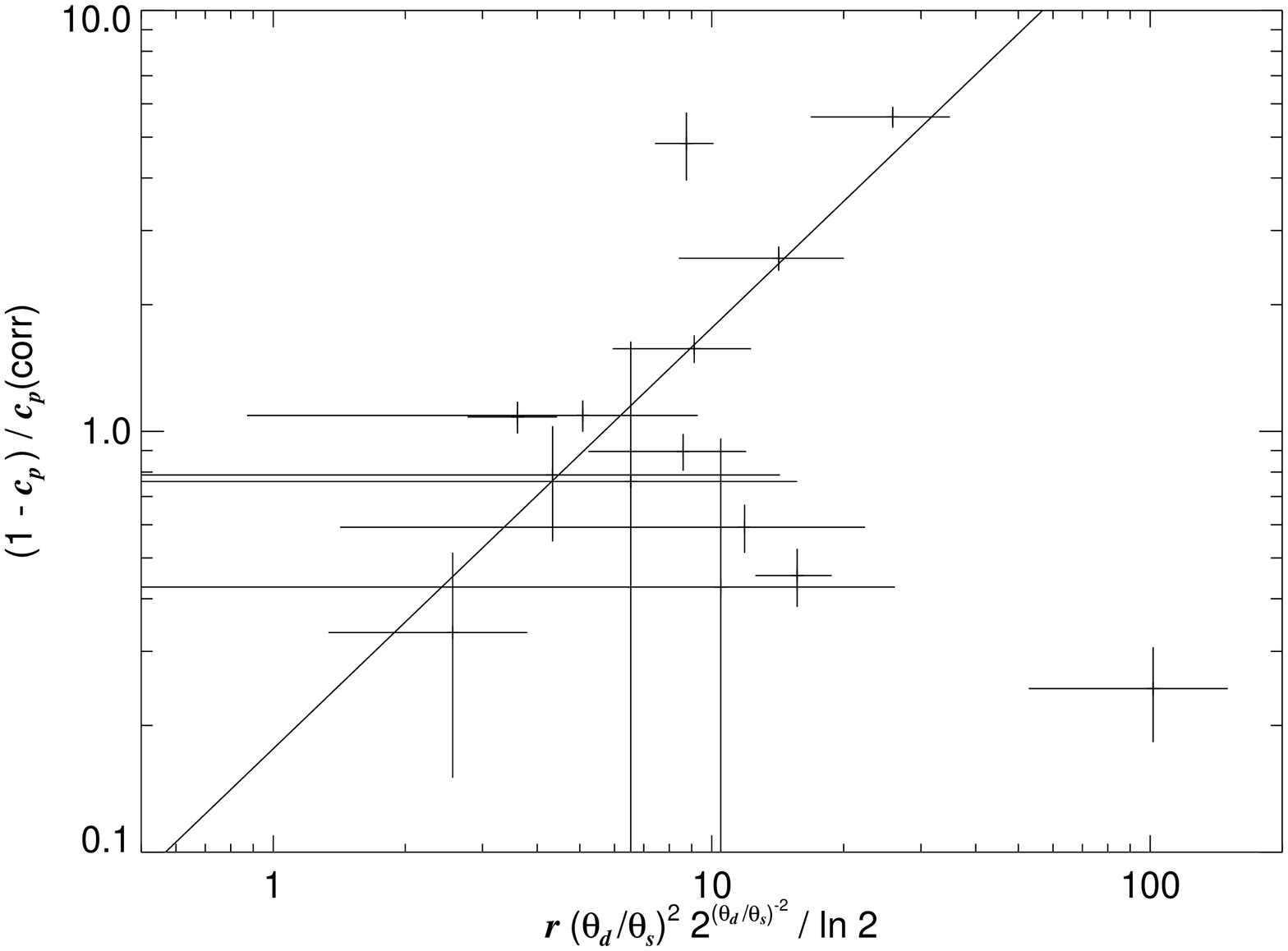}}
\end{center}
\caption[A plot of the ratio of projected disk to stellar
area on the sky as a function of the flux ratio $F_d/F_s$]
{A plot of the approximate disk to star flux ratio
$F_d/F_s = (1 - c_p)/c_p({\rm corr})$ as a function of
a product related to the projected disk to stellar area on the sky.
A linear relationship (unit slope line) may exist for large
optically thick disks.  The scatter is largest among faint disk systems
(the lower, right point represents $\kappa$~Dra = HD~109387).}
\label{fxratio}
\end{figure*}

\clearpage
\begin{figure*}
\begin{center}
{\includegraphics[angle=90,height=7cm]{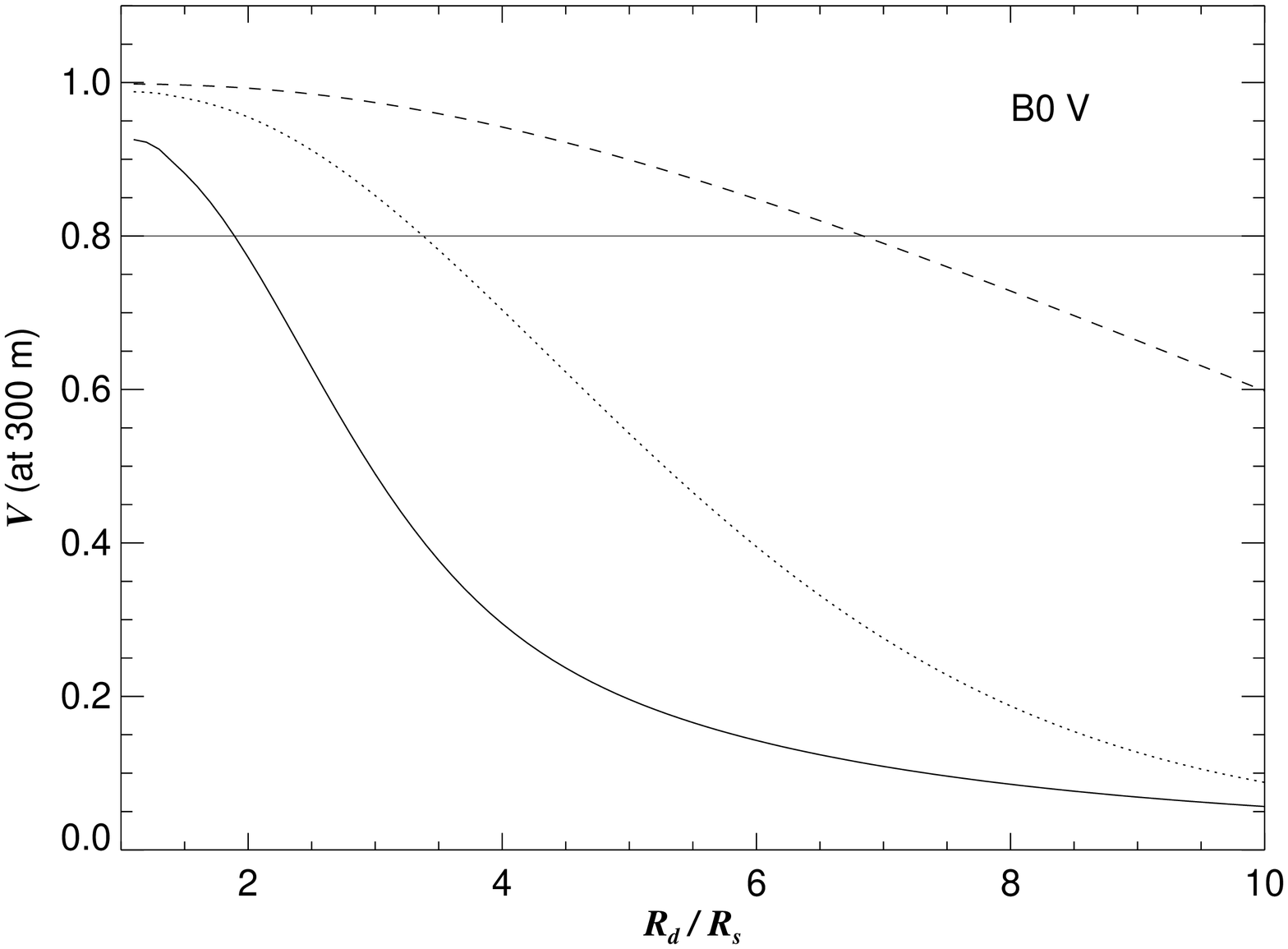}}
{\includegraphics[angle=90,height=7cm]{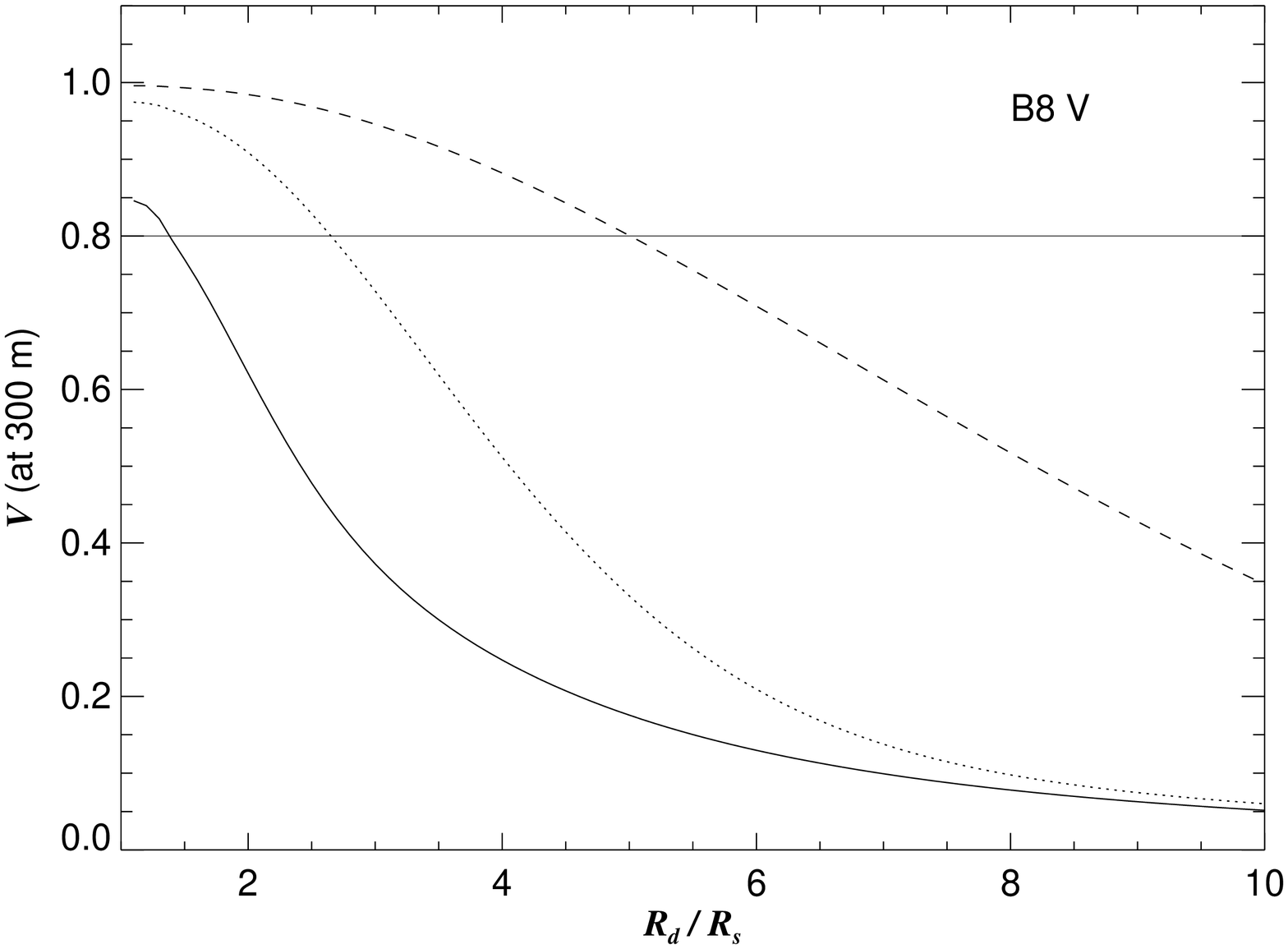}}
\end{center}
\caption[A plot of the expected calibrated visibility
measured at a baseline of 300~m for a Be star of type B0~V and B8~V]
{A plot of the expected calibrated visibility
measured at a baseline of 300~m for a Be star of
type B0~V ({\it above}) and B8~V ({\it below})
as a function of disk to stellar radius along the major axis.
The thick solid, dotted, and dashed lines correspond to
predictions for a star of visual magnitude 3, 5, and 7, respectively.
The thin horizontal line marks the $V=0.8$ criterion, and if
the visibility drops below this line then the disk is detected
with some confidence.}
\label{limit1}
\end{figure*}





\begin{deluxetable}{lcccccc}
\tabletypesize{\scriptsize}
\tablewidth{0pt}
\tablenum{1}
\tablecaption{Adopted Stellar Parameters\label{param}}
\tablehead{

\colhead{HD} &
\colhead{Star} &
\colhead{Spectral} &
\colhead{$T_{\rm eff}$} &
\colhead{$\log g$} &
\colhead{$PA$}&
\colhead{}
\\
\colhead{Number} &
\colhead{Name} &
\colhead{Class.} &
\colhead{(K)} &
\colhead{(cm s$^{-2}$)} &
\colhead{(deg)}&
\colhead{$r$}
}
\startdata

HD 004180 & $o$ Cas      & B2 Ve    & 14438 & 3.284& 164     & 0.582\\
HD 005394 & $\gamma$ Cas & B0 IVe   & 26431 & 3.800& \phn 20 & 0.764\\
HD 010516 & $\phi$ Per   & B0.5 IVe & 25556 & 3.899& 117     & 0.322\\
HD 022192 & $\psi$ Per   & B4.5 Ve  & 15767 & 3.465& 125     & 0.280\\
HD 023630 & $\eta$ Tau   & B5 IIIe  & 12258 & 3.047& 124     & 0.727\\
HD 023862 & $28$ Tau     & B8 Vpe   & 12106 & 3.937& 159     & 0.438\\
HD 025940 & 48 Per       & B4 Ve    & 16158 & 3.572& \phn 55 & 0.798\\
HD 037202 & $\zeta$ Tau  & B1 IVe   & 19310 & 3.732& 122     & 0.071\\
\enddata
\end{deluxetable}


\newpage
\begin{deluxetable}{lcccccccccc}
\tabletypesize{\scriptsize}
\tablewidth{0pt}
\rotate
\tablenum{2}
\tablecaption{Calibrator star angular diameters \label{calib}}
\tablehead{

\colhead{Calibrator}&
\colhead{Object}&
\colhead{$T_{\rm eff}$}&
\colhead{}&
\colhead{$\log g$}&
\colhead{}&
\colhead{Spectral}&
\colhead{$E(B-V)$}&
\colhead{}&
\colhead{$\theta_{\rm LD}$}&
\colhead{$\theta_{\rm UD}$}
\\
\colhead{HD Number}&
\colhead{HD Number}&
\colhead{(K)}	&
\colhead{Ref.} &
\colhead{(cm~s$^{-2}$)}&
\colhead{Ref.}&
\colhead{Class.}&
\colhead{(mag)}&
\colhead{Ref.}&
\colhead{(mas)}&
\colhead{(mas)}
\\
\colhead{(1)}&
\colhead{(2)}&
\colhead{(3)}&
\colhead{(4)}&
\colhead{(5)}&
\colhead{(6)}&
\colhead{(7)}&
\colhead{(8)}&
\colhead{(9)}&
\colhead{(10)}&
\colhead{(11)}
}
\startdata
HD 004222 &HD 004180 &\phn 8970 &   1 & 4.20  &   2 & A2 V   & 0.009 & 3 & $0.303\pm0.017$ & $0.301 \pm 0.021$\\
HD 006210 &HD 005394 &\phn 6065 &   4 & 3.86  &   4 & F6 V   & 0.009 & 5 & $0.516\pm0.025$ & $0.506 \pm 0.035$\\
HD 011151 &HD 010516 &\phn 6405 &   6 & 4.00  &   6 & F5 V   & 0.003 & 7 & $0.424\pm0.011$ & $0.416 \pm 0.029$\\
HD 020675 &HD 022192 &\phn 6577 &   6 & 4.28  &   6 & F6 V   & 0.006 & 7 & $0.418\pm0.018$ & $0.417 \pm 0.028$\\
HD 024167 &HD 023630 &\phn 8200 &   1 & 4.01  &   8 & A5 V   & 0.016 & 8 & $0.246\pm0.015$ & $0.243 \pm 0.017$\\
HD 024357 &HD 023862 &\phn 6890 &   1 & 4.30  &   2 & F4 V   & 0.008 & 8 & $0.374\pm0.014$ & $0.368 \pm 0.026$\\
HD 025948 &HD 025940 &\phn 6440 &   1 & 4.07  &   8 & F5 V   & 0.008 & 7 & $0.374\pm0.014$ & $0.367 \pm 0.025$\\
HD 037147 &HD 037202 &\phn 7200 &   1 & 4.13  &   8 & F0 V   & 0.003 & 8 & $0.403\pm0.021$ & $0.397 \pm 0.028$\\
\enddata
\tablerefs{
1. \citet{wri03};
2. \citet{laf10};
3. from $B-V$ and spectral classification;
4. \citet{bal90};
5. \citet{kar06};
6. \citet{val05};
7. \citet{amm06};
8. \citet{phi80};
9. \citet{das11};
10. \citet{sch09};
11. \citet{edv93};
12. \citet{pru01};
13. \citet{gra01};
14. \citet{mas06};
15. \citet{all99};
16. \citet{boe90};
17. \citet{boe86};
18. \citet{ger07};
19. \citet{fuh98}
}
\end{deluxetable}



\begin{deluxetable}{lcccccccccc}
\tabletypesize{\scriptsize}
\tablewidth{0pt}
\rotate
\tablenum{3}
\tablecaption{Calibrated Visibilities \label{calibvis}}
\tablehead{

\colhead{HD} &
\colhead{Date} &
\colhead{Telescope} &
\colhead{$u$} &
\colhead{$v$} &
\colhead{Baseline} &
\colhead{Effective} &
\colhead{} &
\colhead{} &
\colhead{} &
\colhead{}
\\
\colhead{Number} &
\colhead{(HJD$-$2400000)} &
\colhead{Pair} &
\colhead{(cycles~arcsec$^{-1}$)} &
\colhead{(cycles~arcsec$^{-1}$)} &
\colhead{(m)} &
\colhead{baseline (m)} &
\colhead{$V$} &
\colhead{$\delta V$}&
\colhead{$V_{c}$} &
\colhead{$\delta V_{c}$}
\\
\colhead{(1)} &
\colhead{(2)} &
\colhead{(3)} &
\colhead{(4)} &
\colhead{(5)} &
\colhead{(6)} &
\colhead{(7)} &
\colhead{(8)} &
\colhead{(9)} &
\colhead{(10)} &
\colhead{(11)}
}
\startdata
HD 004180 &       54756.622&S1/W1&\phn       $-$75.187&       $-$627.643&      278.114&     162.970&    0.705&    0.042& 0.689& 0.046\\
HD 004180 &       54756.629&S1/W1&\phn       $-$44.906&       $-$630.398&      278.055&     164.379&    0.693&    0.042& 0.838& 0.045\\
HD 004180 &       54756.637&S1/W1&\phn       $-$15.062&       $-$631.747&      278.024&     166.438&    0.821&    0.047& 0.826& 0.051\\
HD 004180 &       54756.649&S1/W1&\phs\phn      19.818&       $-$631.622&      278.027&     169.637&    0.736&    0.044& 0.865& 0.048\\
HD 004180 &       54756.661&S1/W1&\phs\phn      52.167&       $-$629.866&      278.067&     173.303&    0.767&    0.048& 0.793& 0.053\\
HD 004180 &       54757.610&S1/W1&\phn       $-$91.896&       $-$625.513&      278.157&     162.450&    0.812&    0.045& 0.863& 0.049\\
HD 004180 &       54757.626&S1/W1&\phn       $-$53.847&       $-$629.731&      278.070&     163.890&    0.569&    0.032& 0.670& 0.035\\
HD 004180 &       54757.633&S1/W1&\phn       $-$21.944&       $-$631.555&      278.029&     165.906&    0.724&    0.041& 0.762& 0.045\\
\enddata
\end{deluxetable}



\begin{deluxetable}{lcccccccc}
\tabletypesize{\scriptsize}
\tablewidth{0pt}
\tablenum{4}
\tablecaption{Binary Star Properties\label{binary}}
\tablehead{

\colhead{HD} &
\colhead{Number of} &
\colhead{Speckle} &
\colhead{Binary} &
\colhead{$P$}&
\colhead{$a$}&
\colhead{$\Delta K$}&
\colhead{}&
\colhead{}
\\
\colhead{Number} &
\colhead{Components} &
\colhead{Obs. Ref.} &
\colhead{Designation} &
\colhead{(y)} &
\colhead{(mas)} &
\colhead{(mag)} &
\colhead{Corr.}&
\colhead{Ref.}
}
\startdata
HD 004180 &  3  &  1      & A,B     & 2.824   &    17   &  2.6       &Y&\phn 2 \\
          &     &         & Ba,Bb   & 0.01:   &  0.4:   &  0:        &Y&\phn 3 \\
HD 005394 &  3  &  1      & A,B     & 1800:   & 2070:   &  6.5:      &N&\phn 4 \\
          &     &         & Aa,Ab   & 0.557   &  8.2:   & 3.5 -- 5.5 &N&  5,6  \\
HD 010516 &  2  &  1      & A,B     & 0.347   &  4.9    &  3.1       &Y&\phn 7 \\
HD 022192 &  1  &  1      & \nodata & \nodata & \nodata & \nodata    &N&\phn 8 \\
HD 023630 &  1  &  1      & \nodata & \nodata & \nodata & \nodata    &N&\phn 4 \\
HD 023862 &  3  &  9      & Aa,Ab   &  35:    &  150:   &  2.1:      &Y& 10    \\
\enddata
\tablerefs{
1. \citet{mas97};
2. \citet{kou10};
3. \citet{gru07};
4. \citet{rob07};
5. \citet{har00};
6. \citet{mir02};
7. \citet{gie07};
8. \citet{del11};
9. \citet{mas93};
10. \citet{lut94};
11. \citet{nem10};
12. \citet{ruz09};
13. \citet{saa05};
14. \citet{mas09};
15. \citet{fab00};
16. \citet{kou97};
17. \citet{riv06};
18. \citet{har87};
19. \citet{tok10};
20. \citet{sta82};
21. \citet{bal88};
22.  B. D. Mason, priv. communication;
23. \citet{mai05};
24. \citet{nei05};
25. \citet{kou03};
26. \citet{oud10};
27. \citet{riv03};
28. \citet{zhu10};
29. \citet{dac86}.
}
\end{deluxetable}



\begin{deluxetable}{lcccccccccc}
\tabletypesize{\scriptsize}
\tablewidth{0pt}
\rotate
\tablenum{5}
\tablecaption{Adopted Binary Orbital Elements\label{orbital}}
\tablehead{

\colhead{HD}  &
\colhead{$P$} &
\colhead{$e$} &
\colhead{$T$} &
\colhead{$\omega$}&
\colhead{$a$} &
\colhead{$i$} &
\colhead{$\Omega$}&
\colhead{$\triangle K_{\rm obs}$}&
\colhead{}
\\
\colhead{Number}&
\colhead{(d)}&
\colhead{}&
\colhead{(HJD)}&
\colhead{(deg)}&
\colhead{(mas)}&
\colhead{(deg)}&
\colhead{(deg)}&
\colhead{(mag)}&
\colhead{Ref.}
}
\startdata
HD 004180 &\phn\phn   1031.550  & 0.00 & 2452792.200 &   180.0 &\phn   17.00 &   115.0&   267   & 2.6   & \citet{kou10} \\
HD 010516 &\phn\phn\phn126.673  & 0.00 & 2450091.770 &   180.0 &\phn\phn4.90 &\phn72.5&   117   & 3.1   & \citet{gie98} \\
HD 023862 &\phn      12615.000  & 0.90 & 2453629.000 &   115.0 &      150.47 &   138.0&   357   & 2.1   & This work \\
HD 138749 &\phn      73048.440  & 0.50 & 2433000.000 &\phn30.0 &      581.61 &\phn98.0&\phn21   & 1.8   & This work \\
HD 164284 &\phn\phn   9861.539  & 0.40 & 2451150.000 &   165.0 &      103.23 &\phn52.0&   180   & 2.1   & This work \\
HD 198183 &\phn\phn   4272.603  & 0.52 & 2444797.000 &   253.5 &\phn   48.70 &   135.1&   119   & 0.7   & \citet{bai93}, this work \\
HD 198183 &         168559.275  & 0.35 & 2378057.800 &   322.0 &      770.00 &   146.3&   144   & 1.2   & \citet{bai83} \\
HD 200120 &\phn      58973.320  & 0.26 & 2460416.414 &   265.5 &      207.60 &   145.8&   205   & 2.6   & B.\ Mason, priv.\ comm.\  \\
\enddata
\end{deluxetable}



\begin{deluxetable}{lcccccccc}
\tabletypesize{\scriptsize}
\tablewidth{0pt}
\rotate
\tablenum{6}
\tablecaption{SED fits of B${\rm e}$ Stars\label{SEDinfo}}
\tablehead{

\colhead{HD}&
\colhead{$E(B-V)$}&
\colhead{$\delta E(B-V)$}&
\colhead{$\theta_s$}&
\colhead{$\delta \theta_s$}&
\colhead{$E^{\star}(UV-K)$}&
\colhead{$\delta E^{\star}(UV-K)$}&
\colhead{}&
\colhead{}
\\
\colhead{Number}&
\colhead{(mag)}&
\colhead{(mag)}&
\colhead{(mas)}&
\colhead{(mas)}&
\colhead{(mag)}&
\colhead{(mag)}&
\colhead{$c_p$}&
\colhead{$\delta c_p$}
}
\startdata
HD 004180 &0.118 &0.008 &0.333 &0.009 &0.070 &0.016 &0.938 &0.014 \\
HD 005394 &0.096 &0.008 &0.446 &0.012 &1.442 &0.230 &0.265 &0.057 \\
HD 010516 &0.162 &0.010 &0.235 &0.008 &0.936 &0.283 &0.422 &0.111 \\
HD 022192 &0.099 &0.008 &0.307 &0.008 &0.533 &0.264 &0.612 &0.150 \\
HD 023630 &0.014 &0.008 &0.638 &0.019 &0.662 &0.234 &0.544 &0.118 \\
HD 023862 &0.017 &0.008 &0.229 &0.006 &0.547 &0.023 &0.604 &0.013 \\
HD 025940 &0.104 &0.008 &0.329 &0.009 &0.671 &0.312 &0.539 &0.157\\
HD 037202 &0.044 &0.009 &0.445 &0.015 &0.785 &0.280 &0.485 &0.127 \\
\enddata
\end{deluxetable}


\begin{deluxetable}{lccccccccccccccc}
\tabletypesize{\scriptsize}
\tablewidth{0pt}
\rotate
\tablenum{7}
\tablecaption{Gaussian Elliptical Fits of the Interferometric Visibilities\label{visfit}}
\tablehead{

\colhead{HD} &
\colhead{} &
\colhead{} &
\colhead{$PA$} &
\colhead{$\delta PA^{a}$} &
\colhead{} &
\colhead{} &
\colhead{$\theta_{\rm maj}$}&
\colhead{$\delta \theta_{\rm maj}$}&
\colhead{} &
\colhead{} &
\colhead{} &
\colhead{} &
\colhead{} &
\colhead{}
\\
\colhead{Number} &
\colhead{$r$} &
\colhead{$\delta r^{a}$} &
\colhead{(deg)} &
\colhead{(deg)} &
\colhead{$c_p$} &
\colhead{$\delta c_p^{a}$} &
\colhead{(mas)} &
\colhead{(mas)} &
\colhead{$\chi_{\nu}^2$}&
\colhead{$c_p(\rm corr)$}&
\colhead{$\delta c_p(\rm corr)$}&
\colhead{$R_d/R_s$} &
\colhead{$\delta R_d/R_s$} &
\colhead{Res.}
\\
\colhead{(1)} &
\colhead{(2)} &
\colhead{(3)} &
\colhead{(4)} &
\colhead{(5)} &
\colhead{(6)} &
\colhead{(7)} &
\colhead{(8)} &
\colhead{(9)} &
\colhead{(10)} &
\colhead{(11)} &
\colhead{(12)} &
\colhead{(13)} &
\colhead{(14)} &
\colhead{(15)}
}
\startdata
HD 004180 & 0.583 & 0.101 &  101.4 &     13.8 & 0.500 & 0.000 & 1.027 & 0.173 &\phn  3.03 & 0.558 & 0.019 &  3.243 & 0.493 & Y \\
HD 005394 & 0.722 & 0.038 &\phn38.2&\phn  5.0 & 0.082 & 0.036 & 1.236 & 0.063 &     15.63 & 0.190 & 0.032 &  2.946 & 0.134 & Y \\
HD 010516 & 0.100 & 0.000 &  135.5 &\phn  3.5 & 0.682 & 0.017 & 2.441 & 0.198 &\phn  8.18 & 0.700 & 0.016 & 10.437 & 0.838 & Y \\
HD 022192 & 0.251 & 0.562 &  136.8 &\phn  4.5 & 0.518 & 0.000 & 1.030 & 0.264 &\phn  1.50 & 0.612 & 0.090 &  3.502 & 0.898 & Y \\
HD 023630 & 1.000 & 0.000 &\phn00.0&\phn  0.0 & 0.448 & 0.297 & 0.091 & 0.000 &\phn  5.85 & 1.000 & 0.000 &  1.010 & 0.000 & N \\
HD 023862 & 0.438 & 0.000 &  159.0 &\phn  0.0 & 0.500 & 0.000 & 0.364 & 0.220 &\phn  3.33 & 0.715 & 0.057 &  1.879 & 2.315 & M \\
HD 025940 & 0.866 & 0.000 &  108.0 &\phn  0.0 & 0.412 & 0.000 & 0.597 & 0.245 &\phn  2.55 & 0.539 & 0.009 &  2.072 & 0.676 & Y \\
HD 037202 & 0.148 & 0.027 &  125.4 &\phn  1.3 & 0.422 & 0.016 & 1.790 & 0.073 &\phn  4.30 & 0.534 & 0.013 &  4.146 & 0.159 & Y \\
\enddata
\tablenotetext{a}{Uncertainty of zero indicates cases where the model parameter was fixed in advance of the fit.}
\end{deluxetable}



\begin{deluxetable}{lcccccccc}
\tabletypesize{\scriptsize}
\tablewidth{0pt}
\tablenum{8}
\tablecaption{Other Interferometric Results\label{OLBI}}
\tablehead{
\colhead{HD} &
\colhead{} &
\colhead{} &
\colhead{$PA$} &
\colhead{$\delta PA$} &
\colhead{$\theta_{\rm maj}$} &
\colhead{$\delta \theta_{\rm maj}$} &
\colhead{} &
\colhead{}
\\
\colhead{Number} &
\colhead{$r$} &
\colhead{$\delta r$} &
\colhead{(deg)} &
\colhead{(deg)} &
\colhead{(mas)} &
\colhead{(mas)} &
\colhead{Band}&
\colhead{Ref.}
\\
\colhead{(1)} &
\colhead{(2)} &
\colhead{(3)} &
\colhead{(4)} &
\colhead{(5)} &
\colhead{(6)} &
\colhead{(7)} &
\colhead{(8)} &
\colhead{(9)}
}
\startdata
HD 004180 &   1.00 &\nodata &\nodata  &\nodata  &\phn\phn 1.90 &\phn 0.10  & H$\alpha$ &    1 \\
HD 005394 &   0.70 & 0.02   &\phn 19  &\phn 2   &\phn\phn 3.47 &\phn 0.02  & H$\alpha$ &    2 \\
HD 005394 &   0.58 & 0.03   &\phn 31  &\phn 1   &\phn\phn 3.59 &\phn 0.04  & H$\alpha$ &    3 \\
HD 005394 &   0.74 & 0.05   &\phn 19  &\phn 5   &\phn\phn 0.76 &\phn 0.05  & $R$       &    4 \\
HD 005394 &   0.75 & 0.05   &\phn 12  &\phn 9   &\phn\phn 0.82 &\phn 0.08  & $H$       &    4 \\
HD 010516 &   0.46 & 0.04   &    118  &\phn 5   &\phn\phn 2.67 &\phn 0.20  & H$\alpha$ &    2 \\
HD 010516 &   0.27 & 0.01   &    119  &\phn 1   &\phn\phn 2.89 &\phn 0.09  & H$\alpha$ &    3 \\
HD 022192 &   0.47 & 0.11   &    147  &    11   &\phn\phn 3.26 &\phn 0.23  & H$\alpha$ &    2 \\
\enddata
\tablerefs{
1. \citet{kou10};
2. \citet{qui97};
3. \citet{tyc06};
4. \citet{ste12};
5. \citet{del11};
6. \citet{dou92};
7. \citet{tyc05};
8. \citet{tyc04};
9. \citet{sch10};
10. \citet{kra12};
11. C.\ Tycner, priv. communication;
12. \citet{stf12};
13. \citet{pot10};
14. \citet{tyc08}.
}
\end{deluxetable}


\clearpage

\begin{deluxetable}{lccccc}
\tabletypesize{\scriptsize}
\tablewidth{0pt}
\tablenum{9}
\tablecaption{Linear Rotational and Critical Velocities\label{tabq}}
\tablehead{
\colhead{HD} &
\colhead{$V \sin i$} &
\colhead{$V_{\rm crit}$} &
\colhead{$V_{\rm rot}$} &
\colhead{} &
\colhead{}
\\
\colhead{Number} &
\colhead{(km s$^{-1}$)} &
\colhead{(km s$^{-1}$)} &
\colhead{(km s$^{-1}$)} &
\colhead{$V_{\rm rot} / V_{\rm crit}$} &
\colhead{$\Omega_{\rm rot} / \Omega_{\rm crit}$}
}
\startdata
HD 004180 & 208$\pm$13 & 332$\pm$21 & 249$\pm$26 & 0.75$\pm$0.09 & 0.91$\pm$0.06 \\
HD 005394 & 441$\pm$27 & 577$\pm$36 & 599$\pm$37 & 1.04$\pm$0.09 & $\approx 1$   \\
HD 010516 & 462$\pm$33 & 590$\pm$42 & 487$\pm$ 6 & 0.82$\pm$0.06 & 0.96$\pm$0.03 \\
HD 022192 & 295$\pm$15 & 397$\pm$20 & 310$\pm$70 & 0.78$\pm$0.18 & 0.93$\pm$0.11 \\
HD 023630 & 149$\pm$ 8 & 274$\pm$15 & 216$\pm$20 & 0.79$\pm$0.09 & 0.94$\pm$0.05 \\
HD 025940 & 220$\pm$13 & 386$\pm$21 & 323$\pm$82 & 0.84$\pm$0.22 & 0.96$\pm$0.10 \\
HD 037202 & 326$\pm$ 7 & 466$\pm$13 & 333$\pm$ 6 & 0.71$\pm$0.02 & 0.89$\pm$0.02 \\
HD 058715 & 231$\pm$14 & 380$\pm$24 & 327$\pm$91 & 0.86$\pm$0.25 & 0.97$\pm$0.10 \\
HD 142983 & 407$\pm$22 & 501$\pm$28 & 484$\pm$56 & 0.97$\pm$0.12 & 1.00$\pm$0.01 \\
HD 202904 & 167$\pm$20 & 468$\pm$30 & 171$\pm$ 7 & 0.37$\pm$0.03 & 0.52$\pm$0.04 \\
HD 209409 & 282$\pm$20 & 391$\pm$27 & 288$\pm$ 4 & 0.74$\pm$0.05 & 0.90$\pm$0.04 \\
HD 217891 & 100$\pm$ 6 & 367$\pm$24 & 135$\pm$44 & 0.37$\pm$0.12 & 0.53$\pm$0.16 \\
\enddata
\end{deluxetable}

\end{document}